\documentclass[12pt, a4paper, oneside, openright, onecolumn]{report}
\renewcommand{\baselinestretch}{2}
\author{David M. Wilkins}
\title{A Theoretical Investigation Into Energy Transfer In Photosynthetic Open Quantum Systems}

\usepackage{amsmath}
\usepackage{graphicx}
\usepackage{fancyhdr}
\usepackage{cite}
\usepackage{pict2e}
\usepackage{xfrac}

\newcommand{\bra}[1]{\langle #1 |}
\newcommand{\ket}[1]{| #1 \rangle}
\newcommand{\inner}[2]{\langle #1 | #2 \rangle}
\newcommand{\commut}[1]{\hat{#1}^{\times}}
\newcommand{\anticommut}[1]{\hat{#1}^{\circ}}

\DeclareMathOperator*{\Res}{Res}

\let\tmp\oddsidemargin
\let\oddsidemargin\evensidemargin
\let\evensidemargin\tmp
\reversemarginpar

\begin{document}

\begin{titlepage}
\renewcommand{\baselinestretch}{1}
\begin{center}

\Huge{\textbf{A Theoretical Investigation Into Energy Transfer In Photosynthetic Open Quantum Systems}}

\vspace*{+7.0cm}

\Large{David M. Wilkins}\\[0.5cm]
\Large{St. Edmund Hall}\\[0.5cm]
\Large{Oxford University}

\vspace*{+7.5cm}

\emph{A thesis submitted for the Honour School of Chemistry:}\\[0.5cm]
\emph{Chemistry Part II}\\[0.5cm]
\emph{June 2012}

\end{center}
\end{titlepage}

{\renewcommand{\baselinestretch}{1}\normalsize
\section*{Summary}\label{summary}
}
\pagestyle{empty}

\noindent This thesis concerns exact and approximate treatments of electronic energy transfer in photosynthetic systems. While the methods used are completely general, their application is focused on the Fenna-Matthews-Olson (FMO) pigment-protein antenna complex, found in certain bacteria. \\[0.5cm]
The FMO complex is a trimer consisting of 24 bacteriochlorophyll (BChl) sites, each of which is coupled to a dissipative environment, which renders the energy transfer irreversible, and means that the dynamics of this transfer must be treated using techniques appropriate for open quantum systems. \\[0.5cm]
Recently, experimental evidence of quantum beating effects in the energy transfer in the FMO complex was found, and theoretical and experimental results suggested that quantum coherence might be observed at room temperature (300 K). One of the aims of this thesis is to ascertain how important coherence effects are for this transfer. \\[0.5cm]
Firstly, the Hierarchical Equations of Motion (HEOM) are introduced, which give numerically exact transfer dynamics, albeit with computational cost rising steeply with the size of the system. The efficient numerical implementation of the HEOM is discussed, and a Taylor-series integration is compared to the traditional Runge-Kutta method, the former proving to be more efficient. \\[0.5cm]
The accuracy of the HEOM can also be improved by representing the bath correlation function as a Pad\'e series instead of the more traditional Matsubara series: the former converges very rapidly compared to the latter, and allows more efficient simulation of the dynamics.\\[0.5cm]
Results are then presented for the FMO complex, both for a 7-site subsystem of the monomer and for the trimer. Numerically exact results are calculated to provide a benchmark for cheaper approximate methods including the Redfield and F\"orster theories. It is found that incoherent F\"orster theory describes the overall features of the dynamics well at 300 K, suggesting that coherence has little if any effect on energy transfer efficiency at room temperature. \\[0.5cm]
Finally, F\"orster theory is used to test the effects of two phenomena on the energy transfer dynamics in the FMO complex: that of including vibrational structure in the environment (traditionally modelled as being unstructured), and that of static disorder due to a slowly fluctuating environment. \\[0.5cm]
It is found that energy transfer in the complex is very robust with respect to changes in the environment at room temperature, and that the results are largely the same if structure is introduced into the environment or disorder is accounted for. Rather than electronic coherence, it is this robustness that is advantageous for the complex's biophysical role.

\newpage

{\renewcommand{\baselinestretch}{1}\normalsize
\section*{Acknowledgements}\label{acknowledgements}
}
\pagestyle{empty}

I would like to thank David Manolopoulos for supervising my Part II project and for his unparalleled guidance and inspiration throughout my degree.

The members of the Manolopoulos group, Nike, Josh, Michele, Jon and Kenji, also helped make this year an enjoyable and productive one. Along with members of the Barford and Wilson groups, they have also provided many interesting conversations over many hours of coffee.

I thank my friends for providing a distraction when I needed one, including Karolis, David R., Emily N., Richard, Emily B., Alan, Emma, Michael, Iain, Nat, Paul and Karen.

The support of my parents has been invaluable both during my degree and otherwise, and I would like to thank them very much for this, along with my brother, Nicholas, and my girlfriend Anne, as well as my grandparents and great-aunt.

Finally, this work is dedicated to the memory of my Nan, Joyce Wilkins, whom I miss very much.

\renewcommand\thepage{}
{\renewcommand{\baselinestretch}{1}\normalsize
\tableofcontents{\thispagestyle{empty}}
}
\renewcommand\thepage{\arabic{page}}



{\renewcommand{\baselinestretch}{1}\normalsize
\chapter{Introduction}\pagestyle{fancy}\setcounter{page}{1}\label{chapter_introduction}
}

\fancyhead{}
\renewcommand{\sectionmark}[1]{\markright{\thesection.\ #1}}
\fancyhead[LO,RE]{\rightmark}
\pagestyle{fancy}

The process of photosynthesis is one whose importance cannot be overestimated: photosynthetic organisms capture energy from sunlight and store it in chemical form. This energy is taken up by organisms higher up the food chain \cite{blankenship,renger_review,fleming_lightharv}.

In an antenna, photons of sunlight are absorbed by pigments, generally based on chlorophyll molecules, and the resulting excitation is transported between these pigments, eventually reaching an energy ``funnel'', in which the excitation moves over time towards pigments of lower energy, meaning that the process is irreversible. Its destination is the reaction centre, in which the energy is converted to chemical form \cite{blankenship}.


This work focuses on the excitation energy transfer that occurs in the energy funnel, although the methods used are suitable for much more general problems.

This transfer is extremely efficient, and it is immediately apparent that it would be very useful if we could design systems that were as effective at transferring energy. For this, it is necessary to understand the features of the system that lead to this efficiency.

As an example, we will be interested in the effect of the environment surrounding the pigments: presumably (as with many natural systems), the environment has been finely tuned to facilitate transfer. If we were to alter this environment, what effect would it have on transfer rates?

We will see in Section \ref{quant_coh} that the importance of quantum-mechanical effects in the transfer is currently under question. By modelling the transfer using methods that both include and ignore these quantum effects, we can gain an idea of how vital they really are at physiological temperatures.

This transfer is of current experimental and theoretical interest \cite{fleming_lightharv,coherence,ishizakifleming,schulten_LH2,2d_fmo}, and we now discuss a complex whose study will allow us to benchmark the techniques used, and to learn about the physics of energy transfer.

\section{Fenna-Matthews-Olson Complex}

The Fenna-Matthews-Olson (FMO) complex is a pigment-protein complex in which bacteriochlorophyll (BChl) pigments are bound to a protein. It is found in green sulfur bacteria such as \emph{C. tepidum} and \emph{P. aestuarii} \cite{fleming_lightharv}, and collects electronic excitation from a light-harvesting chlorosome, funnelling it towards the reaction centre.

\begin{figure}[!ht]
\begin{center}
\includegraphics[width=13.5cm, keepaspectratio=true]{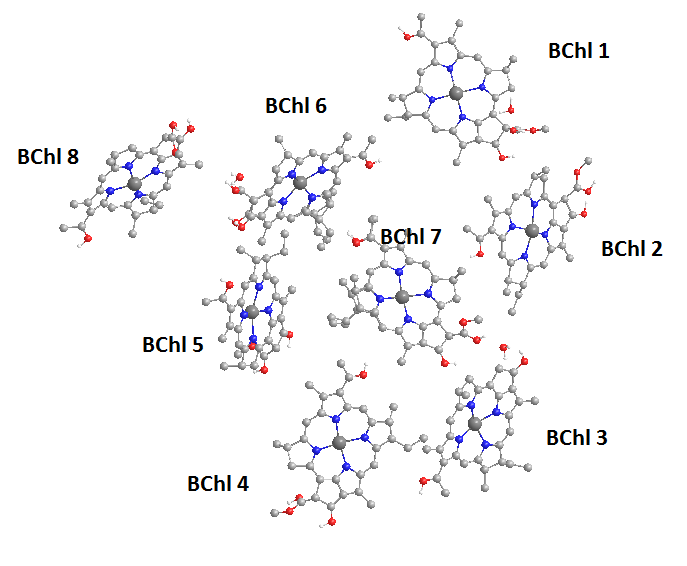}
\caption[Bacteriochlorophyll arrangement in FMO monomer.]{\small{Bacteriochlorophyll arrangement in FMO monomer. Figure created with Chem3D 2010, using PDB entry 3ENI \cite{olbrich}.}}
\label{fmo_figure}
\end{center}
\end{figure}

The complex itself is a trimer, each monomer of which was thought originally to comprise 7 BChl molecules \cite{fenna_matthews_olson,fenna_matthews}, though it is now accepted that there is an 8th, more weakly bound BChl site in addition \cite{eight_site,schmidt_am_busch}. The currently accepted arrangement of bacteriochlorophylls is shown in Fig. \ref{fmo_figure}.

The sites labelled 1, 6 and 8 are those found nearest to the chlorosome, so are the ones that capture the excitation energy, while those labelled 3 and 4 are closest to the reaction centre and lowest in energy, so it is to these sites that the energy is funnelled.

There are several reasons for the popularity of the FMO complex in studies of energy transfer: its structure is well-documented \cite{fenna_matthews,eight_site}, and has been since 1974 \cite{fenna_matthews_olson}, making it ideal for theoretical studies, and its solubility in water means it is convenient for use in experiments \cite{coherence,eight_site}. Because it has been extensively studied in the past, the data required to carry out our simulations is readily available \cite{adolphs_renger}.

Next, we turn to the problem of finding a model Hamiltonian to describe the excitation energy transfer, which will require an introduction to the open quantum system.

\section{Open Quantum Systems}\label{open_qsys}

Consider a system of N ``sites'' (for example, BChl sites) labelled by Latin letters, each of which has a ground and a single excited electronic state. If $\ket{e_{j}}$ refers to site $j$ being in the excited state, and $\ket{g_{j}}$ the ground state, we will focus on the single-excitation manifold, that is, states $\ket{j}$:
\begin{equation}
\ket{j} = \ket{e_{j}}\prod_{k\ne j}^{\text{N}}\ket{g_{k}}.
\end{equation}

Using the states $\lbrace\ket{j}\rbrace$ as a basis, the so-called ``system'' Hamiltonian is given by:
\begin{equation}
\hat{H}_{S} = \sum_{j}^{\text{N}}\hbar\omega_{j}\ket{j}\bra{j} + \sum_{j,k \ne j}^{\text{N}}\hbar J_{jk}(\ket{j}\bra{k} + \ket{k}\bra{j}).
\end{equation}

Here, $\hbar\omega_{j}$ is the energy of site $j$ and $\hbar J_{jk}$ represents the coupling between two sites, due to dipolar interactions \cite{renger_review}. The eigenstates of the Hamiltonian (referred to as excitons) are represented by Greek letters $\ket{\nu}$, such that:
\begin{equation}
\hat{H}_{S}\ket{\nu} = \hbar\omega_{\nu}\ket{\nu}.
\end{equation}

Where:
\begin{equation}
\ket{\nu} = \sum_{j}^{\text{N}}U_{\nu j}\ket{j}.\label{exciton_basis}
\end{equation}

\begin{figure}[!ht]
\begin{center}
\includegraphics[width=13.5cm, keepaspectratio=true]{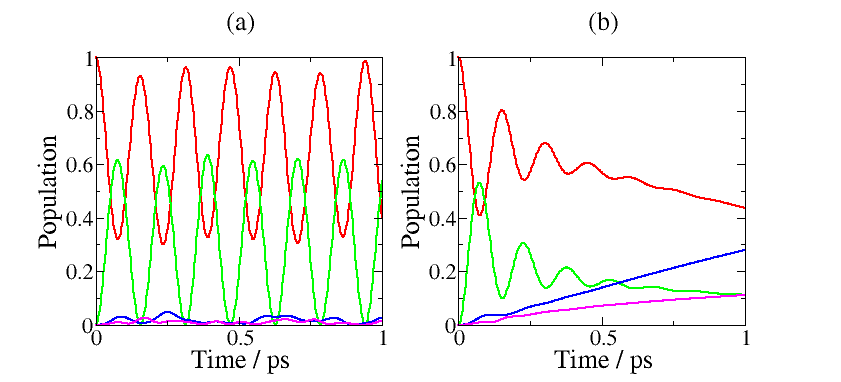}
\caption[Comparison of closed and open quantum system dynamics.]{\small{Comparison of the dynamics of (a) a closed and (b) an open quantum system, showing the effects of dissipation and decoherence.}}
\label{irreversible}
\end{center}
\end{figure}

If $\hat{H}_{S}$ were sufficient to describe the entire process, we see that we could use a time-dependent state vector $\ket{\psi(t)} = e^{i\hat{H}_{S}t/\hbar}\ket{\psi(0)}$ to describe the dynamics of the system, and furthermore that these dynamics would be reversible (see Fig. \ref{irreversible} (a)).

Clearly, this is not the full story, as we are not considering a completely isolated system. Rather, the electronic system represented by $\hat{H}_{S}$ is an open system, surrounded by an environment with which it can exchange energy. We take each site to be associated with its own ``bath'', the vibrations of the bacteriochlorophyll and of the surrounding protein, and if each bath is comprised of harmonic oscillators, then the total bath Hamiltonian is:
\begin{equation}
\hat{H}_{B} = \sum_{j}^{\text{N}}\sum_{\alpha}\left( \frac{\hat{p}_{j\alpha}^{2}}{2m_{j\alpha}} + \frac{1}{2}m_{j\alpha}\omega_{j\alpha}^{2}\hat{q}_{j\alpha}^{2} \right).
\end{equation}

Here, $\hat{q}_{j\alpha}$ is the coordinate of the $\alpha^{th}$ oscillator in the $j^{th}$ bath, $\hat{p}_{j\alpha}$ is the conjugate momentum, $m_{j\alpha}$ the mass and $\omega_{j\alpha}$ the frequency. The system-bath interaction is:
\begin{equation}
\hat{H}_{SB} = \sum_{j}^{\text{N}}\ket{j}\bra{j}\sum_{\alpha}g_{j\alpha}\hat{q}_{j\alpha} \equiv \sum_{j}^{\text{N}}\hat{V}_{j}\hat{\xi}_{j},
\end{equation}

\noindent with $\hat{\xi}_{j} = \sum_{j}g_{j\alpha}\hat{q}_{j\alpha}$ the generalized force exerted on the $j^{th}$ site by its bath and $\hat{V}_{j} \equiv \ket{j}\bra{j}$. Before considering the effect that the environment will have on the dynamics of the excitation transfer, we introduce the spectral density, $J_{j}(\omega)$, for the $j^{th}$ site:
\begin{equation}\label{spectral_density}
J_{j}(\omega) = \sum_{\alpha}\frac{g_{j\alpha}^{2}}{2 m_{j\alpha}\omega_{j\alpha}}\delta(\omega - \omega_{j\alpha}).
\end{equation}

This function has peaks at the frequencies of the bath oscillators, weighted by the strength of their coupling to the site $j$, and quantifies the system-bath interaction. In general, it is replaced by a smooth function of $\omega$ (corresponding to an uncountably infinite number of oscillators). Commonly, for reasons that will be discussed later on in this thesis, the spectral density chosen is the Lorentz-Drude function:
%
\begin{equation}\label{lorentz_drude}
J_{j}(\omega) = \frac{2\hbar}{\pi}\frac{\lambda_{j}\gamma_{j}\omega}{\gamma_{j}^2 + \omega^2}.
\end{equation}

\begin{figure}[!ht]
\begin{center}
\includegraphics[width=13.5cm, keepaspectratio=true]{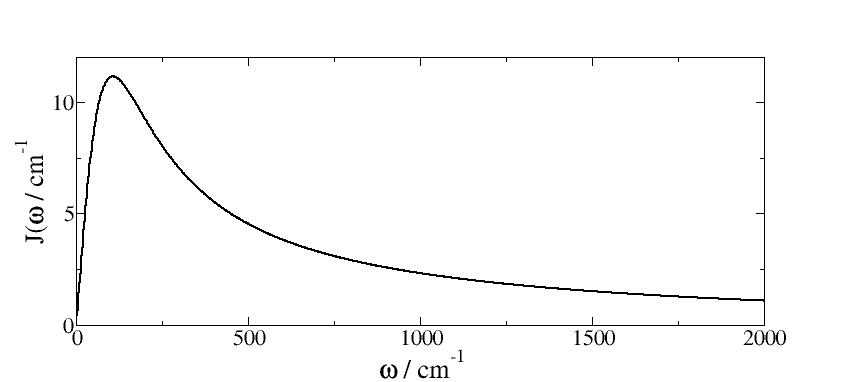}
\caption[Lorentz-Drude spectral density.]{\small{Lorentz-Drude spectral density of Eqn. \eqref{lorentz_drude}. Parameters used: $\lambda_{j} = 35\text{ cm}^{-1}$, $\gamma_{j} = 106.1\text{ cm}^{-1}$.}}
\label{drude_lorentz}
\end{center}
\end{figure}

Here $\lambda_{j} = \int_{0}^{\infty}d\omega J_{j}(\omega)/\omega$ is the reorganization energy, a measure of system-bath coupling strength, and $\gamma_{j}$ is the characteristic frequency of the bath, which satisfies $\dfrac{dJ_{j}(\omega)}{d\omega}(\omega=\gamma_{j}) = 0$. Fig. \ref{drude_lorentz} illustrates this spectral density.

The harmonic oscillators of the bath cause a kind of quantum-mechanical friction, one effect of which is exactly the same as classical friction: energy is dissipated from the electronic system, giving irreversible energy transfer.

There is, in addition, the purely quantum-mechanical effect of decoherence. The vibrational states associated with each electronic state may be excited or de-excited (we refer to them as absorbing or emitting a phonon), which alters the site energy \cite{leegwater_klug}. This in turn causes the coherence between two sites to decay, as the phase $e^{i\theta}$ of one state $\ket{k}$ with respect to another, $\ket{j}$, is altered when the energy of a site changes, and when these phases are randomized, their average value is zero. Over time, quantum-mechanical oscillations will be damped.

Both of these phenomena are illustrated in Fig. \ref{irreversible}, which shows the dynamics of a closed system and of a system with exactly the same Hamiltonian $\hat{H}_{S}$, but coupled to a bath of harmonic oscillators.

Since a larger reorganization energy $\lambda_{j}$ gives a stronger interaction between bath and system \cite{spin_boson}, we will expect stronger dissipation and more decoherence. For the same reason, higher temperatures also lead to greater decoherence.

One further property of the open quantum system must be mentioned: that of Markovianity. Due to the statistical nature of the environment, the force acting on a given site is a stochastic function of time.

If this stochastic process were Markovian, then at any given time no memory of its past behaviour would be required to predict its behaviour in the future; only its current value would be needed. There is no reason to assume \emph{a priori} that the behaviour of the function is Markovian, so that an accurate description of the dynamics should allow for non-Markovian effects.

The techniques used to deal with open quantum systems are applicable not only to biological systems such as the FMO complex, but also to many others, such as electron and proton transfer \cite{filtering1,proton_transfer,spin_boson}, as well as damping of Rabi oscillations due to excitation in quantum dots \cite{variational}. The theoretical tools described in this work are thus very versatile, and useful in many fields of current scientific interest.

\section{Quantum Coherence In Energy Transfer}\label{quant_coh}

We might intuitively expect that for the FMO complex at a physiological temperature of around 300 K, there would be a very strong interaction between environment and system (due to thermal excitation of bath oscillators), leading to very fast decoherence.

Fairly recently Engel \emph{et al.} \cite{coherence} observed quantum beating effects in a 2-dimensional electronic spectroscopy study of energy transfer in the FMO complex at a cryogenic temperature of 77 K, and Collini \emph{et al.} \cite{coherence2} observed similar effects at an ambient temperature of 294 K for a system similar to the FMO complex, leading to the suggestion that quantum coherence is important for energy transfer, for example in tunnelling through energy barriers \cite{ishizakifleming}.

This result was followed by a number of theoretical studies by Ishizaki and Fleming \cite{ishizakifleming,heom1,i&f_redfield} with the aim of accurately modelling the energy transfer dynamics in the FMO complex.

Importantly, using the Hierarchical Equations of Motion (which will be discussed in this thesis), the dynamics were simulated at both 77 K and 300 K for a 7-site subsystem of the monomer \cite{ishizakifleming}. At both temperatures oscillations were observed, suggesting that even at room temperature, quantum coherence was seen in the energy transfer, thus sparking renewed interest in the FMO complex.

One of our aims in the following work will be to appraise whether or not quantum coherent effects are necessary for efficient energy transfer at room temperature.

In the literature, it is conventional to discuss excitation transfer dynamics by showing time-dependent site populations \cite{ishizakifleming} (i.e., in the site basis $\lbrace\ket{j}\rbrace$), and we will use this convention in the work to follow.

A number of timescales are important in the study of the FMO complex: electronic coherence lasts for a number of femtoseconds (around 500 fs at 77 K, and around 200 fs at 300 K), so that a numerical simulation up to 1 ps will capture the effects of coherence, if appropriate.

After the coherence has decayed, the population dynamics show a simple decay, like those predicted by a rate equation, and eventually the populations reach a steady state. This occurs at around 10 ps at 77 K, and around 5 ps at 300 K, and will be observed in a numerical simulation up to 15 ps.

A number of experiments have been carried out which show that the fluorescence lifetime of bacteriochlorophylls in the FMO complex is on the order of 1 ns \cite{fluor_lifetime}. This means that we need not consider loss of excitation energy via spontaneous emission on the timescales used in our simulations, as such events are suitably rare.

\section{Summary}

A brief survey of the remainder of this thesis is now appropriate. Firstly, in Chapter \ref{chapter_heom} the Hierarchical Equations Of Motion (HEOM) for calculation of exact energy transfer dynamics will be introduced, and the advantages and disadvantages of using these equations will be described.

Next, in Chapter \ref{approximate} we will introduce several cheaper methods that can be used to obtain approximate results for the dynamics, the Redfield and F\"orster theories.

Chapter \ref{implementation} contains a description of some methods used to implement the HEOM efficiently, while doing so more accurately than in previous work.

Chapter \ref{numerical_results} is dedicated to numerical results for the 7-site and the 24-site FMO complex (the latter of which has not had any exact results published yet) from the HEOM and from the approximate methods described in Chapter \ref{implementation}. It is found that at 300 K, the incoherent F\"orster theory describes the general features of the energy transfer quite well.

Chapter \ref{applications} then covers applications of F\"orster theory: we find the effects of using a structured spectral density instead of the Lorentz-Drude function, and of static disorder due to slow fluctuations in the protein environment.

Finally, Chapter \ref{conclusions} concludes.

{\renewcommand{\baselinestretch}{1}\normalsize
\chapter{Hierarchical Equations Of Motion}\label{chapter_heom}
}

There are several methods for finding a numerically exact description of the dynamics of an open quantum system, many of which have been in use for some decades. The Feynman-Vernon influence functional \cite{influence} is based on the path integral formulation of quantum mechanics \cite{feynman}, wherein the dynamics of the system alone are considered, having integrated out those of the bath.

A more recent method is the self-consistent hybrid method \cite{spin_boson}, in which the spectral density is not considered as a continuum, but is discretized, giving a number of ``bath modes''. The dynamics of a number of these modes is treated exactly along with the system, while the dynamics of the rest of the modes is treated classically: the number of modes treated exactly is increased until the results converge.

In this Chapter we present the Hierarchical Equations of Motion (HEOM), which were formulated originally by Tanimura and Kubo \cite{tanimura1,tanimura2} using the influence functional as a starting point. We work instead with operators \cite{heom1}, deriving the influence operator from a consideration of the statistical properties of a bath of harmonic oscillators.

Section \ref{RDO} introduces the reduced density operator, and gives the general equation of motion for this operator. In Section \ref{infl}, the influence operator is derived, and in Section \ref{EOM}, a hierarchy of auxiliary density operators is introduced, giving exact equations of motion.

\section{Reduced Density Operator}\label{RDO}

An ensemble of quantum systems is most appropriately described not with wavefunctions but rather, using the density operator formalism. If $\hat{H}$ is the Hamiltonian for a system and $\lbrace|\phi_{j}\rangle\rbrace$ the eigenstates of $\hat{H}$, then the density operator is given by:
%
\begin{equation}\label{densop}
\hat{\rho} = \sum_{n} p_{j} \ket{\phi_{j}}\bra{\phi_{j}},
\end{equation}

\noindent where $p_{j}$ is the probability that a system in this ensemble will be found in the state $\ket{\phi_{j}}$. Particularly important is the situation in which the system is a member of a canonical ensemble, so that $p_{j} = e^{-\beta\epsilon_{j}}/Z_{\beta}$, where $\beta = 1/k_{B}T$, $\epsilon_{j}$ is the energy eigenvalue corresponding to eigenstate $\ket{\phi_{j}}$, and $Z_{\beta} = tr[e^{-\beta\hat{H}}]$ is the canonical partition function. In this case, the canonical density operator is $\hat{\rho}_{\beta} = e^{-\beta\hat{H}}/Z_{\beta}$.

A diagonal matrix element of the density operator, $\bra{\phi_{k}}\hat{\rho}\ket{\phi_{k}}$, gives the population of state $\ket{\phi_{k}}$ \cite{density_matrix}, and by cyclic permutation of operators within a trace, the average value of an operator $\hat{A}$ is given by:
\begin{equation}
\langle A\rangle = \sum_{j,k} p_{j} \bra{\phi_{j}}\hat{A}\ket{\phi_{k}}\inner{\phi_{k}}{\phi_{j}} \equiv tr[\hat{A}\hat{\rho}].
\end{equation}

For a system described by a single state vector, if this state is not an eigenstate of the Hamiltonian then it will evolve in time, and similarly, from the Time-Dependent Schr\"odinger equation, if a density matrix does not commute with the Hamiltonian, we will observe dynamics \cite{density_matrix}, given by $i\hbar d\hat{\rho}(t)/d t = [\hat{H},\hat{\rho}(t)]$.

\subsection{Superoperators And The Interaction Picture}\label{supops}

It proves useful now to introduce some definitions. Firstly, the Liouvillian superoperator $\hat{\mathcal{L}}$ is defined such that:
\begin{equation}
\hat{\mathcal{L}}\hat{O} = [\hat{H},\hat{O}].
\end{equation}

Further, we can define two superoperators $\hat{A}^{\rightarrow}$ and $\hat{A}^{\leftarrow}$ such that $\hat{A}^{\rightarrow}\hat{O} = \hat{A}\hat{O}$ and $\hat{A}^{\leftarrow}\hat{O} = \hat{O}\hat{A}$. Then the commutation and anticommutation superoperators are given by \cite{irreversible}:
\begin{equation}
\commut{A} = \hat{A}^{\rightarrow} - \hat{A}^{\leftarrow} \Rightarrow \hat{A}^{\times}\hat{O} = [\hat{A},\hat{O}],
\end{equation}
\begin{equation}
\anticommut{A} = \hat{A}^{\rightarrow} + \hat{A}^{\leftarrow} \Rightarrow \hat{A}^{\circ}\hat{O} = \lbrace\hat{A},\hat{O}\rbrace.
\end{equation}

A useful relation is the following, which can be verified by differentiating both its left and right hand sides:
\begin{equation}
e^{i\hat{\mathcal{L}}t/\hbar}\hat{O} \equiv e^{i\hat{H}t/\hbar}\hat{O}e^{-i\hat{H}t/\hbar}.
\end{equation}

In order to derive an equation of motion for the density operator, we first split up the total Hamiltonian as $\hat{H} = \hat{H}_{0} + \hat{V}$ (and thus, the Liouvillian as $\hat{\mathcal{L}} = \hat{\mathcal{L}}_{0} + \hat{\mathcal{L}}_{V}$). Then, in the interaction picture with respect to $\hat{H}_{0}$, denoted by a tilde following the notation of \cite{heom1}, the state vectors and operators are given by (pp. 172-173 of \cite{dirac}):
\begin{subequations}\label{interaction}
\begin{equation}\label{int_vec}
\ket{\tilde{\phi}(t)} = e^{i\hat{H}_{0}t/\hbar}\ket{\phi(t)} = e^{i\hat{H}_{0}t/\hbar}e^{-i\hat{H}t/\hbar}\ket{\phi(0)},
\end{equation}
\begin{equation}\label{int_op}
\tilde{O}(t) = e^{i\hat{\mathcal{L}_{0}}t/\hbar}\hat{O}.
\end{equation}
\end{subequations}

Inserting Eqn. \eqref{int_vec} into the time-dependent Schr\"odinger equation gives:
\begin{equation}\label{lipmann-schwinger}
i\hbar\frac{d}{d t}\ket{\tilde{\phi}(t)} = \tilde{V}(t)\ket{\tilde{\phi}(t)},
\end{equation}

\noindent and this gives the expression for the time-derivative of the density operator in the interaction picture:
\begin{equation}\label{von_neumann}
\frac{d}{d t}\tilde{\rho}(t) = -\frac{i}{\hbar}\tilde{\mathcal{L}}_{V}(t)\tilde{\rho}(t).
\end{equation}

Solution of this differential equation will give an expression for the time-dependence of the density operator for the system described by Hamiltonian $\hat{H}$.

\subsection{Time-Evolution And The Partial Trace}\label{partial_trace}

Solution of \eqref{von_neumann} is not as straightforward as it first appears. To understand this, we first integrate the equation to give a recurrence relation:
\begin{equation}\label{recurrence}
\tilde{\rho}(t) = \tilde{\rho}(0) - \frac{i}{\hbar}\int_{0}^{t}dt_{1}\tilde{\mathcal{L}}_{V}(t_1)\tilde{\rho}(t_{1}).
\end{equation}

The density matrix at a given time is determined by an integral over density matrices in the past, and can be expressed as an infinite series by expanding \eqref{recurrence}:
\begin{equation}\label{infinite_series}
\tilde{\rho}(t) = \tilde{\rho}(0) + \sum_{k=1}^{\infty}\left[\left(\frac{-i}{\hbar}\right)^{k}\int_{0}^{t}dt_{1}\dots\int_{0}^{t_{k-1}}dt_{k}\tilde{\mathcal{L}}_{V}(t_{1})\dots\tilde{\mathcal{L}}_{V}(t_{k})\right]\tilde{\rho}(0).
\end{equation}

There is a notable similarity between this and the so-called perturbation expansion \cite{feynman}. By analogy, we can interpret \eqref{infinite_series} as a sum over alternative possibilities.

A general term in the sum represents the effect of $\hat{H}_{SB}$ on the system at  times $t_{1},t_{2},\dots,t_{k}$. Liouvillians at two different times do not necessarily commute, and so the order in which they are applied is important.

We introduce the time-ordering operator $\mathcal{T}$. For a product of two operators, the action of $\mathcal{T}$ is:
\begin{equation}\label{time_order}
\mathcal{T}\hat{O}_{1}(t_{1})\hat{O}_{2}(t_{2}) = \hat{O}_{1}(t_{1})\hat{O}_{2}(t_{2})\Theta(t_{1}-t_{2}) + \hat{O}_{2}(t_{2})\hat{O}_{1}(t_{1})\Theta(t_{2}-t_{1}).
\end{equation}

Here, $\Theta(t)$ is the Heaviside step function (equal to 1 when $t>0$ and 0 when $t<0$). When $\mathcal{T}$ acts on a product of $n$ operators, there will be a total of $n!$ terms on the right hand side (this is the number of ways that the operators could be ordered).

Applying $\mathcal{T}$ to \eqref{infinite_series} so that the influences of the $\tilde{\mathcal{L}}_{V}(t)$ are experienced in chronological order (thus preserving causality) gives:
\begin{align}
\tilde{\rho}(t) & = \hat{\rho}_{I}(0) + \mathcal{T}\sum_{k=1}^{\infty}\left[\frac{1}{k!}\left(\frac{-i}{\hbar}\right)^{k}\int_{0}^{t}dt_{1}\dots\int_{0}^{t}dt_{k}\tilde{\mathcal{L}}_{V}(t_{1})\dots\tilde{\mathcal{L}}_{V}(t_{k}) \right]\hat{\rho}_{I}(0)\nonumber \\
& = \mathcal{T}\sum_{k=0}^{\infty}\frac{1}{k!}\left(\frac{-i}{\hbar}\int_{0}^{t}dt_{1}\tilde{\mathcal{L}}_{V}(t_{1})\right)^{k}\tilde{\rho}(0).
\end{align}

\noindent This is the power series for the exponential function, so can be rewritten:
\begin{equation}\label{density_evolution}
\tilde{\rho}(t) = \mathcal{T}\exp\left(-\frac{i}{\hbar}\int_{0}^{t}dt_{1}\tilde{\mathcal{L}}_{V}(t_{1})\right)\tilde{\rho}(0).
\end{equation}

The above discussion is entirely general, and is suitable for any quantum-mechanical system, including the system-plus-environment supersystem we are considering. However, the environment has an infinite number of degrees of freedom, making it impossible to implement \eqref{density_evolution} in the form shown.

A popular method of dealing with this problem is to take a partial trace; the reduced density operator is a trace, over all bath degrees of freedom, of the total density operator:
\begin{equation}\label{reduced_density}
\tilde{\rho}_{S}(t) = \text{tr}_{B}[\tilde{\rho}(t)].
\end{equation}

Finally, we assume that the initial density matrix is given by $\tilde{\rho}(0) = \tilde{\rho}_{S}(0)\hat{\rho}_{B,\beta}$, where $\hat{\rho}_{B,\beta} = e^{-\beta \hat{H}_{B}}/Z_{B}$ is the canonical density matrix for the bath (note that no tilde is used in this case, since $\tilde{\rho}_{B,\beta} = \hat{\rho}_{B,\beta}$ and the canonical density operator is unchanged in the interaction representation).

This is justified in an electronic excitation process by invoking the Franck-Condon principle \cite{heom1}: when electronic excitation occurs, the baths remain in their equilibrium state because nuclear motion is slow on the timescale of electronic reorganization.

With the definition $\left< \tilde{O}(t) \right>_{\beta} = \text{tr}_{B}[\tilde{O}(t)\hat{\rho}_{B,\beta}]$, the time-evolution of the reduced system density matrix in the interaction picture is given by:
\begin{equation}\label{trace_rdo}
\tilde{\rho}_{S}(t) = \mathcal{T}\left< \exp\left(-\frac{i}{\hbar}\int_{0}^{t}dt_{1}\tilde{\mathcal{L}}_{V}(t_{1})\right)\right>_{\beta}\tilde{\rho}_{S}(0).
\end{equation}

This expression is extremely important, and will be revisited in the work of Chapter \ref{approximate}.

\section{Influence Operator}\label{infl}

Evaluating the bath average in \eqref{trace_rdo} will give an expression that depends not on the bath operators but only on terms that give the influence of the bath on the system. In order to achieve this, some observations must be made about the bath. Firstly, we set $\hat{H}_{0} = \hat{H}_{S}+\hat{H}_{B}$ and $\hat{V} = \hat{H}_{SB}$, so that in the remainder of this Chapter the interaction Liouvillian is $\tilde{\mathcal{L}}_{V}(t) = \tilde{\mathcal{L}}_{SB}(t)$.

\subsection{The Force Operator}\label{force_op}

The operator $\tilde{\xi}_{j}(t) = \sum_{\alpha}g_{j\alpha}\tilde{q}_{j\alpha}(t)$ gives the force exerted on the $j^{th}$ site by the bath oscillators $\alpha$ associated with this site. It is our aim here to find an explicit form for $\tilde{\xi}_{j}(t)$, and to show that the force exerted on a site by its bath is a random, Gaussian force, which will allow the bath average to be performed simply.

In the interaction picture, differentiating Eqn. \eqref{int_op} gives:
\begin{equation}
\frac{d}{d t}\tilde{O}(t) = \frac{i}{\hbar}\hat{\mathcal{L}}_{0}\tilde{O}(t).
\end{equation}

Using $\hat{H}_{0} = \hat{H}_{S}+\hat{H}_{B}$ gives $d\tilde{q}_{j\alpha}(t)/d t = \tilde{p}_{j\alpha}(t)/m_{j\alpha}$, and $d\tilde{p}_{j\alpha}(t)/d t = -m_{j\alpha}\omega_{j\alpha}^2\tilde{q}_{j\alpha}(t)$, whose solution is:
\begin{equation}\label{bath_position}
\tilde{q}_{j\alpha}(t) = \tilde{q}_{j\alpha}(0)\cos\left(\omega_{j\alpha}t\right) + \frac{\tilde{p}_{j\alpha}(0)}{m_{j\alpha}\omega_{j\alpha}}\sin\left(\omega_{j\alpha}t\right).
\end{equation}

From this we infer \cite{quantum_langevin} that the force operator at time $t$ is determined by the positions and momenta of the bath oscillators at time $0$, and so we now show that these positions and momenta are Gaussian variables. That is:
\begin{equation}\label{posn_gauss}
\left< f(\hat{q}_{j\alpha})\right>_{\beta} \propto \int_{-\infty}^{\infty}dq_{j\alpha}f(q_{j\alpha})e^{-\kappa q_{j\alpha}^2},
\end{equation}

\noindent for any function $f(\hat{q}_{j\alpha})$ of $\hat{q}_{j\alpha}$ in the Schr\"odinger representation, and that a similar relation will be true for functions of momentum.

Recalling the definition of the bath average, the trace can be evaluated in the position basis. The trace is then given by $\int\prod_{j,\alpha} dq_{j\alpha}\bra{q_{j\alpha}}\dots\ket{q_{j\alpha}}$. Since the canonical density matrix is a product $\hat{\rho}_{B,\beta}= \prod_{j}\hat{\rho}_{j,\beta} = \prod_{j,\alpha}\hat{\rho}_{j\alpha,\beta}$, integration over each coordinate $q_{j^{\prime}\alpha^{\prime}}$ gives the partition function $Z_{j^{\prime}\alpha^{\prime}}$, except for coordinate $q_{j\alpha}$:
\begin{align}
\left<f(\hat{q}_{j\alpha})\right>_{\beta} & = \int_{-\infty}^{\infty}dq_{j\alpha} \langle q_{j\alpha} | f(\hat{q}_{j\alpha})\hat{\rho}_{j\alpha,\beta}|q_{j\alpha}\rangle / Z_{j\alpha} \nonumber\\
& = \frac{1}{Z_{j\alpha}} \int_{-\infty}^{\infty}dq_{j\alpha}f(q_{j\alpha})\langle q_{j\alpha} | \hat{\rho}_{j\alpha,\beta}|q_{j\alpha}\rangle.
\end{align}

The canonical density matrix element $\bra{q_{j\alpha}}\hat{\rho}_{j\alpha,\beta}\ket{q_{j\alpha}}$ is perhaps most easily evaluated by the path integral formulation of quantum mechanics \cite{feynman}, which gives\footnote{This result is obtained by noting that the canonical density matrix $e^{-\beta\hat{H}}$ is analogous to the propagator $e^{-i\hat{H}t/\hbar}$, for ``thermal time'' $t=-i\beta\hbar$, and treating it using the path integral method for evaluating such propagators.}:
\begin{equation}
\left<f(\hat{q}_{j\alpha})\right>_{\beta} = \frac{1}{Z_{j\alpha}}\int_{-\infty}^{\infty}dq_{j\alpha}f(q_{j\alpha})\exp\left( -\frac{m_{j\alpha}\omega_{j\alpha}}{\hbar}\tanh\left(\frac{\beta\hbar\omega_{j\alpha}}{2}\right) q_{j\alpha}^2 \right).
\end{equation}

The positions and momenta of thermal bath oscillators are stochastic variables (due to the statistical nature of the problem) with Gaussian distributions. Since a linear combination of variables with Gaussian distributions is also Gaussian \cite{kubo}, we have the result that the force operator $\tilde{\xi}_{j}(t)$ acting on site $j$ has such a distribution.

\subsection{Evaluation Of Bath Average}\label{bath_average}

We can now eliminate the dependency of Eqn. \eqref{trace_rdo} on the operators of the bath. Using the identity $(\hat{A}\hat{B})^{\times} = \frac{1}{4}\lbrace \hat{A}^{\times},\hat{B}^{\circ}\rbrace + \frac{1}{4}\lbrace \hat{A}^{\circ},\hat{B}^{\times}\rbrace + \frac{1}{2}[\hat{A},\hat{B}]^{\times}$ for the commutator of a product of operators \cite{irreversible} gives:
\begin{align}
\tilde{\mathcal{L}}_{SB}(t) & = \sum_{j}^{\text{N}}\left(\tilde{\xi}_{j}(t)\tilde{V}_{j}(t)\right)^{\times} \nonumber\\
& \equiv \frac{1}{2}\sum_{j}^{\text{N}}\left( \tilde{\xi}_{j}(t)^{\circ}\tilde{V}_{j}(t)^{\times} +  \tilde{\xi}_{j}(t)^{\times}\tilde{V}_{j}(t)^{\circ}\right).
\end{align}

We can infer that the interaction Liouvillian also has a Gaussian distribution with respect to a bath average, as it will also be a linear combination of the force operators. This distribution means that the statistical properties of $\tilde{\mathcal{L}}_{SB}(t)$ are fully determined by its autocorrelation function \cite{kubo}:
\begin{align}
\left<\tilde{\mathcal{L}}_{SB}(t)\tilde{\mathcal{L}}_{SB}(s)\right>_{\beta} = \frac{1}{2}\sum_{j}^{\text{N}} & \left( \tilde{V}_{j}(t)^{\times}\left<\tilde{\xi}_{j}(t)^{\circ}\tilde{\xi}_{j}(s)^{\circ}\right>_{\beta}\tilde{V}_{j}(s)^{\times} +\right.\nonumber\\
& \tilde{V}_{j}(t)^{\times}\left<\tilde{\xi}_{j}(t)^{\circ}\tilde{\xi}_{j}(s)^{\times}\right>_{\beta}\tilde{V}_{j}(s)^{\circ} +\nonumber\\
& \tilde{V}_{j}(t)^{\circ}\left<\tilde{\xi}_{j}(t)^{\times}\tilde{\xi}_{j}(s)^{\circ}\right>_{\beta}\tilde{V}_{j}(s)^{\times} +\nonumber\\
& \left.\tilde{V}_{j}(t)^{\circ}\left<\tilde{\xi}_{j}(t)^{\times}\tilde{\xi}_{j}(s)^{\times}\right>_{\beta}\tilde{V}_{j}(s)^{\circ}\right).
\end{align}

Noting that, for example, $\left<\tilde{\xi}_{j}(t)^{\circ}\tilde{\xi}_{j}(s)^{\circ}\right> = \text{tr}_{B,j}[\tilde{\xi}_{j}(t)^{\circ}\tilde{\xi}_{j}(s)^{\circ}\hat{\rho}_{j,\beta}]$, this trace will have four terms, and can be simplified (by cyclic permutation of the operators within the trace) to give $2\left<\lbrace\tilde{\xi}_{j}(t),\tilde{\xi}_{j}(s)\rbrace\right>_{\beta}$. The remaining three bath averages can be carried out (with the third and fourth vanishing) to give:
\begin{multline}\label{autocorr_lvl}
\left<\tilde{\mathcal{L}}_{SB}(t)\tilde{\mathcal{L}}_{SB}(s)\right>_{\beta} =\dots\\
 \sum_{j}^{\text{N}} \tilde{V}_{j}(t)^{\times}\left(\alpha_{r,j}(t-s)\tilde{V}_{j}(s)^{\times} + i\alpha_{i,j}(t-s)\tilde{V}_{j}(s)^{\circ}\right).
\end{multline}

Here $\alpha_{r,j}(t) = \frac{1}{2}\left<\lbrace \tilde{\xi}_{j}(t),\tilde{\xi}_{j}(0)\rbrace\right>_{\beta}$ and $i\alpha_{i,j}(t) = \frac{1}{2}\left<\left[\tilde{\xi}_{j}(t),\tilde{\xi}_{j}(0)\right]\right>_{\beta}$, so that $\alpha_{j}(t) = \alpha_{r,j}(t) + i\alpha_{i,j}(t)$ is equal to $\left<\tilde{\xi}_{j}(t)\tilde{\xi}_{j}(0)\right>_{\beta}$, the autocorrelation function of the force exerted on the system by the bath.

Using the identity for a Gaussian process $x(t)$ \cite{kubo},
\begin{equation}\label{gaussian}
\left<\exp\left(\int_{0}^{t}dt_{1} f(t_{1})x(t_{1})\right)\right> = \exp\left(\frac{1}{2}\int_{0}^{t}dt_{1}\int_{0}^{t}dt_{2}f(t_{1})f(t_{2})\left<x(t_{1})x(t_{2})\right>\right),
\end{equation}

we finally obtain \cite{heom1}\footnote{Note that Eqn. \eqref{gaussian} has been adapted here due to the time-ordering; the operator $\mathcal{T}$ allows us to change the integration limits and removes the factor of $\frac{1}{2}$.}:
\begin{multline}\label{influence}
\tilde{\rho}_{S}(t) = \mathcal{T}\prod_{j}^{\text{N}}\exp\left(-\frac{1}{\hbar^2}\int_{0}^{t}dt_{1}\int_{0}^{t_{1}}dt_{2} \tilde{V}_{j}(t_{1})^{\times}\left[\alpha_{r,j}(t_{1}-t_{2})\tilde{V}_{j}(t_{2})^{\times}\right.\right.\\
+\left.\left.i\alpha_{i,j}(t_{1}-t_{2})\tilde{V}_{j}(t_{2})^{\circ}\right]\right)\tilde{\rho}_{S}(0).
\end{multline}

The operator acting on $\tilde{\rho}_{S}(0)$ is the influence operator \cite{heom1}. In the position representation, this becomes the influence functional of Feynman and Vernon \cite{influence,weiss}, which expresses the time-propagation of the reduced density matrix in terms only of statistical quantities of environment variables.

It is shown in Appendix \ref{QFDT} that the bath correlation function $\alpha_{j}(t)$ can be expressed in terms of the spectral density, such that $J_{j}(-\omega)=-J_{j}(\omega)$:
\begin{equation}\label{bath_corrfunc}
\alpha_{j}(t) = \int_{0}^{\infty}J_{j}(\omega)\left[\coth(\beta\hbar\omega/2)\cos(\omega t) - i\sin(\omega t)\right]d\omega.
\end{equation}

Thus, the influence operator encodes the effects of the baths (through the spectral densities) and of the temperature (through the $\coth(\beta\hbar\omega/2)$ term) on the system.

\section{Equations Of Motion}\label{EOM}

With an explicit form for the bath correlation function $\alpha_{j}(t)$, we can now find a differential equation to evolve the reduced density matrix. We take:
\begin{equation}\label{corrfunc_exp}
\alpha_{j}(t) = p_{j0}\delta_{+}(t) + \sum_{k=1}^{K}p_{jk}e^{-\gamma_{jk}t}.
\end{equation}

Here, $\delta_{+}(t)$ is a one-sided delta function, which has all of the properties of the  delta function (pp. 58-60 of \cite{dirac}), but the integration limits in its definition are $(0,\infty)$ instead of $(-\infty,\infty)$. For example (and most importantly), $\int_{0}^{\infty}\delta_{+}(t)dt = 1$. We also use the notation $p_{jk} = a_{jk} + ib_{jk}$, where $a_{jk}$ and $b_{jk}$ are real.

The reasoning behind this expression is given in Chapter \ref{implementation}, and the advantage of this sum of exponentials will quickly become apparent in the following work.

\subsection{Auxiliary Density Operators}

Using \eqref{corrfunc_exp}, we can write the time-evolution of the density operator as \cite{tanimura2}
\begin{multline}\label{hierarchy1}
\tilde{\rho}_{S}(t) = \mathcal{T}\prod_{j}\exp\left(\frac{1}{\hbar^2}\int_{0}^{t}dt_{1}\tilde{\phi}_{j}(t_{1})\tilde{\theta}_{j,0}(t_{1})\right)\times\dots\\
\prod_{k}\exp\left(\frac{1}{\hbar^2}\int_{0}^{t}dt_{1}\int_{0}^{t_{1}}dt_{2}\tilde{\phi}_{j}(t_{1})e^{-\gamma_{jk}(t_{1}-t_{2})}\tilde{\theta}_{j,k}(t_{2})\right)\tilde{\rho}_{S}(0),
\end{multline}

\noindent where we have introduced the notation,
\begin{subequations}\label{notation}
\begin{equation}
\tilde{\phi}_{j}(t) = i\tilde{V}_{j}(t)^{\times},
\end{equation}
\begin{equation}
\tilde{\theta}_{j,k}(t) = ia_{jk}\tilde{V}_{j}(t)^{\times} - b_{jk}\tilde{V}_{j}(t)^{\circ}.
\end{equation}
\end{subequations}

We define a set of auxiliary density operators (ADOs) $\hat{\sigma}_{\textbf{n}}(t)$, indexed by the matrix $\textbf{n}$ whose elements are $n_{jk}$:
\begin{multline}\label{ADOs}
\tilde{\sigma}_{\textbf{n}}(t) = \mathcal{T}\prod_{j}\exp\left(\frac{1}{\hbar^2}\int_{0}^{t}dt_{1}\tilde{\phi}_{j}(t_{1})\tilde{\theta}_{j,0}(t_{1})\right)\times\dots\\
\prod_{k}\left(\frac{1}{\hbar^2} \int_{0}^{t}dt_{1}e^{-\gamma_{jk}(t-t_{1})}\tilde{\theta}_{j,k}(t_{1}) \right)^{n_{jk}}\times\dots\\
\exp\left(\frac{1}{\hbar^2}\int_{0}^{t}dt_{1}\int_{0}^{t_{1}}dt_{2}\tilde{\phi}_{j}(t_{1})e^{-\gamma_{jk}(t_{1}-t_{2})}\tilde{\theta}_{j,k}(t_{2})\right)\tilde{\rho}_{S}(0).
\end{multline}

The ADO with $\textbf{n} = \textbf{0}$ is the reduced density operator (RDO). The reason for these ADOs being introduced is made clear when the time-derivative of Eqn. \eqref{ADOs} is taken \cite{heom1}:
\begin{multline}\label{timederiv_ADO}
\frac{d}{d t}\tilde{\sigma}_{\textbf{n}}(t) = \sum_{j,k}\left( \tilde{\phi}_{j}(t)\tilde{\theta}_{j,0}(t) - n_{jk}\gamma_{jk}\right)\tilde{\sigma}_{\textbf{n}}(t) + \dots\\
\sum_{j,k}\left(\tilde{\phi}_{j}(t)\tilde{\sigma}_{\textbf{n}_{jk+}}(t) + n_{jk}\tilde{\theta}_{j,k}(t)\tilde{\sigma}_{\textbf{n}_{jk-}}(t)\right).
\end{multline}

The ADOs $\hat{\sigma}_{\textbf{n}}(t)$ and $\hat{\sigma}_{\textbf{n}_{jk\pm}}(t)$ have matrices $\textbf{n}$ that are identical, except that for the latter the element $n_{jk}$ is replaced by $n_{jk}\pm1$. For each ADO, we define:
\begin{equation}
\mathcal{M}_{\textbf{n}} = \sum_{j}^{\text{N}}\sum_{k}^{K}n_{jk}.
\end{equation}

And now, noting that $d\tilde{O}(t)/d t = e^{i\mathcal{L}_{0}t/\hbar}d\hat{O}/d t + \frac{i}{\hbar}\mathcal{L}_{0}e^{i\mathcal{L}_{0}t/\hbar}\hat{O}$, the time-derivative \eqref{timederiv_ADO} can be rearranged and rewritten in the Schr\"odinger picture:
\begin{multline}\label{heom_eqs}
\frac{d}{d t}\hat{\sigma}_{\textbf{n}}(t)	= -\frac{i}{\hbar}\hat{\mathcal{L}}_{S}\hat{\sigma}_{\textbf{n}}(t) - \sum_{j,k}\left(a_{j0}\hat{V}_{j}^{\times}\hat{V}_{j}^{\times} - ib_{j0}\hat{V}_{j}^{\times}\hat{V}_{j}^{\circ} + n_{jk}\gamma_{jk} \right)\hat{\sigma}_{\textbf{n}}(t) + \dots\\
\sum_{j,k}\left(i\hat{V}_{j}^{\times}\hat{\sigma}_{\textbf{n}_{jk+}}(t) + in_{jk}p_{jk}\hat{V}_{j}\hat{\sigma}_{\textbf{n}_{jk-}}(t) + in_{jk}p_{jk}^{\ast}\hat{\sigma}_{\textbf{n}_{jk-}}(t)\hat{V}_{j} \right).
\end{multline}

We see now that these ADOs form a hierarchy: a given ADO's level in the hierarchy is given by its value of $\mathcal{M}_{\textbf{n}}$, with only one operator (the reduced density operator) in the lowest level, $\text{N}\times K$ operators in the first level, and so on. Each ADO is coupled to operators on the levels above and below.

An interesting qualitative picture of the HEOM is given by reference to creation and annihilation operators (pp. 136-139 of \cite{dirac}): to each element of $\textbf{n}$ we can assign a ``bath mode'' labelled by $jk$. Then, a given ADO $\hat{\sigma}_{\textbf{n}}$ has $n_{jk}$ quanta in the bath mode $jk$.

It is then possible to see that a given level of the hierarchy corresponds to a certain total number of quanta, and that each ADO is connected to the ADOs that can be formed from it by creating or annihilating one quantum.

By propagating these ADOs through time (we see in \ref{comments} that a finite number of these operators can be used), the time-evolution of the diagonal elements of the RDO (the site populations) can be found.

\subsection{Comments}\label{comments}

The remainder of this Chapter is given to a discussion of several important points about the HEOM.

\begin{itemize}

\item We have made no approximation in this derivation, so have a method that is numerically exact and able to describe the effects of quantum-mechanical coherence in energy transfer and of non-Markovian system-bath interactions.

\item For all ADOs but the RDO, the initial condition is that all elements of these matrices be equal to zero.
\item Other than the RDO, the ADOs cannot be system density matrices, as their traces are not conserved.
\item In principle, the hierarchy continues to infinity. However, this is not possible in practice. Integrating \eqref{heom_eqs}:
\begin{multline}
\hat{\sigma}_{\textbf{n}}(t) = \int_{0}^{t}dt_{1}\cdot e^{-\left(\frac{i}{\hbar}\hat{\mathcal{L}}_{S}+\sum_{j,k}\left(a_{j0}\hat{V}_{j}^{\times}\hat{V}_{j}^{\times} -ib_{j0}\hat{V}_{j}^{\times}\hat{V}_{j}^{\circ} + n_{jk}\gamma_{jk} \right)\right)(t-t_{1})}\times\\
\sum_{j,k}\left(i\hat{V}_{j}^{\times}\hat{\sigma}_{\textbf{n}_{jk+}}(t_{1}) + in_{jk}p_{jk}\hat{V}_{j}\hat{\sigma}_{\textbf{n}_{jk-}}(t_{1}) + in_{jk}p_{jk}^{\ast}\hat{\sigma}_{\textbf{n}_{jk-}}(t_{1})\hat{V}_{j} \right).
\end{multline}

For ADOs above some level $\mathcal{N}$ of the hierarchy, the values of $n_{jk}$ will be so large that $\exp\left(-\sum{j,k}n_{jk}\gamma_{jk}(t-t_{1})\right)$ will decay very rapidly and thus will be proportional to $\delta_{+}(t-t_{1})$. This gives the result \cite{tanimura2,heom1} that for these ADOs, the time derivative does not depend on ADOs in any other levels, so their elements will remain at their initial value of zero.

\item We obtain converged results by carrying out calculations with an increasing number of levels $\mathcal{N}$, until increasing this number has no further effect on the resulting dynamics.

\item For a calculation using $\mathcal{N}$ levels of the hierarchy, the number of ADOs required is \cite{ishizakifleming}:
\begin{equation}\label{num_ados}
\frac{\left(\mathcal{N}+(\text{N}K)\right)!}{\mathcal{N}!(\text{N}K)!}.
\end{equation}

For some calculations, this number can be extremely large, meaning that the HEOM can be very computationally expensive, requiring a lot of time and memory to run. We will wish to implement the equations as efficiently as possible; this will be discussed further in Chapter \ref{implementation}.

\end{itemize}


{\renewcommand{\baselinestretch}{1}\normalsize
\chapter{Approximate Methods}\label{approximate}
}

Although the HEOM provide a very powerful method for calculating energy transfer dynamics, the large computational cost incurred means that we will need to seek an approximate method with a much lower cost, but with dynamics that are as accurate as possible, since we will be unable to use the HEOM for several of the physical features of the FMO complex that we wish to investigate.

There are a great number of methods that are suitable candidates, both historical and modern. Some of these, such as the F\"orster (1948) \cite{forster1} and Redfield (1957) \cite{redfield1} theories, treat some physical quantity as a perturbation; these two methods will be presented in this Chapter.

Other methods include the Zeroth Order Functional Expansion quantum master equation (ZOFE, 2011) \cite{zofe1}, in which a key operator is treated as being independent of the bath, and which has recently been applied both to the 7-site FMO system \cite{zofe_fmo1} and the 24-site trimer \cite{zofe_fmo2}.

This more modern method will not be considered in this thesis, as preliminary calculations (not reported here) showed the ZOFE to be less accurate than simple F\"orster theory for the FMO complex at 300 K.

The existence of so many methods, all with different conditions for applicability, means that it is necessary for us to decide which we will use in our investigations, and which we need not consider any further. We turn now to the task of introducing two techniques whose validity  we will test in Chapter \ref{numerical_results}.

\section{Redfield Theory}\label{redfield_theory}

The perturbative nature of both the Redfield and F\"orster theories \cite{redfield_forster} means that the validity of each depends on some parameter being small enough to justify it being treated as a perturbation. In the case of the Redfield theory, the coupling between the system and bath is assumed to be weak, characterized by reorganization energies $\lambda_{j}$ that are small (compared to dipolar couplings $J_{jk}$). Thus, $\hat{H}_{SB}$ is taken as the perturbation to the Hamiltonian.

In qualitative terms, we might consider an electronic system evolving under its own Hamiltonian $\hat{H}_{S}$, which is then weakly coupled to an environment. The phonons destroy phase information \cite{leegwater_klug}, assisting the relaxation of the system to an equilibrium state.

\subsection{Time-Dependent Redfield Theory}

We take as a starting point Eqn. \eqref{trace_rdo}, with $\tilde{\mathcal{L}}_{V}(t) = \tilde{\mathcal{L}}_{SB}(t)$. Taking the time-derivative gives:
\begin{align}
\frac{d}{d t}\tilde{\rho}_{S}(t) & = \left<-\frac{i}{\hbar}\mathcal{T}\tilde{\mathcal{L}}_{SB}(t)\exp\left(-\frac{i}{\hbar}\int_{0}^{t}dt_{1}\tilde{\mathcal{L}}_{SB}(t_{1})\right)\right>_{\beta}\tilde{\rho}_{S}(0)\nonumber\\
& \approx -\frac{i}{\hbar} \left<\tilde{\mathcal{L}}_{SB}(t)\right>_{\beta}\tilde{\rho}_{S}(0) - \frac{1}{\hbar^2}\int_{0}^{t}dt_{1}\left<\tilde{\mathcal{L}}_{SB}(t)\tilde{\mathcal{L}}_{SB}(t_{1})\right>_{\beta}\tilde{\rho}_{S}(0),\label{redfield_approx}
\end{align}

\noindent with the second line being the series expansion up to second order in the perturbation $\hat{H}_{SB}$. The first term on the right vanishes, since it is an average of a Gaussian variable, which is zero \cite{kubo}.

The correlation function appearing in the integrand is given explicitly by \eqref{autocorr_lvl}. We also replace $\tilde{\rho}_{S}(0)$ on the right hand side by $\tilde{\rho}_{S}(t)$, assuming that the weak perturbation causes only a negligible change in $\tilde{\rho}_{S}(t)$ between time $0$ and time $t$. Then, expanding the commutators and anticommutators \cite{heom1}:
\begin{multline}\label{redfield_timederiv}
\frac{d}{d t}\tilde{\rho}_{S}(t) = -\frac{1}{\hbar^2}\sum_{j}\int_{0}^{t}dt_{1}\left\lbrace \alpha_{j}(t-t_{1})\tilde{V}_{j}(t)\tilde{V}_{j}(t_{1})\tilde{\rho}_{S}(t)\right. +\\
-\alpha_{j}(t-t_{1})\tilde{V}_{j}(t_{1})\tilde{\rho}_{S}(t)\tilde{V}_{j}(t) - \alpha_{j}^{\ast}(t-t_{1})\tilde{V}_{j}(t)\tilde{\rho}_{S}(t)\tilde{V}_{j}(t_{1}) +\\
\left.\alpha_{j}^{\ast}(t-t_{1})\tilde{\rho}_{S}(t)\tilde{V}_{j}(t_{1})\tilde{V}_{j}(t)\right\rbrace.
\end{multline}

We choose to evaluate the density matrix in the exciton basis, where $\hat{H}_{S}\ket{\mu} = \hbar\omega_{\mu}\ket{\mu}$, because this means that the matrix elements of $\tilde{V}_{j}(t)$ have the form:
\begin{equation}
\bra{\mu}e^{i\hat{H}_{S}t/\hbar}\hat{V}e^{-i\hat{H}_{S}t/\hbar}\ket{\nu} = e^{i\omega_{\mu}t}\bra{\mu}\hat{V}\ket{\nu}e^{-i\omega_{\nu}t}.
\end{equation}

The density matrix can be transformed between exciton and site representation by using the matrix $\hat{U}$ of Eqn. \eqref{exciton_basis} that diagonalizes $\hat{H}_{S}$ \cite{leegwater_klug}.

In finding the matrix elements of \eqref{redfield_timederiv}, the completeness relation is used: $\sum_{\mu'}\ket{\mu'}\bra{\mu'}=\hat{1}$. For the first term on the right-hand side, the matrix elements are given by:
\begin{multline}\label{multiline_redfield}
-\frac{1}{\hbar^2}\sum_{j}\int_{0}^{t}dt_{1}\alpha_{j}(t-t_{1})\bra{\mu}\tilde{V}_{j}(t)\tilde{V}_{j}(t_{1})\tilde{\rho}_{S}(t)\ket{\nu} = \\
-\frac{1}{\hbar^2}\sum_{j}\sum_{\kappa}\sum_{\mu'}\int_{0}^{t}dt_{1}\cdot \alpha_{j}(t-t_{1})e^{i(\omega_{\mu}-\omega_{\nu})t}e^{i(\omega_{\mu'}-\omega_{\kappa})(t-t_{1})}\times\\
\bra{\mu}\hat{V}_{j}\ket{\kappa}\bra{\kappa}\hat{V}_{j}\ket{\mu'}\bra{\mu'}\hat{\rho}_{S}(t)\ket{\nu}.
\end{multline}

\noindent Defining a ``partial Fourier transform'',
\begin{equation}\label{partial_fourier}
\alpha_{j}[\omega;t] = \int_{0}^{t}\alpha_{j}(t_{1})e^{i\omega t_{1}}dt_{1},
\end{equation}

\noindent and changing the variable of integration, Eqn. \eqref{multiline_redfield} can be rewritten as:
\begin{equation}
-\frac{e^{i(\omega_{\mu}-\omega_{\nu})t}}{\hbar^2}\sum_{j}\sum_{\mu'}\sum_{\nu'}\delta_{\nu\nu'}\sum_{\kappa}\alpha_{j}[\omega_{\mu'}-\omega_{\kappa};t]\inner{\mu}{j}\inner{j}{\kappa}\inner{\kappa}{j}\inner{j}{\mu'}\rho_{\mu',\nu'}(t).
\end{equation}

\noindent Here, $\rho_{\mu',\nu'}(t) = \bra{\mu'}\hat{\rho}(t)\ket{\nu'}$, and $\inner{j}{\kappa} = U_{\kappa j}$, as in Eqn. \eqref{exciton_basis}.

Treating the other three terms in \eqref{redfield_timederiv} in the same manner gives the Time-Dependent Redfield equation \cite{leegwater_klug,i&f_redfield}:
\begin{equation}\label{tdredf}
\frac{d}{d t}\rho_{\mu,\nu}(t) = -\frac{i}{\hbar}(\omega_{\mu}-\omega_{\nu})\rho_{\mu,\nu}(t) - \sum_{\mu'}\sum_{\nu'}R_{\mu,\nu,\mu',\nu'}(t)\rho_{\mu',\nu'}(t).
\end{equation}

The key quantity in the Redfield theory is the matrix $R_{\mu,\nu,\mu',\nu'}(t)$, which governs the relaxation of the electronic system to equilibrium. It is given by:
\begin{multline}\label{relax_matr}
R_{\mu,\nu,\mu',\nu'}(t) = \Gamma_{\nu',\nu,\mu,\mu'}(t) + \Gamma_{\mu',\mu,\nu,\nu'}^{\ast}(t) \\ - \delta_{\nu\nu'}\sum_{\kappa}\Gamma_{\mu,\kappa,\kappa,\mu'}(t) - \delta_{\mu\mu'}\sum_{\kappa}\Gamma_{\nu,\kappa,\kappa,\nu'}^{\ast}(t).
\end{multline}

\noindent Here $\Gamma_{\mu,\nu,\mu',\nu'}(t)$ is the time-dependent damping matrix, and its elements are \cite{i&f_redfield,tanimura_redfield}:
\begin{equation}
\Gamma_{\mu,\nu,\mu',\nu'}(t) = \frac{1}{\hbar^2}\sum_{j}\inner{\mu}{j}\inner{j}{\nu}\inner{\mu'}{j}\inner{j}{\nu'}\alpha_{j}[\omega_{\nu'} - \omega_{\mu'};t].
\end{equation}

The time-dependence of the relaxation matrix \eqref{relax_matr} highlights the non-Markovian nature of the Time-Dependent Redfield theory: Eqn. \eqref{partial_fourier} is responsible for memory effects \cite{ishizaki_review,tanimura_redfield}, and using Eqn. \eqref{corrfunc_exp}, this function is given by \cite{tanimura_redfield}:
\begin{equation}
\alpha_{j}[\omega;t] = p_{j0} + \sum_{k=1}^{K}p_{jk}\frac{1-e^{(i\omega - \gamma_{jk}) t}}{\gamma_{jk}-i\omega}.
\end{equation}

\subsection{Discussion}

The following points are salient when considering Redfield theory:

\begin{itemize}
\item Due to the assumption of weak system-bath coupling in Eqn. \eqref{redfield_approx}, the full dynamics are not captured: the perturbation $\hat{H}_{SB}$ is proportional to a linear combination of bath coordinates, so can only induce single-phonon transitions in the bath \cite{redfield_forster} (there will only be non-zero matrix elements between bath states whose energy levels are adjacent).

\item In its original form \cite{redfield1}, the Redfield theory was Markovian. This theory can be derived simply from the Time-Dependent Redfield theory by assuming that $e^{-\gamma t}\approx \gamma^{-1}\delta_{+}(t)$ in Eqn. \eqref{partial_fourier}: qualitatively, by time $t$, the integrand has already decayed to zero, so that the upper bound of the integral can be set to infinity without appreciating its value.

\item The only change made to the Time-Dependent theory to derive the Markovian theory is to replace $\alpha_{j}[\omega;t]$ by $\alpha_{j}[\omega] = \alpha_{j}[\omega;t\rightarrow\infty]$ \cite{ishizaki_review}, giving a relaxation matrix $R_{\mu,\nu,\mu',\nu'}$ that is time-independent.

\item When implementing the Time-Dependent Redfield theory to evolve the density matrix through time, the relaxation matrix \eqref{relax_matr} must be calculated at each timestep, whereas it is not dependent on time in the Markovian Redfield case so only needs to be calculated once.

\item More recently, a modified Redfield theory has been developed \cite{redfield_forster,modified_redfield}, which can interpolate between the Redfield and F\"orster theories. However, this theory ignores off-diagonal density matrix elements in the exciton basis, so cannot be used to describe dynamics in the site basis \cite{fleming_lightharv}.
\end{itemize}

\section{F\"orster Theory}\label{forster_theory}

F\"orster theory, or F\"orster-Dexter theory \cite{forster1,forster2,dexter}, was originally formulated to describe resonant energy transfer (RET) between two electronic states (a donor and an acceptor) \cite{RET}.

More recently, Jang \emph{et al.} introduced a generalized F\"orster theory \cite{silbey}, which takes into account situations in which environmental phonons have not relaxed to equilibrium before energy transfer occurs.

In this Section, we derive the original F\"orster theory, and then explain how to calculate the resulting energy transfer rate constants.

\subsection{Resonant Energy Transfer Rate}

For clarity, we begin by considering two sites, labelled 1 and 2, each of whose electronic states are associated with a continuum of vibrational modes, so that when there is no interaction between the sites, the vibronic states are,
\begin{equation}\label{vibronic_states}
\ket{\psi(E_{1},E_{2})},
\end{equation}

\noindent where $E_{j}$ is the vibronic energy of site $j$. $\hat{H}_{0}$ is the unperturbed Hamiltonian:
\begin{equation}
\hat{H}_{0} = \sum_{j=1}^{2}\hbar\omega_{j}\ket{j}\bra{j} + \hat{H}_{B} + \hat{H}_{SB},
\end{equation}

\noindent such that $\ket{\psi(E_{1},E_{2})}$ is an eigenstate of $\hat{H}_{0}$:
\begin{equation}
\bra{\psi(E_{1},E_{2})}\hat{H}_{0}\ket{\psi(E_{1},E_{2})} = E_{1} + E_{2}.
\end{equation}

Now, we add an interaction $\hat{J}$ (which has no diagonal matrix elements between vibronic states) to the Hamiltonian (note that this gives $\hat{H}_{0} + \hat{J} = \hat{H}$, the full Hamiltonian):
\begin{equation}
\hat{J} = \sum_{j,k\ne j}\hbar J_{jk}(\ket{j}\bra{k} + \ket{k}\bra{j}).
\end{equation}

By construction, this interaction is weak (its matrix elements are smaller than the reorganization energies $\lambda_{j}$) and treating it as a perturbation, it will induce vibronic transitions, leading to energy transfer between sites.

We take site 1 to be the ``donor'' and site 2 the ``acceptor'', with the former being electronically excited initially and the latter being unexcited, giving an initial state denoted $\ket{\psi(\epsilon_{1},\epsilon_{2})}$ (where $\epsilon_{1}$ is a vibrational level of the excited electronic state of site 1 and likewise $\epsilon_{2}$ for the ground electronic state of site 2).

Invoking Fermi's Golden Rule (\cite{forster_original}, p. 178 of \cite{dirac}) gives the probability of being in some state $\ket{\psi(E_{1},E_{2})}$ as a function of time. Since this probability is the square modulus of a single probability amplitude, we are ignoring phase information and any theory built on this framework will be incoherent. The probability is given by:
\begin{equation}
P(E_{1},E_{2},t;\epsilon_{1},\epsilon_{2},0) = |\bra{\psi(\epsilon_{1},\epsilon_{2})}\hat{J}\ket{\psi(E_{1},E_{2})}|^{2} \cdot \frac{4\sin^{2}(\Delta E t/2\hbar)}{\Delta E^{2}},
\end{equation}
\noindent where $\Delta E = (\epsilon_{1}+\epsilon_{2}) - (E_{1} + E_{2})$ is the energy difference between the initial and final vibronic states. The total rate of transfer from site 1 to site 2 is given by integrating over all final energies $E_{1}$ and $E_{2}$ (where $E_{1}$ is a vibrational level of the ground electronic state of site 1 and likewise $E_{2}$ for the excited electronic state of site 2):
\begin{multline}
P_{1\rightarrow 2}(t;\epsilon_{1},\epsilon_{2},0) = \\ \int_{0}^{\infty}dE_{1}\int_{0}^{\infty}dE_{2}|\bra{\psi(\epsilon_{1},\epsilon_{2})}\hat{J}\ket{\psi(E_{1},E_{2})}|^{2} \cdot \frac{4\sin^{2}(\Delta E t/2\hbar)}{\Delta E^{2}}.
\end{multline}

The integrand contains a function of the form $\sin^{2}(\kappa x t)/x^{2}$ (where $\kappa$ is a constant), which in the limit as $t\rightarrow \infty$ is proportional to the delta-function. We thus make the approximation \cite{forster_original}:
\begin{equation}
\frac{4\sin^{2}(\Delta E t/2\hbar)}{\Delta E^{2}} \approx \frac{\pi t}{\hbar^{2}}\delta\left(\frac{\Delta E}{2\hbar}\right) = \frac{2\pi t}{\hbar}\delta(\Delta E),
\end{equation}\label{delta_approx}

\noindent and use the property (p. 60 of \cite{dirac}):
\begin{align}
\delta(\Delta E) = \delta(\epsilon_{1}+\epsilon_{2}-E_{1}-E_{2}) & = \int_{-\infty}^{\infty}d(\hbar\omega)\cdot\delta(\hbar\omega+\epsilon_{2}-E_{2})\delta(\hbar\omega-\epsilon_{1}+E_{1}) \nonumber\\
& = \hbar\int_{-\infty}^{\infty}d\omega\cdot\delta(E_{2}-[\epsilon_{2}+\hbar\omega])\delta(E_{1}-[\epsilon_{1}-\hbar\omega]).
\end{align}

This gives:
\begin{multline}\label{delta_fns}
P_{1\rightarrow 2} = 2\pi t\int_{-\infty}^{\infty}d\omega\int_{0}^{\infty}dE_{1}\int_{0}^{\infty}dE_{2} |\bra{\psi(\epsilon_{1},\epsilon_{2})}\hat{J}\ket{\psi(E_{1},E_{2})}|^{2}\times \\
\delta(E_{2}-[\epsilon_{2}+\hbar\omega])\delta(E_{1}-[\epsilon_{1}-\hbar\omega]),
\end{multline}

\noindent and we see, since $\Delta E = 0$ (due to the approximation in Eqn. \eqref{delta_approx}), that energy is conserved, and from Eqn. \eqref{delta_fns}, that $\hbar\omega$ is the energy transferred from donor to acceptor \cite{forster_original}.

Finally, the rate constant for energy transfer is found by using the definition $\int_{-\infty}^{\infty}f(x)\delta(x-a)dx = f(a)$ and taking the time-derivative of the probability:
\begin{equation}\label{forster_rate}
k_{1\rightarrow 2}^{F}(\epsilon_{1},\epsilon_{2}) = 2\pi\int_{-\infty}^{\infty}d\omega|\bra{\psi(\epsilon_{1},\epsilon_{2})}\hat{J}\ket{\psi(\epsilon_{1}-\hbar\omega,\epsilon_{2}+\hbar\omega)}|^{2}.
\end{equation}

Since the probabilities depend linearly on time, the rate constants are time-independent. Next, we will rewrite the integrand in terms of the more familiar functions of open quantum systems.

\subsection{Spectral Overlap}\label{spectral_overlap}

We now assume that the system is initially at equilibrium, so that we do not precisely know the initial vibronic energy; rather, we know that the probability density of being in state $\ket{\psi(\epsilon_{1},\epsilon_{2})}$ is given by $g(\epsilon_{1})g(\epsilon_{2})$, where $g(\epsilon) = e^{-\beta\epsilon}/Z$ and $Z = \int_{0}^{\infty}e^{-\beta\epsilon}d\epsilon$ is the canonical partition function. The total rate of transfer is given by:
\begin{multline}
k_{1\rightarrow 2}^{F} =2\pi\int_{-\infty}^{\infty}d\omega\int_{0}^{\infty}d\epsilon_{1}\int_{0}^{\infty}d\epsilon_{2}\cdot g(\epsilon_{1})g(\epsilon_{2})\times\\
|\bra{\psi(\epsilon_{1},\epsilon_{2})}\hat{H}\ket{\psi(\epsilon_{1}-\hbar\omega,\epsilon_{2}+\hbar\omega)}|^{2}.
\end{multline}

Within the Born-Oppenheimer approximation, the vibronic state is given as a product of electronic and vibrational states:
\begin{equation}
\ket{\psi(E_{1},E_{2})} = \ket{\psi_{\text{elec}}}\ket{\chi_{1}(E_{1})}\ket{\chi_{2}(E_{2})},
\end{equation}

\noindent where $\ket{\chi_{j}(E_{j})}$ is the vibrational state of site $j$. We also take $\ket{1}$ as the initial electronic state, with only the donor excited, and $\ket{2}$ as the final electronic state, with only the acceptor excited. Then, using $\langle 1|\hat{J}|2\rangle = \hbar J_{12}$:
\begin{multline}
k_{1\rightarrow 2}^{F} = 2\pi\hbar^{2}|J_{12}|^{2}\int_{-\infty}^{\infty}d\omega \left(\int_{0}^{\infty}d\epsilon_{1}\cdot g(\epsilon_{1}) |\langle\chi_{1}(\epsilon_{1})|\chi_{1}(\epsilon_{1}-\hbar\omega)\rangle|^{2}\right)\times\\
\left(\int_{0}^{\infty}d\epsilon_{2}\cdot g(\epsilon_{2}) |\langle\chi_{2}(\epsilon_{2})|\chi_{2}(\epsilon_{2}+\hbar\omega)\rangle|^{2}\right).
\end{multline}

The fluorescence and absorption spectral line shapes, $F_{1}[\omega]$ and $A_{2}[\omega]$ respectively, are defined in terms of Franck-Condon factors $S^{2}(\epsilon_{j},\epsilon_{j}\pm\hbar\omega) = |\langle\chi_{j}(\epsilon_{j})|\chi_{j}(\epsilon_{j}\pm\hbar\omega)\rangle|^{2}$ \cite{forster_original}:
\begin{subequations}
\begin{equation}
F_{1}[\omega] = 2\pi\hbar\int_{0}^{\infty}d\epsilon_{1}\cdot g(\epsilon_{1}) S^{2}(\epsilon_{1},\epsilon_{1}-\hbar\omega),
\end{equation}
\begin{equation}
A_{2}[\omega] = 2\pi\hbar\int_{0}^{\infty}d\epsilon_{2}\cdot g(\epsilon_{2}) S^{2}(\epsilon_{2},\epsilon_{2}+\hbar\omega).
\end{equation}
\end{subequations}

\noindent The rate constant is given by:
\begin{equation}\label{overlap_equation}
k_{1\rightarrow 2}^{F} = \frac{1}{2\pi}|J_{12}|^{2}\int_{-\infty}^{\infty}F_{1}[\omega]A_{2}[\omega]d\omega.
\end{equation}

All that is needed now is a method to calculate these spectra using known parameters such as the bath correlation functions $\alpha_{j}(t)$, site energies $\hbar\omega_{j}$ and reorganization energies $\hbar\lambda_{j}$.

This connection is made in \cite{redfield_forster,chromophore_solvent,mukamel}, where the fluorescence and absorption spectra are given by $F_{1}[\omega] = \int_{-\infty}^{\infty}F_{1}(t)e^{i\omega t}dt$ and $A_{2}[\omega] = \int_{-\infty}^{\infty}A_{2}(t)e^{i\omega t}dt$. In the time-domain, the fluorescence and absorption line-shape functions are evaluated using the cumulant expansion method \cite{mukamel,cumulant}, giving:
\begin{subequations}
\begin{equation}
F_{1}(t) = e^{-i(\omega_{1}-\lambda_{1})t-g_{1}^{\ast}(t)},
\end{equation}
\begin{equation}
A_{2}(t) = e^{-i(\omega_{2}+\lambda_{1})t-g_{2}(t)}.
\end{equation}
\noindent The function $g_{j}(t)$ is defined:
\begin{equation}
g_{j}(t) = \frac{1}{\hbar^{2}}\int_{0}^{t}dt_{1}\int_{0}^{t_{1}}dt_{2}\cdot\alpha_{j}(t_{2}).
\end{equation}
\end{subequations}

$\hbar\lambda_{j}$ is the reorganization energy of site $j$, the energy difference between the excited vibrational level reached by a Franck-Condon transition and the ground vibrational level, in a given electronic state \cite{ishizaki_review}.

Parseval's theorem states that the overlap integral in Eqn. \eqref{overlap_equation} can be carried out in either domain, so that with some rearrangement, and assuming that energy transfer in a donor/acceptor pair is not affected by the presence of other sites, the F\"orster rate constant for transfer between two electronic states $\ket{j}$ and $\ket{k}$ is given by the equation ($\Re$ denotes the real part) \cite{redfield_forster}:
\begin{equation}\label{forster_rateconst}
k_{j\rightarrow k}^{F} = 2|J_{jk}|^{2}\Re\left( \int_{0}^{\infty}F_{j}(t)A_{k}(t)dt\right).
\end{equation}

\begin{figure}[!ht]
\begin{center}
\includegraphics[width=13.5cm, keepaspectratio=true]{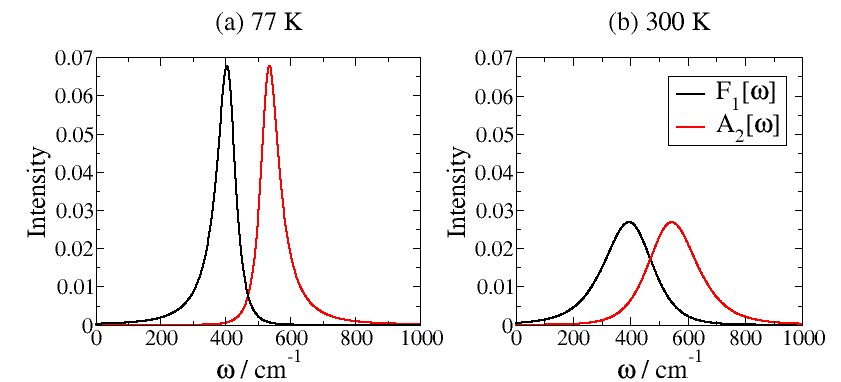}
\caption[Overlap of fluorescence spectrum of site 1 and absorption spectrum of site 2.]{\small{Overlap of the fluorescence spectrum of site 1 and the absorption spectrum of site 2 in the FMO complex at (a) 77 K and (b) 300 K.}}
\label{overlap}
\end{center}
\end{figure}

Fig. \ref{overlap} illustrates the spectra $F_{1}[\omega]$ and $A_{2}[\omega]$, showing their overlap at both 77 K and 300 K. The effect of the function $g_{j}(t)$ in $F_{1}(t)$ and $A_{2}(t)$ is to broaden the spectra, from the delta-functions observed if $\alpha_{j}(t) = 0$ (that is, in the absence of baths).

At the higher temperature, the bath correlation function is larger, the spectral lines are broader and there is more overlap between them, thus increasing the rate of transfer from site 1 to site 2.

Chapter \ref{numerical_results} will test the applicability of the F\"orster theory by comparing the predicted dynamics to the exact dynamics of the HEOM. The quality of the predictions of F\"orster theory will allow us to determine whether or not coherence is important in the physics of the FMO.


{\renewcommand{\baselinestretch}{1}\normalsize
\chapter{Implementation Of The HEOM}\label{implementation}
}

The number of ADOs required for a HEOM calculation, Eqn. \eqref{num_ados}, can be very large, and the computational cost of propagating these matrices through time is high. The first Section of this Chapter explains how the HEOM are solved numerically, with an eye towards maximizing efficiency.

The second Section explains why the bath correlation function $\alpha(t)$ can be represented as a sum of exponentials (which is essential for the HEOM), and explores both the Matsubara and the Pad\'e spectral decompositions, highlighting the greater accuracy that can be achieved by using the latter.

\section{Numerical Propagation}

Eqn. \eqref{heom_eqs} provides a set of coupled differential equations for all of the ADOs. If $\underline{\sigma}(t)$ is a vector containing all of the elements of all ADOs, the equations can be rewritten:
\begin{equation}\label{de_a_sigma}
\frac{d}{d t}\underline{\sigma}(t) = f\left(\underline{\sigma}(t),t\right) \equiv \underline{\underline{\mathcal{A}}}\cdot\underline{\sigma}(t).
\end{equation}

The second form follows from the fact that the differential equations do not contain the elements of the ADOs to any order greater than the first: $\underline{\underline{\mathcal{A}}}$ is a linear operator. We now look at the numerical solution of these equations.

Originally, the well-established fourth-order Runge-Kutta (RK4) method was chosen for numerical integration (Sec. 15.1 of \cite{numerical_recipes}). If timestep $\delta t$ is used for the integration, then the error is of the order of $\delta t^{5}$. However, an alternative numerical solution was noted. The formal solution to Eqn. \eqref{de_a_sigma} is $\underline{\sigma}(t+\delta t) = \exp(\underline{\underline{A}}\delta t)\underline{\sigma}(t)$ which can be Taylor-series expanded:
\begin{equation}
\underline{\sigma}(t+\delta t) = \sum_{m=0}^{M}\frac{\delta t^{m}}{m!}\underline{\underline{A}}^{m}\underline{\sigma}(t) + \mathcal{O}(\delta t^{M+1}).
\end{equation}

If $M = 4$ is chosen, the error will again be on the order of $\delta t^{5}$. The Taylor series method has an advantage that is not immediately obvious: it leads to a reduction in the memory required to run a simulation.

The reason for this is that the RK4 method involves calculation of four vectors $\underline{k}_{1}$, $\underline{k}_{2}$, $\underline{k}_{3}$ and $\underline{k}_{4}$. It is possible to use only two of these vectors, with one used for permanent storage of $\underline{k}_{1}$ and the other updated so that it will variously store the remaining three vectors.

On the other hand, the Taylor-series method requires that only one such vector, $\underline{k}_{1}$, is used. The program will loop through four steps, each time calculating $\frac{\delta t}{m}\underline{\underline{A}}\cdot\underline{k}_{1}$ and adding this to the current $\underline{\sigma}(t)$.

The RK4 method requires that the program for implementing the HEOM stores one more vector that contains as many elements as there are in all of the ADOs (that is, $\mathcal{N}\times\text{N}^{2}$, where $\mathcal{N}$ is as in Eqn. \eqref{num_ados}). This can lead to a significant reduction in memory usage.

The time-derivative of each ADO depends on the values of operators on the levels above and below its own. The number of ADOs in a given level is potentially huge, so that an efficient indexing system is vital: before the time-evolution begins, an indexing matrix is set up, each of whose rows corresponds to a single ADO, and whose columns contain references to the operators to which the ADO of this row is coupled, for ease of use at runtime.

For a given problem, we will have a system Hamiltonian $\hat{H}_{S}$ and a bath characterized by $\alpha(t)$, a sum of exponentials. By increasing the number of these exponentials, our $\alpha(t)$ will approach the true function. This number and the number of hierarchical levels may be increased until the dynamics converge and are numerically exact.

Within the program to carry out the HEOM, both the indexing routine and the calculation of the time-derivatives were parallelized: running on multiple CPUs allowed a faster, more efficient program.

The routine calculating the time-derivatives involves (when $\alpha(t)$ contains no $\delta_{+}$-function) the calculation of 6 matrix products for each ADO, or $6 N^3$ scalar multiplications. However, for the commutator and anticommutator of $\hat{V}_{j}$ with a general operator $\hat{O}$, a general matrix element is given by:
\begin{subequations}
\begin{equation}
\bra{m}\hat{V}_{j}^{\times}\hat{O}\ket{n} = \delta_{m j}\bra{n}\hat{O}\ket{n} - \bra{m}\hat{O}\ket{j}\delta_{j n},
\end{equation}
\begin{equation}
\bra{m}\hat{V}_{j}^{\circ}\hat{O}\ket{n} = \delta_{m j}\bra{j}\hat{O}\ket{n} + \bra{m}\hat{O}\ket{j}\delta_{j n}.
\end{equation}
\end{subequations}

Using this formula instead of explicitly calculating the commutators and anticommutators will give only $2 N^{3}$ scalar multiplications (in finding the commutator $\hat{\mathcal{L}}\hat{\sigma}_{\textbf{n}}(t)$), and led to a substantial speeding up of the program, thus increasing efficiency. 

\section{Spectral Decomposition}

The Lorentz-Drude spectral density is frequently used for calculations because it allows the bath correlation function to be written analytically as a sum of decaying exponentials, meaning that the HEOM can be used to benchmark our calculations.

Traditionally, the integral in Eqn. \eqref{bath_corrfunc} (or equivalently \eqref{bath_corrfunc_infty}) is carried out as described below to give the Matsubara series \cite{ishizakifleming,tanimura2,filtering1,filtering2}; this series converges very slowly to an exact result, and so we describe an alternative, the Pad\'e series, which is found to converge very much faster.

\subsection{Matsubara Series}

Inserting the definition of $J(\omega)$ from Eqn. \eqref{drude_lorentz}, with subscripts dropped for clarity, into \eqref{bath_corrfunc} and making use of the fact that the integrand is an even function gives the integral:
\begin{equation}
\alpha(t) = \frac{\gamma\lambda\hbar}{\pi}\int_{-\infty}^{\infty}d\omega \left[ \frac{\omega\coth(\beta\hbar\omega/2)e^{-i\omega t}}{\omega^2 + \gamma^2} + \frac{\omega e^{-i\omega t}}{\omega^2 + \gamma^2}\right].
\end{equation}

This can be solved straightforwardly using contour integration: the integrand has poles at $\omega = \pm i\gamma$, and (for the first term) at $\sinh(\beta\hbar\omega/2) = 0 \Rightarrow \omega = \pm i 2\pi j / \beta\hbar$, with $j$ an integer. As we are only interested in $\alpha(t)$ for $t > 0$, we use the contour below, with $R\rightarrow\infty$:

\setlength{\unitlength}{7cm}
\begin{center}
\begin{picture}(1,1)
\put(-0.25,0.5){\vector(1,0){1.5}}
\put(0.5,0){\vector(0,1){1}}
\put(0.55,1){Im$(\omega)$}
\put(1.25,0.43){Re$(\omega)$}
\linethickness{0.5mm}
\put(0.05,0.5){\line(1,0){0.9}}
\put(0.6,0.5){\line(-1,1){0.03}}
\put(0.6,0.5){\line(-1,-1){0.03}}
\put(0.5,0,5){\arc[180,360]{0.45}}
\put(0.94,0.55){$R$}
\put(0.01,0.55){$-R$}
\end{picture}
\end{center}

The integral over the semi-circle vanishes (pp. 113-115 of \cite{complex_analysis}), so that the integral over the real line is given by a sum of residues:
\begin{align}
\alpha(t) & = -2\gamma\lambda\hbar i  \left\lbrace\Res_{\omega = -i\gamma} + \sum_{k=1}^{\infty}\Res_{\omega = -i2\pi k / \beta\hbar}\right\rbrace \cdot \left\lbrace \frac{\omega\coth(\beta\hbar\omega/2)e^{-i\omega t}}{\omega^2 + \gamma^2} + \frac{\omega e^{-i\omega t}}{\omega^2 + \gamma^2}\right\rbrace \nonumber\\
& = \gamma\lambda\hbar \left\lbrace \cot\left(\frac{\beta\hbar\gamma}{2}\right) - i\right\rbrace e^{-\gamma t} + \sum_{k=1}^{\infty}\frac{4\gamma\lambda\nu_{k}}{\beta\left(\nu_{k}^2 - \gamma^2\right)}e^{-\nu_{k}t}.
\end{align}

Here, $\nu_{k} = 2\pi k/\beta\hbar$ is called a Matsubara frequency and we have a sum of decaying exponentials, as desired. it is possible to rewrite $\cot(\theta)$ as an infinite sum by applying the residue theorem once again (pp. 131-133 of \cite{complex_analysis}), giving the form used by Ishizaki and Fleming \cite{ishizakifleming}:
\begin{equation}\label{matsubara2}
\alpha(t) = \frac{2\lambda}{\beta}\left\lbrace 1-\sum_{k=1}^{\infty}\frac{2\gamma^{2}}{\nu_{k}^2 - \gamma^2} - i\frac{\beta\hbar\gamma}{2} \right\rbrace e^{-\gamma t} + \frac{2\lambda}{\beta}\sum_{k=1}^{\infty}\frac{2\gamma\nu_{k}}{\nu_{k}^2 - \gamma^2}e^{-\nu_{k}t}.
\end{equation}

It is from this expression that a truncation is suggested: a certain number $K$ of exponential terms are retained, and for $k > K$, a Markovian approximation is used: it is assumed that $\nu_{k}$ is large enough that $\nu_{k}e^{-\nu_{k}t} \approx \delta_{+}(t)$, or equivalently that these terms decay quickly enough that they are essentially zero for $t > 0$, and only contribute at $t = 0$.

\begin{figure}[!ht]
\begin{center}
\includegraphics[width=14cm, keepaspectratio=true]{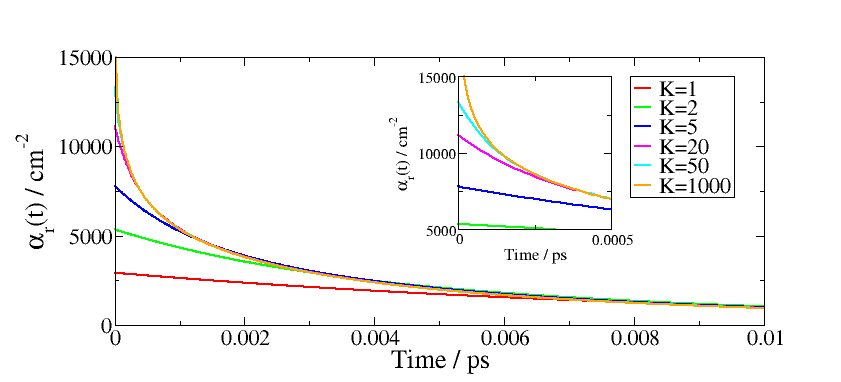}
\caption[Convergence of Matsubara series.]{\small{Convergence of the Matsubara series at 77 K with varying number of exponential terms $K$, for the Lorentz-Drude spectral density. \emph{Inset: Magnified bath correlation functions at short times.}}}
\label{matsubara_figure}
\end{center}
\end{figure}

In practice, one would successively increase $K$ until converged results were found for the dynamics. Fig. \ref{matsubara_figure} shows $\alpha(t)$ for some values of $K$.

We see that a large value of $K$ would be required to fully converge the HEOM calculations, and so we next turn our attention to a better method of representing $\alpha(t)$.

\subsection{Pad\'e Series}\label{section_pade}

Faster convergence can be obtained by finding the Pad\'e approximant to the Bose-Einstein function, $f_{Bose}(x) = (1-e^{-x})^{-1}$ \cite{pade1,pade2}. That is, we find a rational function that approximates $f_{Bose}(x)$. We will focus on the [$N$-1/$N$] approximant (for reasons that will be discussed at the end of this Section), which uses a fraction of the form:
\begin{equation}\label{n-1/n}
\frac{\sum_{j=0}^{N-1}a_{j}x^{j}}{\sum_{k=0}^{N}b_{k}x^{k}}.
\end{equation}

In Appendix \ref{pade_approximant}, it is shown that the Bose-Einstein function can be approximated \cite{pade2}:
\begin{equation}\label{pade_bose}
f_{Bose}(x) \approx \frac{1}{x} + \frac{1}{2} + 2x\sum_{j=1}^{N}\frac{\eta_{j}}{x^2 + \xi_{j}^2},
\end{equation}

\noindent where the $\eta_{j}$ can be found using:
\begin{equation}
\eta_{j} = \left(N^2 + \frac{3}{2}N\right)\frac{\prod_{k=1}^{N-1}(\zeta_{k}^2 - \xi_{j}^2)}{\prod_{k\neq j}^{N} (\xi_{k}^2 - \xi_{j}^2)},\qquad j = 1,2,\dots,N
\end{equation}

Here, $\xi_{j}=2/c_{j}$ and $\zeta_{j}=2/\tilde{c}_{j}$, where $c_{j}$ are the positive eigenvalues of matrix $\underline{\underline{\Lambda}}$ and $\tilde{c}_{j}$ the positive eigenvalues of matrix $\underline{\underline{\tilde{\Lambda}}}$. $\underline{\underline{\Lambda}}$ is a $2N\times 2N$ matrix, and $\underline{\underline{\tilde{\Lambda}}}$ a $2N-1\times 2N-1$ matrix, with elements \cite{pade2}:
\begin{equation}\label{lambda_matrices}
\Lambda_{mn} = \frac{\delta_{m,n\pm 1}}{\sqrt{(2m+1)(2n+1)}} \hspace{2cm} \tilde{\Lambda}_{mn} = \frac{\delta_{m,n\pm 1}}{\sqrt{(2m+3)(2n+3)}}.
\end{equation}

\noindent There will be $N$ parameters $\xi_{j}$ and $N-1$ parameters $\zeta_{j}$.

Inserting \eqref{pade_bose} into \eqref{bath_corrfunc_infty} gives, with $\nu_{j} = \xi_{j}/\beta\hbar$ (hereafter, the Pad\'e frequencies):
\begin{equation}
\alpha(t) = \frac{2\gamma\lambda\hbar}{\pi}\int_{-\infty}^{\infty}d\omega \frac{\omega e^{-i\omega t}}{(\omega^2 + \gamma^2)}\cdot\left\lbrace \frac{1}{\beta\hbar\omega} + \frac{1}{2} + \frac{2\omega}{\beta\hbar}\sum_{j=1}^{N}\frac{\eta_{j}}{\omega^2 + \nu_{j}^2}\right\rbrace.
\end{equation}

This integral can be evaluated using the same contour used to derive the Matsubara series, to give, by analogy with Eqn. \eqref{matsubara2}:
\begin{equation}\label{pade_alpha}
\alpha(t) = \frac{2\lambda}{\beta}\left\lbrace 1 - \sum_{j=1}^{N}\frac{2\eta_{j}\gamma^2}{\nu_{j}^2 - \gamma^2} - i\frac{\gamma\beta\hbar}{2}\right\rbrace e^{-\gamma t} + \frac{2\lambda}{\beta}\sum_{j=1}^{N}\frac{2\eta_{j}\nu_{j}\gamma}{\nu_{j}^2 - \gamma^2}e^{-\nu_{j}t}.
\end{equation}

\begin{figure}[!ht]
\begin{center}
\includegraphics[width=14cm, keepaspectratio=true]{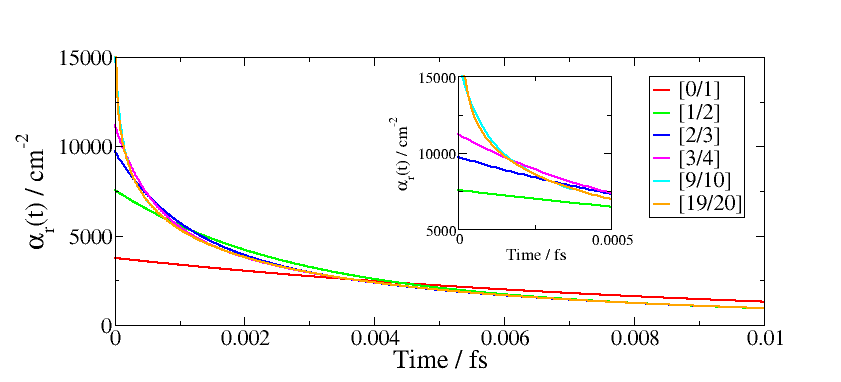}
\caption[Convergence of Pad\'e series.]{\small{Convergence of the [$N$-1/$N$] Pad\'e series at 77 K with varying $N$, for the Lorentz-Drude spectral density. \emph{Inset: Magnified bath correlation functions at short times.} The code used to produce this figure was supplied by David Manolopoulos.}}
\label{pade_figure}
\end{center}
\end{figure}

Fig. \ref{pade_figure} shows how the Pad\'e series for $\alpha(t)$ changes as the $N$ of [$N$-1/$N$] is varied. The improved convergence is very noticeable: while the Matsubara series required a large number of terms to give a good approximation, the Pad\'e series requires a much smaller number.

Using the Pad\'e decomposition, we can hope to predict the dynamics of energy transfer with an accuracy that might not be afforded to us by the Matsubara decomposition: HEOM simulations can be run with increasing number of exponentials until convergence is achieved.

On the other hand, because the Matsubara series requires so many more exponential terms to achieve convergence, it is likely that at some point, the computational expense will prohibit more exponentials from being used, before convergence is achieved.

\subsection{Discussion}

The Matsubara and Pad\'e series are not the only methods of representing $\alpha(t)$ as a sum of exponentials: there are several alternatives, including the Meier-Tannor decomposition \cite{meier_tannor}, in which the spectral density is fitted to an expression of the form:
\begin{equation}
J(\omega) = \frac{\pi}{2}\sum_{k=1}^{n}\frac{\omega}{((\omega+\Omega_{k})^{2} + \Gamma_{k}^{2})((\omega-\Omega_{k})^{2}+\Gamma_{k}^{2})}.
\end{equation}

Another possibility is to numerically integrate Eqn. \eqref{bath_corrfunc}, and then to fit the numerical data to an expression of the form \eqref{corrfunc_exp}. It may be appropriate, in this case, to allow both the prefactors and the frequencies in this expression to be complex.

Both of these methods involve numerical fitting, but since the Pad\'e series allows efficient analytical convergence to the exact correlation function (and efficient numerical convergence to the exact energy transfer dynamics) for a Lorentz-Drude spectral density, it is this method that we use.

As well as the [$N$-1/$N$] approximant, the [$N$/$N$] and [$N$+1/$N$] approximants were also considered in the original literature \cite{pade2}. The latter two require that a $\delta_{+}$-function approximation be introduced, whereas the former involves no such approximation.

We also briefly consider the high-temperature limit in order to make the connection with literature in this field \cite{ishizakifleming,tanimura2}. The imaginary part of the bath correlation function is independent of temperature, $\alpha_{i}(t) = -\lambda\gamma\hbar e^{-\gamma t}$. However, for the real part, using $\lim_{\beta\hbar\rightarrow 0}\coth(\beta\hbar\omega/2)=\frac{1}{\beta\hbar\omega}$:
\begin{equation}
\lim_{\beta\rightarrow 0}\alpha_{r}(t) = \frac{2\lambda\gamma}{\pi\beta}\int_{0}^{\infty}d\omega \frac{\cos(\omega t)}{\omega^2 + \gamma^2} = \frac{2\lambda}{\beta}e^{-\gamma t}.
\end{equation}

This high-temperature expression is purely classical, and we might consider further terms (due to the Pad\'e or Matsubara expansion) as ``quantum corrections'' to this \cite{tanimura2}.

Interestingly, even at high temperatures $\alpha_{i}(t)$ contains a factor of $\hbar$, so is inherently quantum-mechanical. This term can be expressed as the Fourier transform of $J(\omega)$, so can be related to quantum dissipation, whereas the $\beta$-dependence of $\alpha_{r}(t)$ relates it to equilibrium fluctuations, which are classical at high temperatures.


{\renewcommand{\baselinestretch}{1}\normalsize
\chapter{Numerical Results}\label{numerical_results}
}

We have now laid the theoretical foundations necessary to carry out calculations for the 7-site FMO system and the 24-site trimer, at a temperature of 300 K. Calculations at a temperature of 77 K have also been carried out, but are not shown here due to the length constraint for this thesis, as well as the fact that this will allow us to focus on results at physiological temperature, which are more relevant to this work.

We will use the same parameters as in the literature \cite{ishizakifleming}, so that $\hbar\lambda_{n} = 35 \text{~cm}^{-1}$ and $\hbar\gamma_{n} = 106.1 \text{~cm}^{-1}$ for each site. These values are found by fluorescent Stokes shift experiments \cite{ishizakifleming,chromophore_solvent}.

For the 7-site FMO, the system Hamiltonian used is that of Adolphs and Renger \cite{adolphs_renger}, given in Appendix \ref{supplementary}, Eqn. \eqref{7site_Hsys}. The couplings $\hbar J_{jk}$ between sites were found using the transition dipole moments of these sites and assuming the protein to provide a dielectric medium, while the site energies $\hbar\omega_{j}$ were calculated using the interaction between the sites and charged amino acid residues \cite{adolphs_renger}.

For the 24-site FMO, we follow the theoretical study of Ritschel \emph{et al.} \cite{zofe_fmo2}, whose ZOFE quantum master equation calculations used two different sets of site energies. One of these was found by Schmidt am Busch \emph{et al.} \cite{schmidt_am_busch} using a method similar to that of Adolphs and Renger \cite{adolphs_renger}, and the other by Olbrich \emph{et al.} \cite{olbrich} using molecular dynamics simulations and electronic structure calculations. The Hamiltonian is given in Appendix \ref{supplementary}, Eqn. \eqref{24site_full}, with Eqn. \eqref{24site_Hsys} giving the intra-monomer couplings, \eqref{24site_sites} giving the site energies for the two different cases \cite{schmidt_am_busch,olbrich} and \eqref{24site_inter} the inter-monomer couplings.

The initial conditions we use will reflect the physics of the complex: bacteriochlorophylls 1, 6 and 8 are closest to the chlorosome, and it is these that will be most likely to be excited initially \cite{schmidt_am_busch,zofe_fmo2}. Thus, for the 7-site system we will carry out simulations with excitation beginning on sites 1 or on 6, and for the trimer, on sites 1, 6 or 8.

\begin{figure}[!ht]
\begin{center}
\includegraphics[width=3cm, keepaspectratio=true]{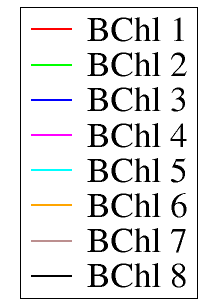}
\caption[Key for numerical results.]{\small{Key for numerical results.}}
\label{key}
\end{center}
\end{figure}

The results are colour-coded according to the key in Fig. \ref{key}, with the population of each site denoted by the colours shown.

We set these results out as follows: firstly, those for the 7-site system are presented, both for 1 ps and 15 ps (i.e., on the timescale of the electronic coherence and on the timescale of equilibration), and then those for the trimer.

These results will then allow conclusions to be drawn about which approximate method is most appropriate for our further work.
\section{7-Site Subsystem}

\begin{figure}[!p]
\begin{center}
\includegraphics[width=13.5cm,keepaspectratio=true]{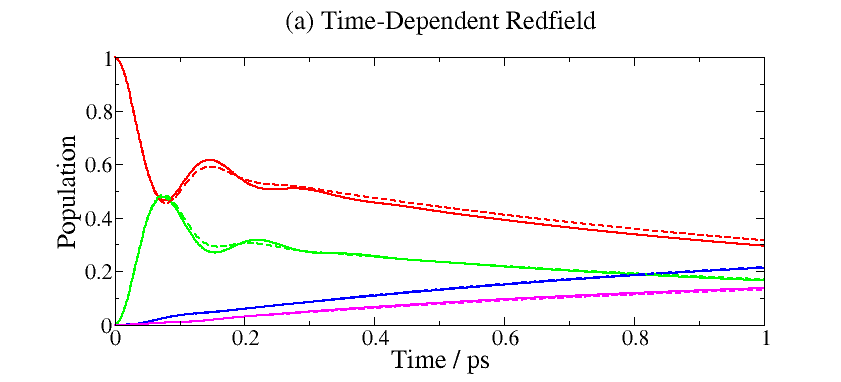}
\includegraphics[width=13.5cm,keepaspectratio=true]{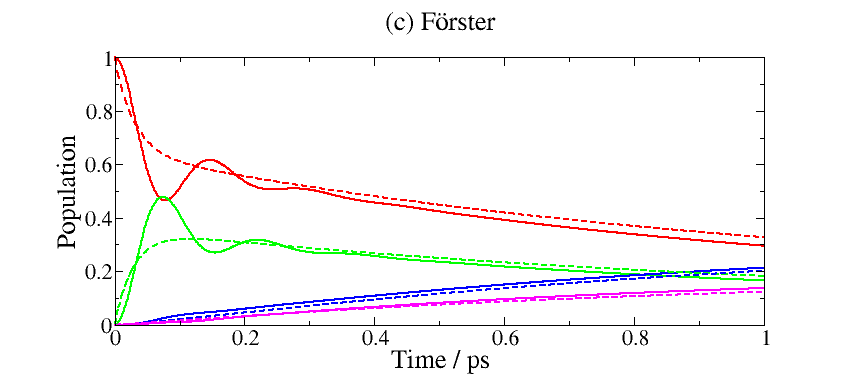}
\caption[Energy transfer dynamics for the 7-site FMO subsystem at 300 K, with initial excitation on site 1.]{\small{Energy transfer dynamics for the 7-site FMO subsystem at 300 K, with initial excitation on site 1. In each case, the solid lines show the HEOM result and the dashed lines show (a) Time-Dependent Redfield, (b) F\"orster results.}}
\label{300K_7site_e1}
\end{center}
\end{figure}

\begin{figure}[!p]
\begin{center}
\includegraphics[width=13.5cm,keepaspectratio=true]{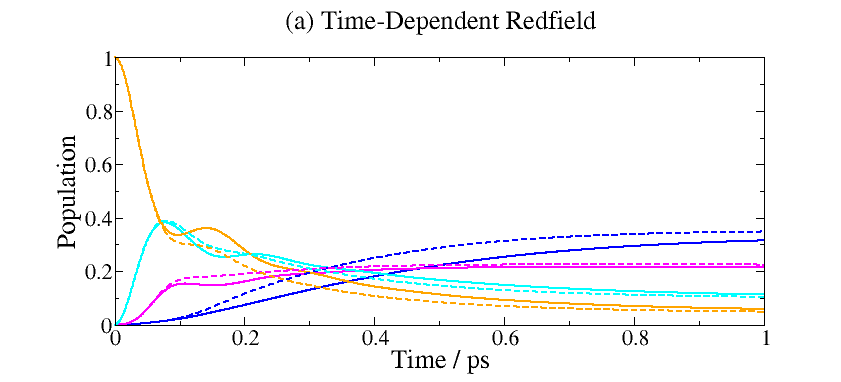}
\includegraphics[width=13.5cm,keepaspectratio=true]{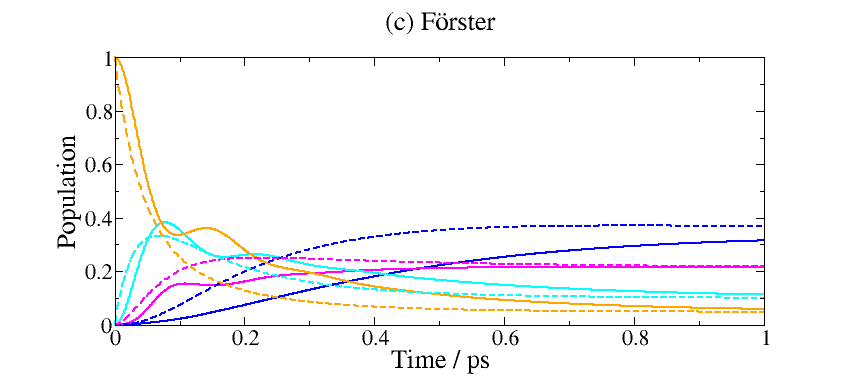}
\caption[Energy transfer dynamics for the 7-site FMO subsystem at 300 K, with initial excitation on site 6.]{\small{Energy transfer dynamics for the 7-site FMO subsystem at 300 K, with initial excitation on site 6. In each case, the solid lines show the HEOM result and the dashed lines show (a) Time-Dependent Redfield, (b) F\"orster results.}}
\label{300K_7site_e6}
\end{center}
\end{figure}

\begin{figure}[!p]
\begin{center}
\includegraphics[width=13.5cm,keepaspectratio=true]{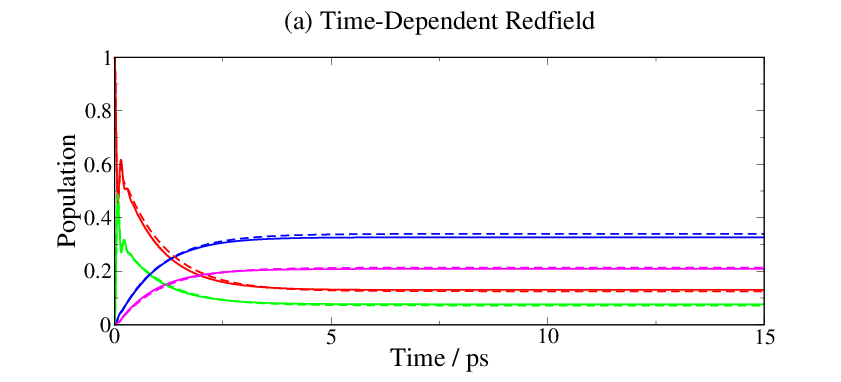}
\includegraphics[width=13.5cm,keepaspectratio=true]{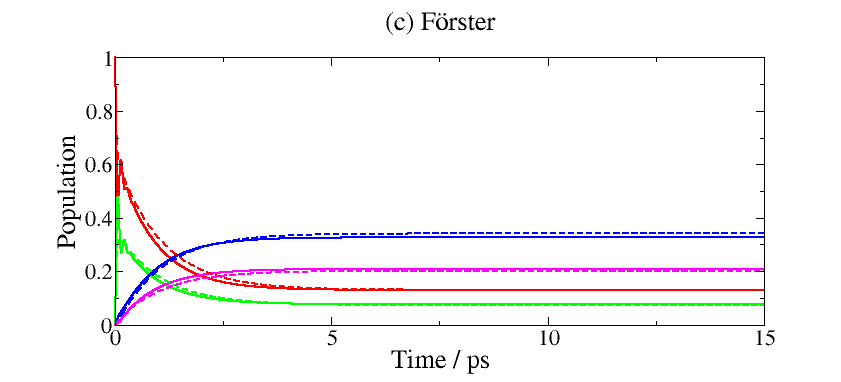}
\caption[Long-time energy transfer dynamics for the 7-site FMO subsystem at 300 K, with initial excitation on site 1.]{\small{Long-time energy transfer dynamics for the 7-site FMO system at 300 K, with initial excitation on site 1. In each case, the solid lines show the HEOM result and the dashed lines show (a) Time-Dependent Redfield, (b) F\"orster results.}}
\label{SteadyState_300K_e1}
\end{center}
\end{figure}

\begin{figure}[!p]
\begin{center}
\includegraphics[width=13.5cm,keepaspectratio=true]{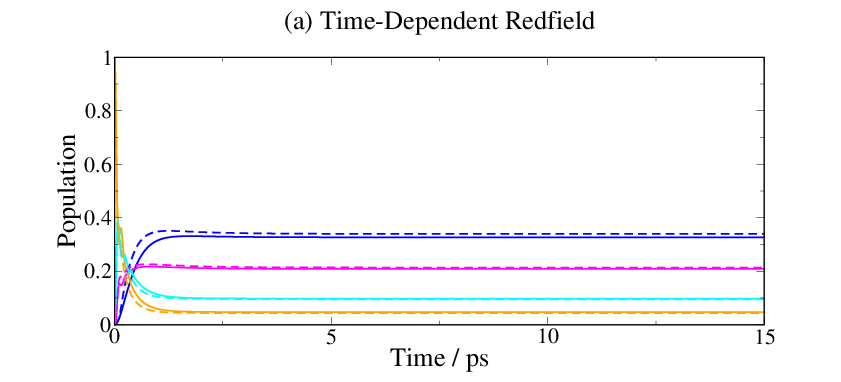}
\includegraphics[width=13.5cm,keepaspectratio=true]{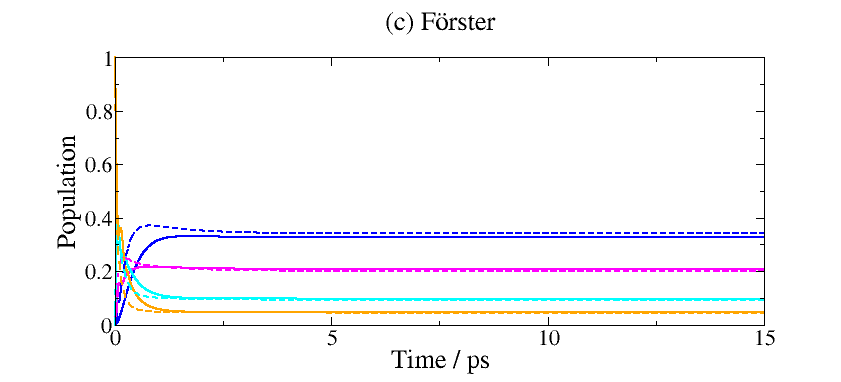}
\caption[Long-time energy transfer dynamics for the 7-site FMO subsystem at 300 K, with initial excitation on site 6.]{\small{Long-time energy transfer dynamics for the 7-site FMO system at 300 K, with initial excitation on site 6. In each case, the solid lines show the HEOM result and the dashed lines show (a) Time-Dependent Redfield, (b) F\"orster results.}}
\label{SteadyState_300K_e6}
\end{center}
\end{figure}

Figs. \ref{300K_7site_e1} and \ref{300K_7site_e6} compare the dynamics of Time-Dependent Redfield and F\"orster theories to those of the HEOM at 300 K, up to 1 ps, with  the initial excitation on site 1 and on site 6 respectively. Figs. \ref{SteadyState_300K_e1} and \ref{SteadyState_300K_e6} show the same dynamics up to 15 ps. In each case, the populations of only four sites are shown, as the rest of the populations stay close to zero.

At 300 K, the HEOM dynamics fully converged with 2 Pad\'e exponential terms in $\alpha(t)$ and 4 levels of the hierarchy. At lower temperatures, the weaker system-bath interaction means that fewer levels of the hierarchy are required, but more exponential terms 

In these results, electronic coherence is observed on short timescales. The formation of a coherent superposition between two states $\ket{j}$ and $\ket{k}$ is expected when the magnitude of $J_{jk}$, the coupling between them, is comparable to $|\omega_{j}-\omega_{k}|$, their energy gap. Consulting Eqn. \eqref{7site_Hsys}, it is for this reason that coherence is observed between states $\ket{1}$ and $\ket{2}$, as well as between $\ket{6}$ and $\ket{5}$.

Whether the first or the sixth site is initially excited, the steady state is the same, but if site 6 is initially excited then this state is reached more quickly. Of the 7 sites, BChl 6 has the highest energy, and we might thus intuitively expect that transfer away from this site is faster, which is borne out by the results.

Calculations were also carried out with the Markovian Redfield theory, and compared to the Time-Dependent Redfield results, the only difference was that the short-time oscillations were less pronounced using the Markovian theory. After the envelope of these oscillations decayed, both theories were quantitatively identical.

F\"orster theory's performance tends to be worse in general than that of Time-Dependent Redfield theory, especially at shorter timescales, as is particularly noticeable in Fig. \ref{300K_7site_e6}.

Both Redfield and F\"orster theories predict the equilibrium populations very well at 300 K. However, as seen in Fig. \ref{SteadyState_300K_e6}, the perturbative methods predict a ``bump'' in the dynamics that is not seen in the HEOM simulation.

Overall, the reasonable accuracy of F\"orster theory's predictions at 300 K suggests that even if coherent effects are important at low temperatures, they do not seem to be very much so at room temperature: energy barriers can be surmounted thermally and therefore tunnelling contributions have little effect.

\section{24-Site Trimer}

In this Section, we present accurate results for the full FMO trimer up to 1 ps, as well as results up to 15 ps that are converged with respect to number of hierarchy levels, but not fully converged with respect to number of exponentials in $\alpha(t)$ (however, it is known that these results are very accurate). These results have not previously been published.

The reason that the 15 ps results are not converged with respect to number of exponentials is one of computation: up to the shorter time, converged results using the HEOM required a large amount of time and memory, and a prohibitively larger time would be required to observe the fully converged steady states.

However, by using the same bath correlation function for all methods, we can carry out a rigorous comparison, which will be illuminating.

\begin{figure}[!p]
\begin{center}
\includegraphics[width=13.1cm,keepaspectratio=true]{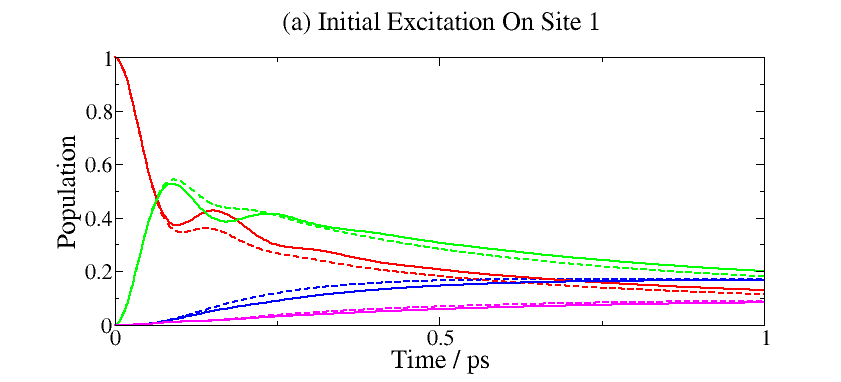}
\includegraphics[width=13.1cm,keepaspectratio=true]{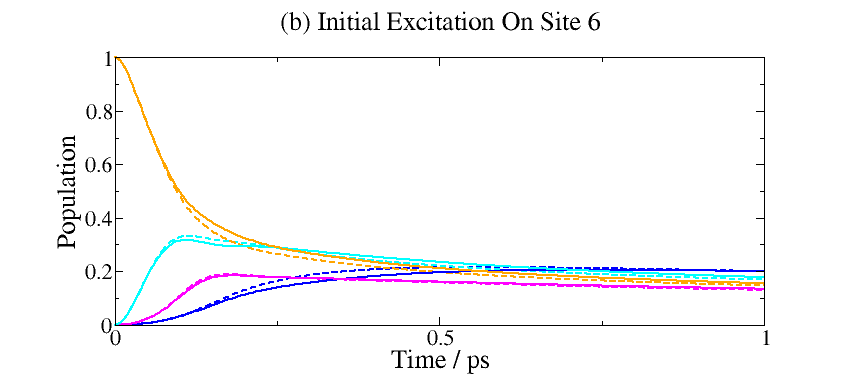}
\includegraphics[width=13.1cm,keepaspectratio=true]{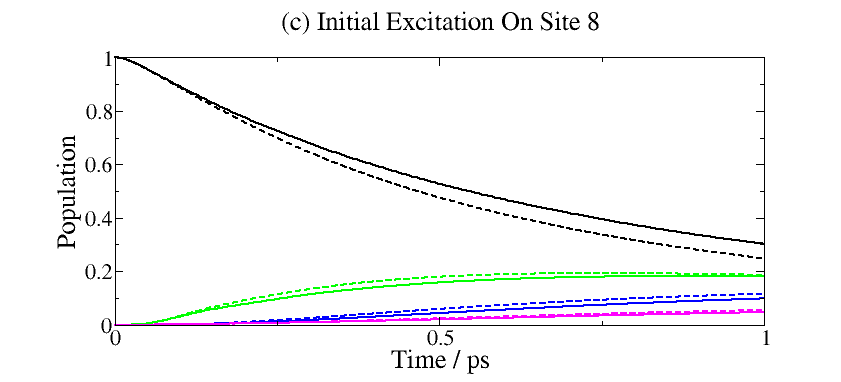}
\caption[Comparison of energy transfer dynamics to the Redfield approximation for the 24-site FMO trimer at 300 K, using Olb site energies.]{\small{Energy transfer dynamics for the FMO trimer at 300 K using Olb site energies, with HEOM (solid lines) compared to Time-Dependent Redfield (dashed lines). Initial excitation on (a) site 1, (b) site 6, (c) site 8.}}
\label{300K_nmkv_24site_Olb}
\end{center}
\end{figure}

\begin{figure}[!p]
\begin{center}
\includegraphics[width=13.1cm,keepaspectratio=true]{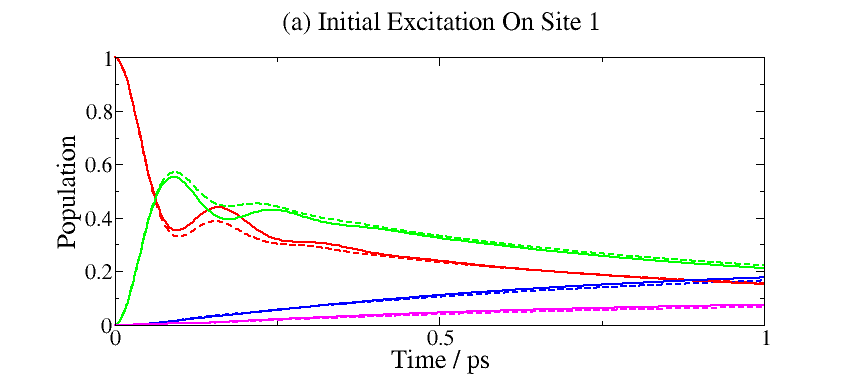}
\includegraphics[width=13.1cm,keepaspectratio=true]{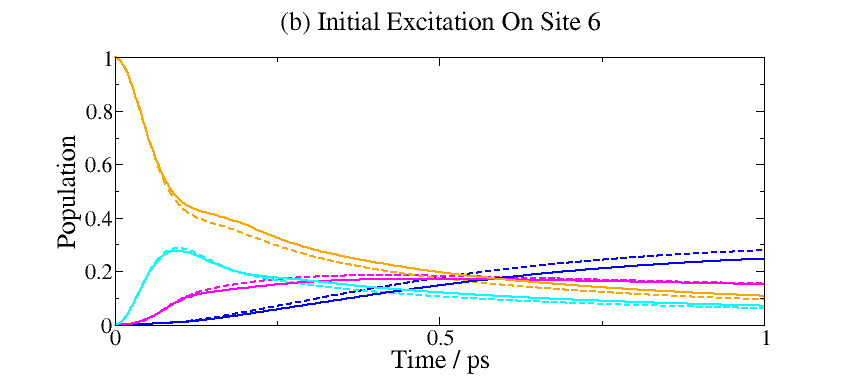}
\includegraphics[width=13.1cm,keepaspectratio=true]{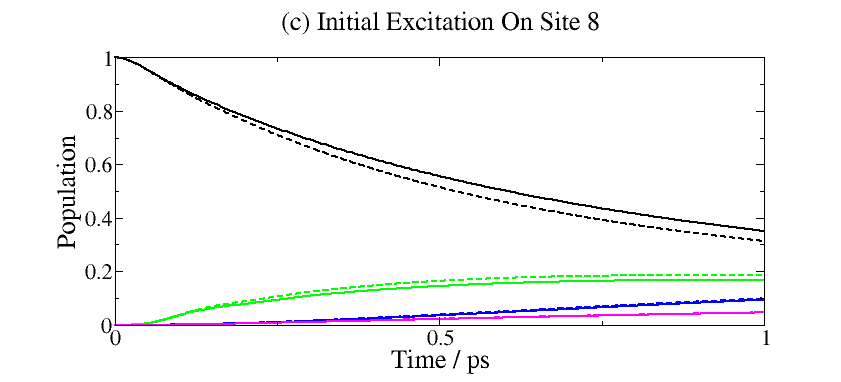}
\caption[Comparison of energy transfer dynamics to the Redfield approximation for the 24-site FMO trimer at 300 K, using SaB site energies.]{\small{Energy transfer dynamics for the FMO trimer at 300 K using SaB site energies, with HEOM (solid lines) compared to Time-Dependent Redfield (dashed lines). Initial excitation on (a) site 1, (b) site 6, (c) site 8.}}
\label{300K_nmkv_24site_SaB}
\end{center}
\end{figure}

\begin{figure}[!p]
\begin{center}
\includegraphics[width=13.1cm,keepaspectratio=true]{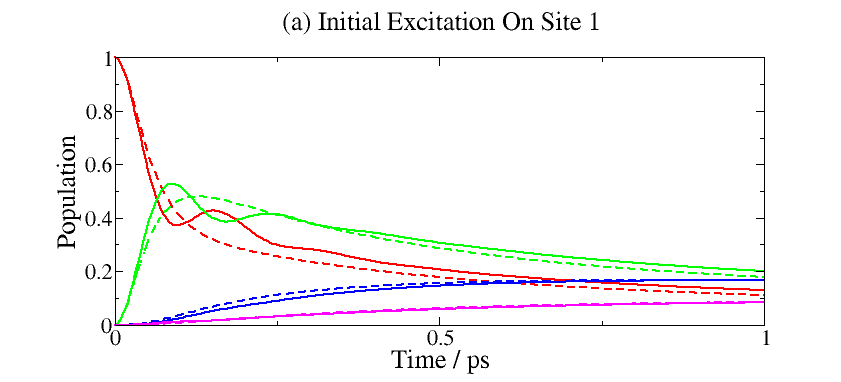}
\includegraphics[width=13.1cm,keepaspectratio=true]{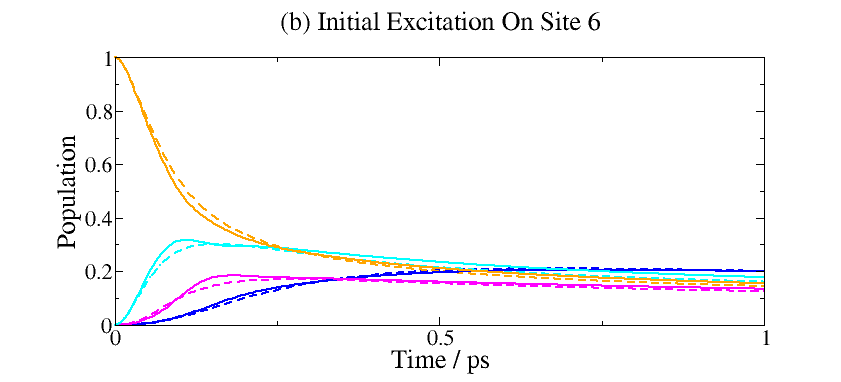}
\includegraphics[width=13.1cm,keepaspectratio=true]{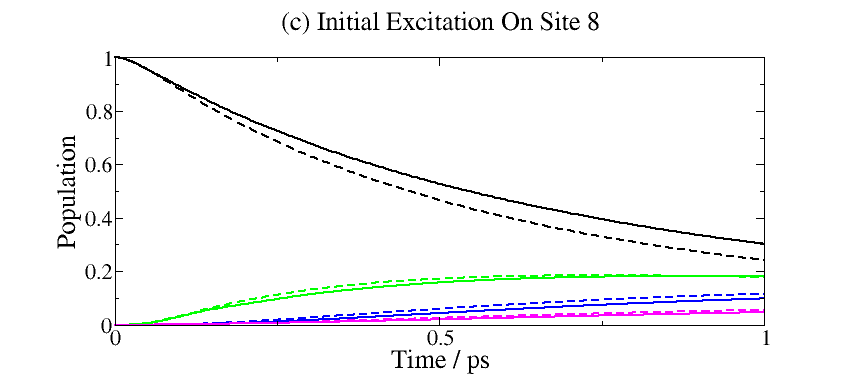}
\caption[Comparison of energy transfer dynamics to the F\"orster approximation for the 24-site FMO trimer at 300 K, using Olb site energies.]{\small{Energy transfer dynamics for the FMO trimer at 300 K using Olb site energies, with HEOM (solid lines) compared to F\"orster theory (dashed lines). Initial excitation on (a) site 1, (b) site 6, (c) site 8.}}
\label{300K_frst_24site_Olb}
\end{center}
\end{figure}

\begin{figure}[!p]
\begin{center}
\includegraphics[width=13.1cm,keepaspectratio=true]{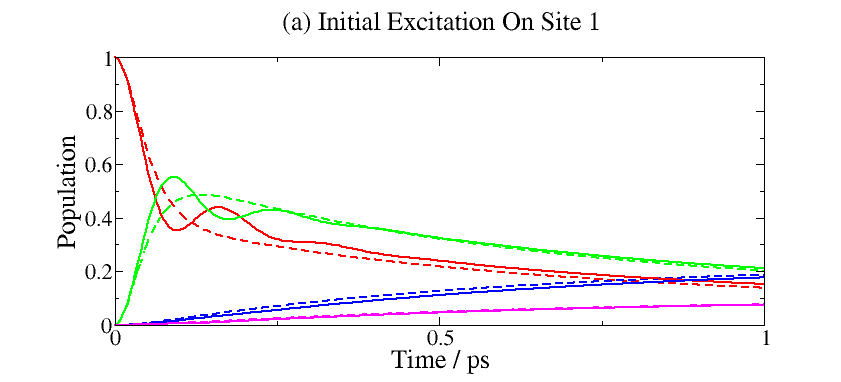}
\includegraphics[width=13.1cm,keepaspectratio=true]{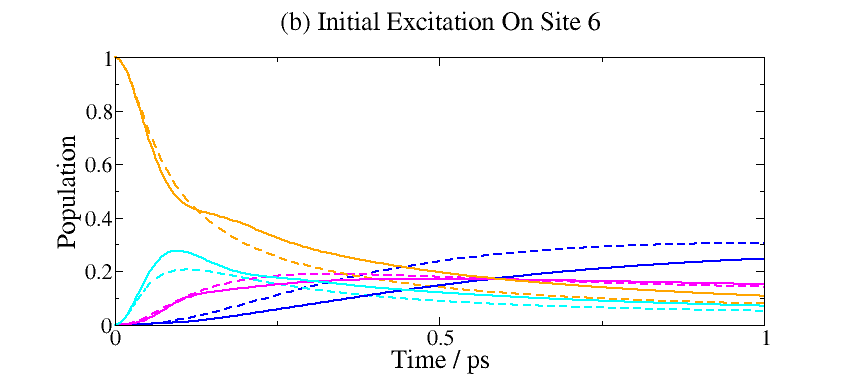}
\includegraphics[width=13.1cm,keepaspectratio=true]{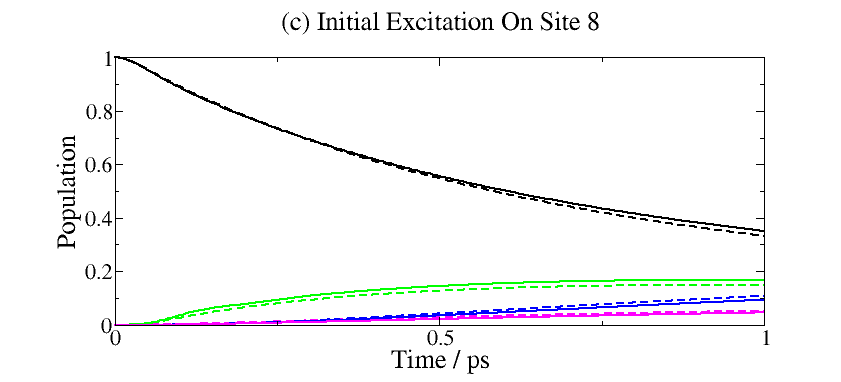}
\caption[Comparison of energy transfer dynamics to the F\"orster approximation for the 24-site FMO trimer at 300 K, using SaB site energies.]{\small{Energy transfer dynamics for the FMO trimer at 300 K using SaB site energies, with HEOM (solid lines) compared to F\"orster theory (dashed lines). Initial excitation on (a) site 1, (b) site 6, (c) site 8.}}
\label{300K_frst_24site_SaB}
\end{center}
\end{figure}

Figs. \ref{300K_nmkv_24site_Olb} and \ref{300K_nmkv_24site_SaB} compare the Time-Dependent Redfield and the HEOM results, using respectively the site energies of Olbrich \emph{et al.} (Olb) and of Schmidt am Busch \emph{et al.} (SaB), while Figs. \ref{300K_frst_24site_Olb} and \ref{300K_frst_24site_SaB} repeat this comparison for the F\"orster and the HEOM results.

\begin{figure}[!p]
\begin{center}
\includegraphics[width=13.1cm,keepaspectratio=true]{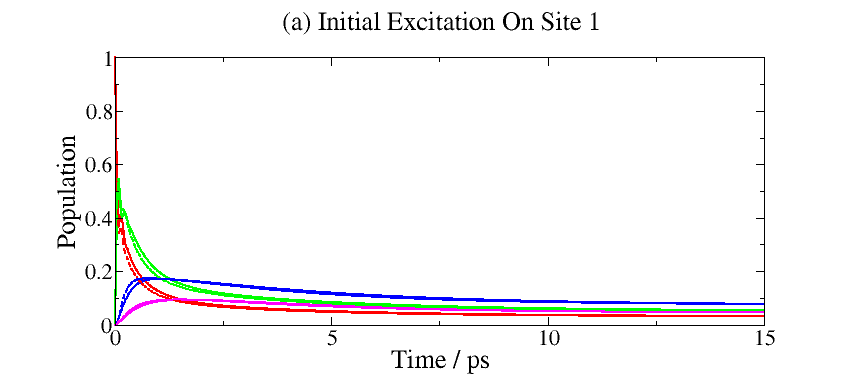}
\includegraphics[width=13.1cm,keepaspectratio=true]{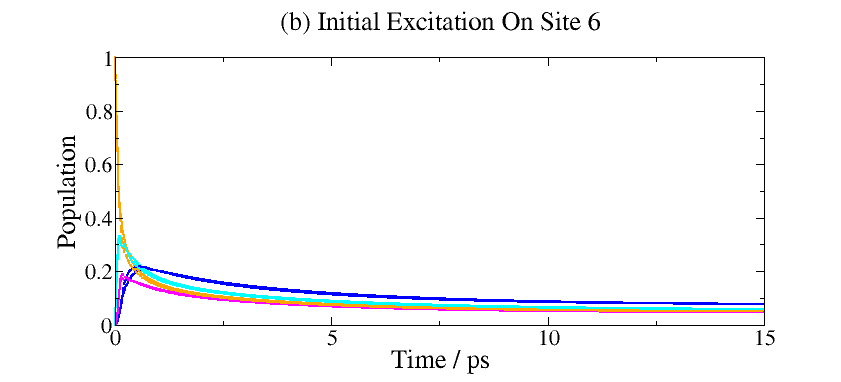}
\caption[Long-time energy transfer dynamics for the 24-site FMO trimer at 300 K, compared to Time-Dependent Redfield predictions]{\small{Comparison of long-time dynamics predicted by the HEOM (solid lines) and Time-Dependent Redfield (dashed lines) theories for (a) initial excitation on site 1, (b) initial excitation on site 2. The Olb site energies are used.}}
\label{300K_24site_steady_nmkv}
\end{center}
\end{figure}

\begin{figure}[!p]
\begin{center}
\includegraphics[width=13.1cm,keepaspectratio=true]{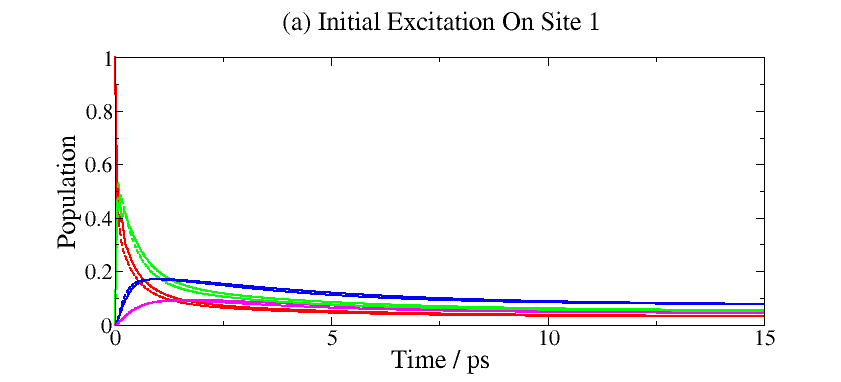}
\includegraphics[width=13.1cm,keepaspectratio=true]{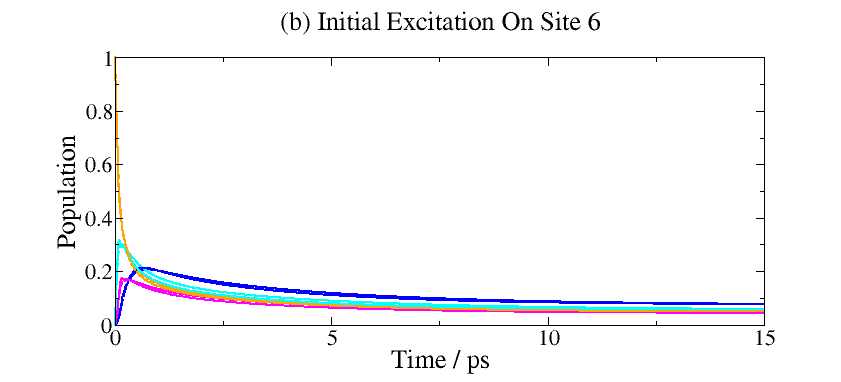}
\caption[Long-time energy transfer dynamics for the 24-site FMO trimer at 300 K, compared to F\"orster predictions.]{\small{Comparison of long-time dynamics predicted by the HEOM (solid lines) and F\"orster (dashed lines) theories for (a) initial excitation on site 1, (b) initial excitation on site 2. The Olb site energies are used.}}
\label{300K_24site_steady_frst}
\end{center}
\end{figure}

Fig. \ref{300K_24site_steady_nmkv} compares the results of the HEOM up to 15 ps to those of the Time-Dependent Redfield, and Fig. \ref{300K_24site_steady_frst} compares the results of the HEOM to those of F\"orster theory on the same timescale. Both of these Figures use the Olb site energies.

When BChl 8 is initially excited, the decay is largely exponential, with no coherence effects, due to the fact that population transferred to BChl 1 is rapidly transferred onwards to BChl 2, and sufficient population for a noticeable coherent superposition is not built up \cite{zofe_fmo2}.



Considering the Time-Dependent Redfield simulations, we can see that this method performs quite well for both sets of site energies, capturing the features of the transfer dynamics including the oscillations.

At 300 K, F\"orster theory is quite accurate overall, and in cases where it is not as much so, the differences are generally over-predictions of transfer rates rather than under-predictions. The results are somewhat worse than those of Redfield theory: aside from its neglect of the oscillations (whose importance we consider in the next Section), the agreement with exact dynamics tends to be poorer for F\"orster theory. Both theories nevertheless predict the long-time dynamics well, as was seen for the 7-site model.

\newpage \section{Conclusions}

In this Chapter we have presented numerically exact electronic energy transfer dynamics for both the 7-site and the 24-site FMO models, and have used them to benchmark approximate calculations using the Time-Dependent Redfield and F\"orster theories, in order to identify the method most suitable for further use.


Comparing the F\"orster and Redfield theories, although the F\"orster theory predicted the general shape of the dynamics and tended to give reasonably good quantitative agreement, the predictions of the Redfield theory were more accurate. Both methods, however, predicted the long-time decays observed, and the steady state, well.

The fluorescence lifetime of the bacteriochlorophyll sites is on the order of nanoseconds, whereas, as we have seen in this Chapter, steady states are reached in a timescale on the order of tens of picoseconds. Whether or not electronic coherence effects are included in the model of the energy transfer, this steady state is reached in the same amount of time.

The fact that F\"orster theory predicted the dynamics at 300 K so well, in terms of the important features (transfer rates, equilibrium populations and the timescale at which a steady state is reached) is extremely interesting, and allows us to conclude that although quantum coherence effects may be present in the energy transfer at physiological temperature, they appear to be unnecessary for efficient transfer.

With this conclusion, we are prepared to further investigate the physics of the FMO complex, using the F\"orster theory, which has been shown to give fairly good results at temperatures of biophysical interest and is computationally extremely cheap compared to all other methods described in this thesis, including the Time-Dependent Redfield theory. This will be the subject of the following Chapter.


{\renewcommand{\baselinestretch}{1}\normalsize
\chapter{Applications Of F\"orster Theory}\label{applications}
}

There are a number of physical features of the FMO complex which cannot be investigated using the HEOM, due to the computational demand. F\"orster theory, on the other hand, requires very little in terms of computational resources, so can be used for these investigations with ease.

The popularity of the Lorentz-Drude spectral density utilized thus far, within the modelling of open quantum systems, is due in large part to the fact that it allows the bath correlation function to be written as a sum of exponentials.

In reality, there is no reason to expect that this should give us physically correct results: in a pigment-protein complex, the featureless spectral density we have been using does not describe the vibrational environment of a bacteriochlorophyll pigment: for example, vibrations due to certain bonds (for example, O-H) might be expected to couple strongly to a site.

A biological system is generally exquisitely tailored, with its physical parameters striking a balance which, if disturbed past their tolerance, could have drastic effects on its functionality. The spectral density contains all information about the environment, and so altering this function is equivalent to altering the environment that interacts with the bacteriochlorophyll.

In Section \ref{section_specdens}, we explore the effect of using structured spectral densities, and in doing so we hope to learn how robust the energy transfer dynamics are to a change in environment.

The site energies given for the FMO complex are in fact averages over some distribution: due to the slow fluctuation of the protein environment, the actual set of site energies for a given complex will differ from this average, a phenomenon known as static disorder \cite{static_disorder}.

The effect of this disorder, simulated by drawing the site energies from a Gaussian distribution, on the energy transfer dynamics of the complex will be the subject of Section \ref{section_static}.

In this Chapter, we carry out simulations only at a physiological temperature of 300 K, as this is more biologically meaningful, and it is at this temperature that we have shown the F\"orster theory to give reasonably accurate results for the important features of the energy transfer dynamics.

\section{Structured Spectral Density}\label{section_specdens}

\begin{figure}[!ht]
\begin{center}
\includegraphics[width=13cm, keepaspectratio=true]{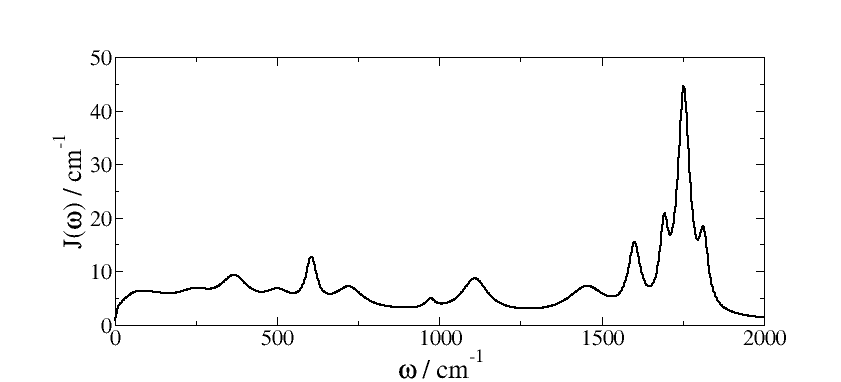}
\caption[Structured spectral density for 7-site FMO complex.]{\small{Structured spectral density for 7-site FMO complex, using parameters given in \cite{kleinekathoefer}.}}
\label{klein_specdens}
\end{center}
\end{figure}

The molecular dynamics simulations used by Olbrich \emph{et al.} to find the site energies of the FMO trimer were also used to calculate a spectral density for this complex \cite{olbrich,kleinekathoefer}. Fig. \ref{klein_specdens} shows an average spectral density for the 7-site subsystem, normalized to give a reorganization energy of $\lambda = 35 \text{~cm}^{-1}$, in agreement with experiment.

The simulations carried out gave the real part of the bath correlation function in the form:
\begin{equation}\label{kl_corrfunc}
\alpha_{r,j}(t) = \sum_{k=1}^{K}c_{jk}\cos(\tilde{\omega}_{jk}t)e^{-\gamma_{jk} t},
\end{equation}

\noindent where $K=15$ and the values of the $c_{jk}$, $\omega_{jk}$ and $\gamma_{jk}$ are given in Appendix \ref{supplementary}, Eqn. \eqref{klein_spectrum_data}. From this, Fourier inversion of the real part of \eqref{bath_corrfunc} gives the spectral density\footnote{A factor of $\frac{2}{\pi}$ is included to give agreement with \cite{kleinekathoefer}, although the scaling to give the correct $\lambda$ means that this factor is unimportant} $J_{j}(\omega) = \frac{2}{\pi}\tanh(\beta\hbar\omega/2)\int_{0}^{\infty}\alpha_{r,j}(t)\cos(\omega t)dt$, or:
\begin{equation}\label{kl_spectral}
J_{j}(\omega) = \frac{\tanh(\beta\hbar\omega / 2)}{\pi} \sum_{k=1}^{K}\left( \frac{c_{jk}\gamma_{jk}}{\gamma_{jk}^{2} + (\omega - \tilde{\omega}_{jk})^{2}} + \frac{c_{jk}\gamma_{jk}}{\gamma_{jk}^{2} + (\omega + \tilde{\omega}_{jk})^{2}} \right).
\end{equation}

In order to use this spectral density in our calculations, we need an expression for the full bath correlation function, so must find $\alpha_{i,j}(t)$:
\begin{equation}
\alpha_{i,j}(t) = -\int_{0}^{\infty}d\omega J_{j}(\omega)\sin(\omega t).
\end{equation}

Using $\tanh(\beta\hbar\omega/2) = \frac{1}{1+e^{-\beta\hbar\omega}}-\frac{1}{1+e^{\beta\hbar\omega}}$ gives an integrand that is an odd function, so that the integral can be rewritten:
\begin{align}
\alpha_{i,j}(t) & = \frac{1}{\pi}\sum_{k=1}^{K}\int_{-\infty}^{\infty}d\omega\cdot \frac{\sin(\omega t)}{1+e^{\beta\hbar\omega}} \left( \frac{c_{jk}\gamma_{jk}}{\gamma_{jk}^{2} + (\omega - \tilde{\omega}_{jk})^{2}} + \frac{c_{jk}\gamma_{jk}}{\gamma_{jk}^{2} + (\omega + \tilde{\omega}_{jk})^{2}} \right) \nonumber\\
& = \frac{2}{\pi}\sum_{k=1}^{K}c_{jk}\gamma_{jk}\int_{-\infty}^{\infty}d\omega \frac{(\omega^{2} + \tilde{\omega}_{jk}^{2} + \gamma_{jk}^{2})\sin(\omega t)f_{Fermi}(\beta\hbar\omega)}{[\gamma_{jk}^{2}+(\omega-\tilde{\omega}_{jk})^{2}][\gamma_{jk}^{2}+(\omega+\tilde{\omega}_{jk})^{2}]}.
\end{align}

Here, $f_{Fermi}(x) = (1+e^{x})^{-1}$ is the Fermi-Dirac function. As with the Bose-Einstein function in \ref{section_pade}, there is an [$N$-1/$N$] Pad\'e approximant that can be used for $f_{Fermi}(x)$ \cite{pade2}:
\begin{equation}
f_{Fermi}(x) \approx \frac{1}{2} - 2x\sum_{l=1}^{N}\frac{\eta_{l}}{x^{2}+\xi_{l}^{2}}.
\end{equation}

For the Fermi-Dirac case, the matrices $\underline{\underline{\Lambda}}$ and $\underline{\underline{\tilde{\Lambda}}}$ are given by:
\begin{equation}
\Lambda_{mn} = \frac{\delta_{m,n\pm 1}}{\sqrt{(2m-1)(2n-1)}} \hspace{2cm} \tilde{\Lambda}_{mn} = \frac{\delta_{m,n\pm 1}}{\sqrt{(2m+1)(2n+1)}}.
\end{equation}

Using the Pad\'e expression gives the following integral, where $\Im$ denotes the imaginary part:
\begin{multline}
\alpha_{i,j}(t) = \frac{4}{\pi\beta\hbar}\sum_{k=1}^{K}\sum_{l=1}^{N}c_{jk}\gamma_{jk}\eta_{l}\Im\left( \int_{-\infty}^{\infty}d\omega\cdot e^{-i\omega t}\times\right.\\
\left. \frac{\omega^{3} + \tilde{\omega}_{jk}^{2}\omega + \gamma_{jk}^{2}\omega}{[\gamma_{jk}^{2}+(\omega-\tilde{\omega}_{jk})^{2}][\gamma_{jk}^{2}+(\omega+\tilde{\omega}_{jk})^{2}][\omega^{2} + \nu_{l}^{2}]} \right).
\end{multline}

This integral can be carried out using the usual semicircular contour. The result is, with $\Omega_{jk} = i\tilde{\omega}_{jk} - \gamma_{jk}$:
\begin{multline}
\alpha_{i,j}(t) = \frac{2}{\beta\hbar}\sum_{k=1}^{K}\sum_{l=1}^{N} \left( \frac{c_{jk}\eta_{l}\Omega_{jk}}{\nu_{l}^{2} - \Omega_{jk}^{2}}e^{\Omega_{jk} t} + \frac{c_{jk}\eta_{l}\Omega_{jk}^{\ast}}{\nu_{l}^{2} - (\Omega_{jk}^{\ast})^{2}}e^{\Omega_{jk}^{\ast} t} \right. \\
\left. -\frac{2\gamma_{jk}c_{jk}\eta_{l}(|\Omega_{jk}|^{2} - \nu_{l}^{2})}{|\nu_{l} - \Omega_{jk}^{2}|^{2}}e^{-\nu_{l} t}\right).
\end{multline}

The real part can also be written in terms of exponentials:
\begin{equation}
\alpha_{r,j}(t) = \sum_{k=1}^{K}\left( \frac{c_{jk}}{2}e^{\Omega_{jk} t} + \frac{c_{jk}}{2}e^{\Omega_{jk}^{\ast} t} \right).
\end{equation}

Since each term in \eqref{kl_spectral} gives rise to a peak in the spectral density, by excluding certain terms we can now find out whether removing the corresponding peaks has an effect on the energy transfer dynamics.

\subsection{Results}

\begin{figure}[!ht]
\begin{center}
\includegraphics[width=13.5cm, keepaspectratio=true]{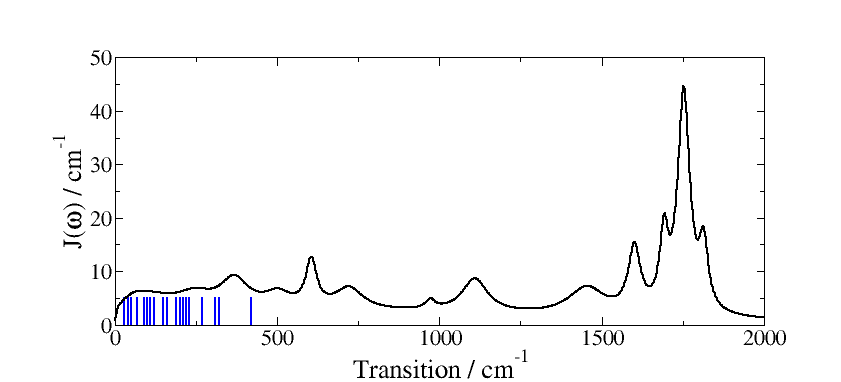}
\caption[Inter-site transition energies for the 7-site FMO complex.]{\small{Inter-site transition energies for the 7-site FMO complex.}}
\label{transitions}
\end{center}
\end{figure}

In order to inform the cases we investigate in this Section, we will wish to know the electronic transition frequencies between sites. Fig. \ref{transitions} shows these frequencies along with the structured spectral density.

The investigation of Olbrich \emph{et al.} using the ZOFE \cite{zofe_fmo1} led to the conclusion that high-energy vibrational modes do not have any effect on the energy transfer dynamics, since the energy of these modes is far from being in resonance with any transitions.

\begin{figure}[!hp]
\begin{center}
\includegraphics[width=13.5cm, keepaspectratio=true]{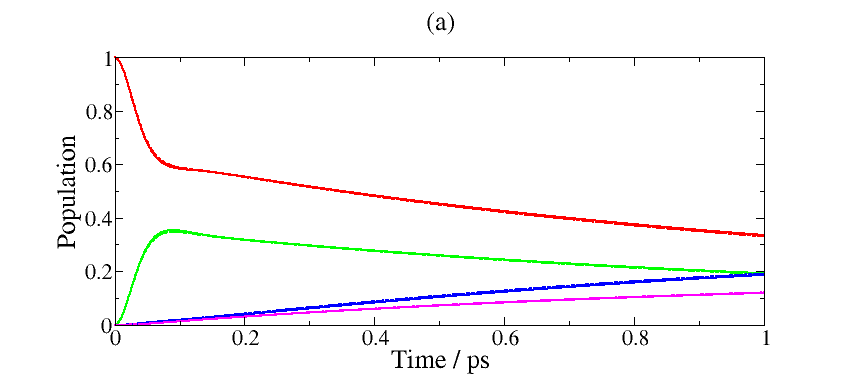}
\includegraphics[width=13.5cm, keepaspectratio=true]{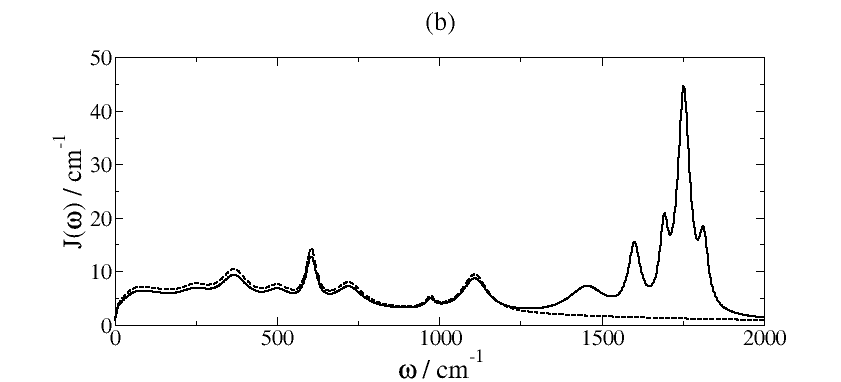}
\caption[Dynamics predicted by F\"orster theory using structured spectral density with and without high-frequency modes.]{\small{(a) Dynamics predicted using structured spectral density with (solid lines) and without (dashed lines) high-frequency modes. (b) Spectral densities used in the calculation of (a).}}
\label{highfreq_1ps}
\end{center}
\end{figure}

The spectral density in \ref{klein_specdens} shows several modes above 1300 cm$^{-1}$. By neglecting a number of terms in Eqn. \eqref{kl_spectral}, these modes can be removed from the spectral density with a negligible effect at smaller frequencies. Fig. \ref{highfreq_1ps} compares the dynamics both with and without the high-frequency modes In both cases, the spectral densities have been scaled to give $\lambda = $ 35 cm$^{-1}$.

\begin{figure}[!hp]
\begin{center}
\includegraphics[width=13.5cm, keepaspectratio=true]{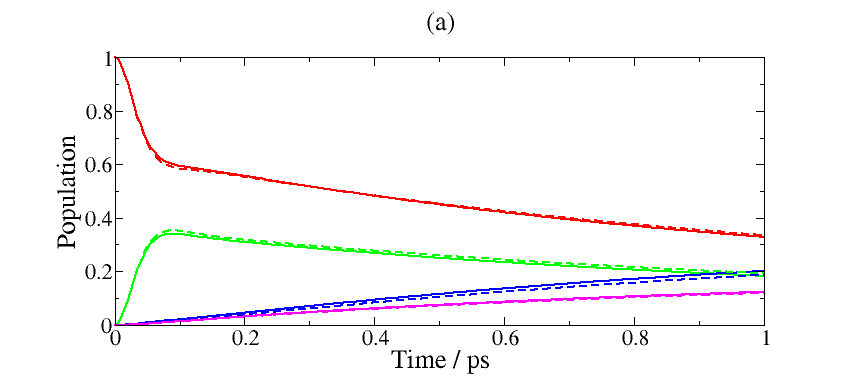}
\includegraphics[width=13.5cm, keepaspectratio=true]{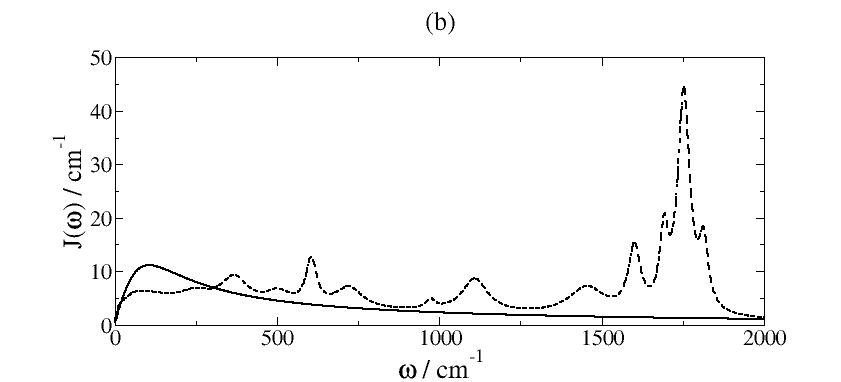}
\caption[Dynamics predicted by F\"orster theory using Lorentz-Drude and structured spectral density.]{\small{(a) Dynamics predicted using Lorentz-Drude (solid lines) and structured (dashed lines) spectral density. (b) Spectral densities used in the calculation of (a).}}
\label{1ps_LDD}
\end{center}
\end{figure}

On this short timescale, there are negligible differences between the predicted dynamics. Another interesting comparison is given by Fig. \ref{1ps_LDD}, which shows the dynamics using the Lorentz-Drude spectral density and the full structured spectral density, with a slightly greater discrepancy than Fig. \ref{highfreq_1ps}, though still a very small difference between the dynamics.

\newpage It is possible that even a small difference in the short-time dynamics could lead to a larger difference in the longer-time dynamics, so it is important to follow the simulations to this limit.

\begin{figure}[!ht]
\begin{center}
\includegraphics[width=13.5cm, keepaspectratio=true]{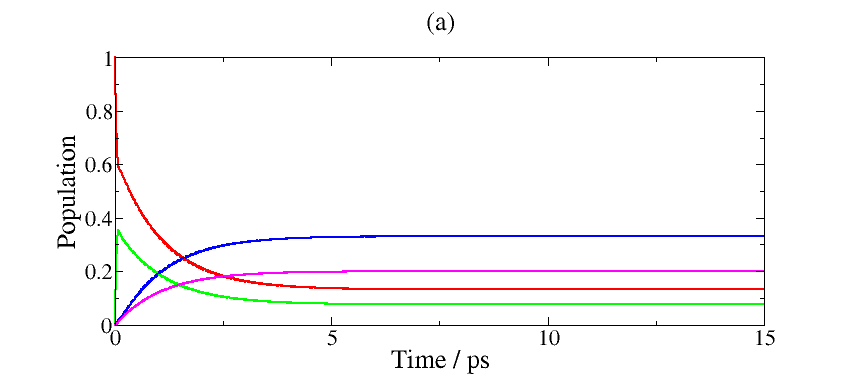}
\includegraphics[width=13.5cm, keepaspectratio=true]{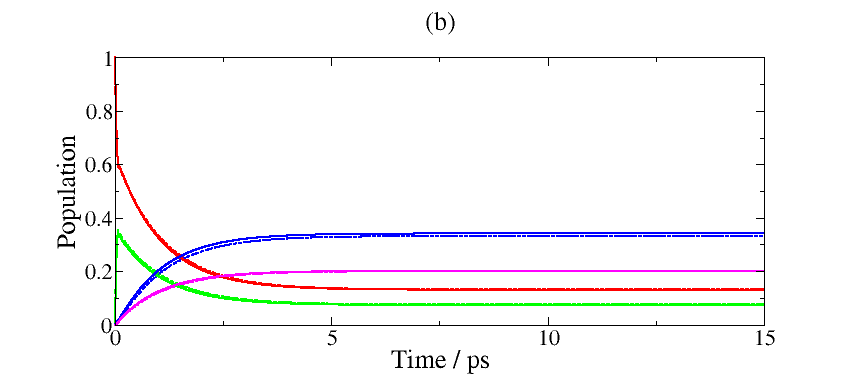}
\caption[Long-time dynamics predicted by F\"orster theory using different spectral densities.]{\small{Long-time dynamics for (a) fully structured (solid lines) vs. omitted high-frequency peaks (dashed lines) spectral densities. (b) Lorentz-Drude (solid lines) vs. fully structured (dashed lines) spectral densities.}}
\label{15ps_full_semi_ldd}
\end{center}
\end{figure}

Fig. \ref{15ps_full_semi_ldd} repeats the two comparisons, but with long-time simulations. As before, there is excellent agreement for the structured spectral density and the same with high-frequency peaks omitted, and good (though less so than the former) agreement for the structured spectral density and the Lorentz-Drude. We now analyse these results.

\subsection{Discussion}

The effect of spectral structure on the dynamics of energy transfer in the FMO complex is very little at room temperature.

We have seen in Fig. \ref{highfreq_1ps} that the high-frequency structure appears to be unnecessary in determining the efficiency of the energy transfer, and in Fig. \ref{1ps_LDD} that the fully structured spectral density of Olbrich \emph{et al.} \cite{kleinekathoefer} can be replaced with the Lorentz-Drude spectral density with only a small effect.

This latter result is remarkable because it suggests that the actual character of the vibrational modes is unimportant, and that it is some other property of the bath that determines the energy transfer. It is interesting to look further into what this property is.

Since we have shown that F\"orster theory provides a reasonable description of the dynamics, an investigation of absorption and fluorescence spectra, as in Section \ref{spectral_overlap}, will be informative.

\begin{figure}[!h]
\begin{center}
\includegraphics[width=13.5cm, keepaspectratio=true]{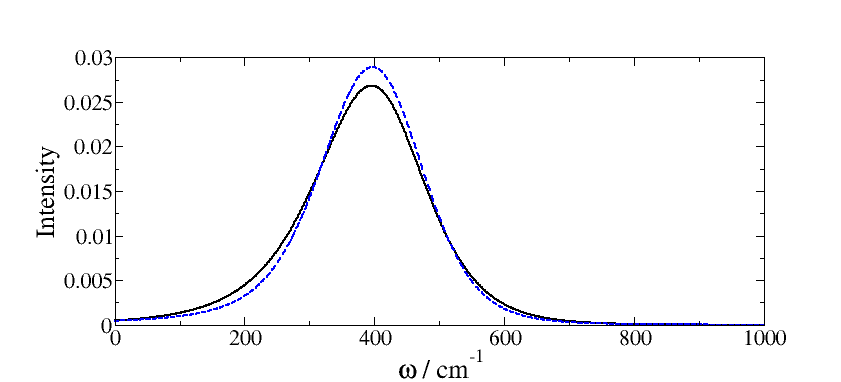}
\caption[Site 1 fluorescence spectra for Lorentz-Drude and structured spectral densities.]{\small{Site 1 fluorescence spectra, $F_{1}[\omega]$, for Lorentz-Drude (solid black line) and fully structured (dashed blue line) spectral densities.}}
\label{spectral_comparison}
\end{center}
\end{figure}

\newpage Fig. \ref{spectral_comparison} shows the fluorescence spectra of site 1 for both the Lorentz-Drude spectral density and the fully structured spectral density of Fig. \ref{kl_spectral}.

The spectrum is narrowed to a small degree, but the effect of this on the spectral overlap is negligible compared to either of the following effects: the much greater narrowing that comes from decreasing the temperature (see Fig. \ref{overlap}) or the shift that comes about from increasing the reorganization energy.

\begin{figure}[!h]
\begin{center}
\includegraphics[width=13.5cm, keepaspectratio=true]{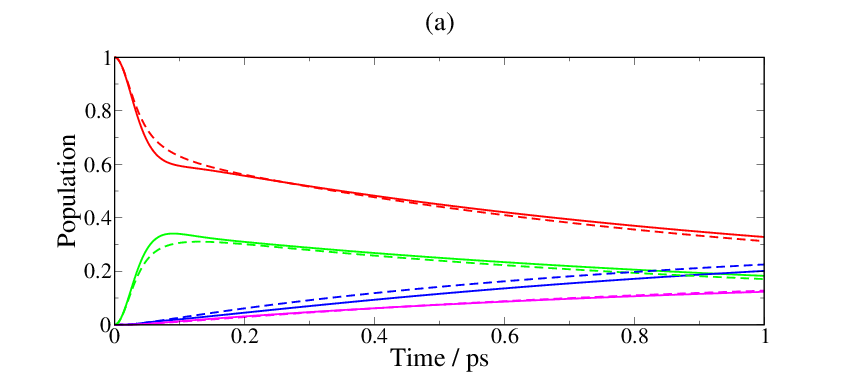}
\includegraphics[width=13.5cm, keepaspectratio=true]{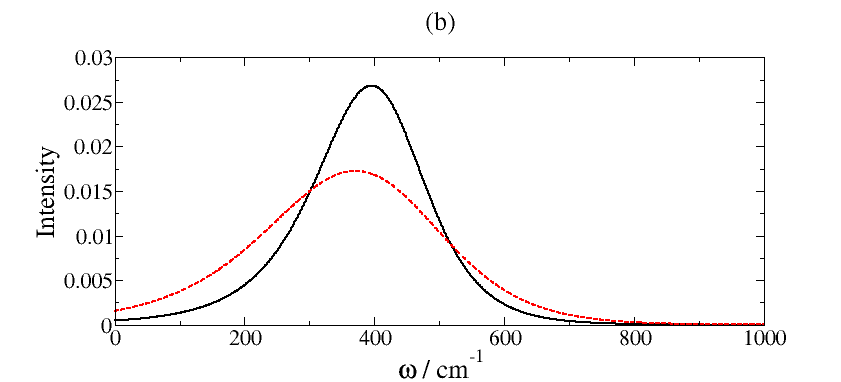}
\caption[Site 1 fluorescence spectra for two different reorganization energies.]{\small{(a) Energy transfer dynamics for the 7-site FMO complex using a Lorentz-Drude spectral density with $\lambda = 35$ cm$^{-1}$ (solid lines) and $\lambda = 70$ cm$^{-1}$ (dashed lines). (b) Site 1 fluorescence spectra, $F_{1}[\omega]$, for the two reorganization energies.}}
\label{shifted_spectrum}
\end{center}
\end{figure}

\newpage Fig. \ref{shifted_spectrum} shows the effect both on the dynamics and on $F_{1}[\omega]$ of increasing the reorganization energy from 35 cm$^{-1}$ to 70 cm$^{-1}$, while using the same Lorentz-Drude spectral density: there is a more significant change in the dynamics, reflecting the fact that the change results in a larger effect on the position and width of the spectral line, and thus a larger effect on its overlap with other lines (and via Eqn. \eqref{overlap_equation}, on the energy transfer rate constants).


These observations lead us to the conclusion that it is the temperature and the environmental reorganization energy that have the greatest effect on the dynamics of excitation energy transfer, rather than the precise shape of the spectral density $J(\omega)$.

\newpage\section{Static Disorder}\label{section_static}

The phenomenon of static disorder causes inhomogeneous line-broadening in electronic spectroscopy. That is, because each FMO complex in a sample is experiencing a different protein environment, giving different electronic transition energies, the spectral properties of each individual complex will differ and the superimposition of different spectral lines gives a broadened line.

In modelling the effects of inhomogeneous broadening, we treat the site energies as random variables with a Gaussian distribution \cite{static_disorder}.

Thus, from the original input data files for our numerical simulations, a large number of other input files are created, in which a random value is added to each site energy, drawn from a Gaussian distribution. The dynamics are found for each input file, and the results are averaged.

In order to carry out this procedure, we require a method of producing Gaussian random numbers. The standard Box-Muller algorithm was chosen for this purpose (Sec. 7.2 of \cite{numerical_recipes}).

The HEOM would be more difficult to use in finding the effect of this disorder, because a converged calculation for a single realization of Gaussian disorder would take a good fraction of an hour, whereas a F\"orster calculation is completed in a matter of seconds. We may require a large number of realizations to find the average effects, so it is preferable to use the faster technique, whose results have been shown to be accurate enough for this purpose at 300 K.


The effect of inhomogeneous broadening on 2-dimensional electronic spectra of the FMO complex has been the subject of a number of studies \cite{static_disorder,2d_fmo}, but in the following, we hope to find the effects, if any, on energy transfer dynamics in the hope once again of understanding the effect of the protein environment.

\subsection{Results}

An implementation of static disorder is characterized by the standard deviation, $\sigma$, of the Gaussian distribution. Often, the full-width at half-maximum (FWHM), equal to $2\sqrt{2\ln 2}\sigma$, is specified instead. Several studies have used a FWHM on the order of 50 cm$^{-1}$ \cite{static_disorder,2d_fmo,read_fleming}, and so this value will be used in this Section.

\begin{figure}[!hp]
\begin{center}
\includegraphics[width=12.3cm, keepaspectratio=true]{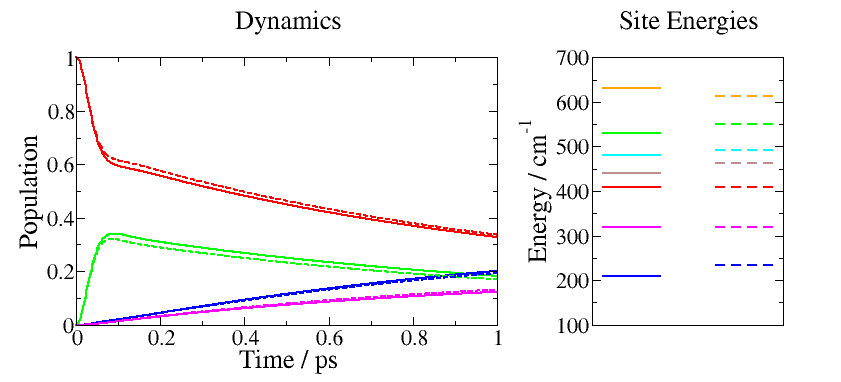}
\includegraphics[width=12.3cm, keepaspectratio=true]{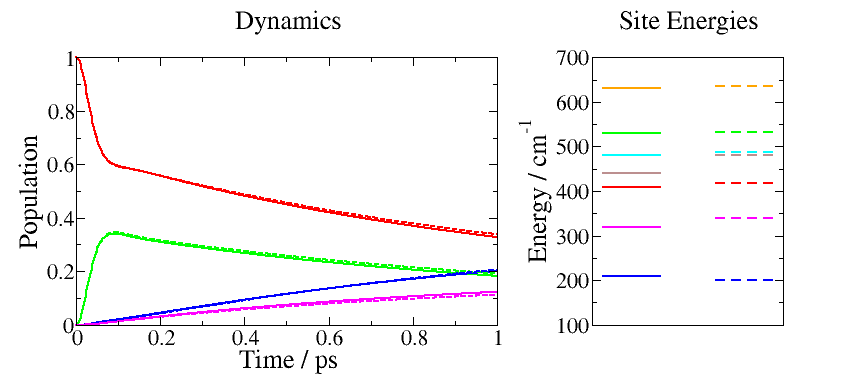}
\includegraphics[width=12.3cm, keepaspectratio=true]{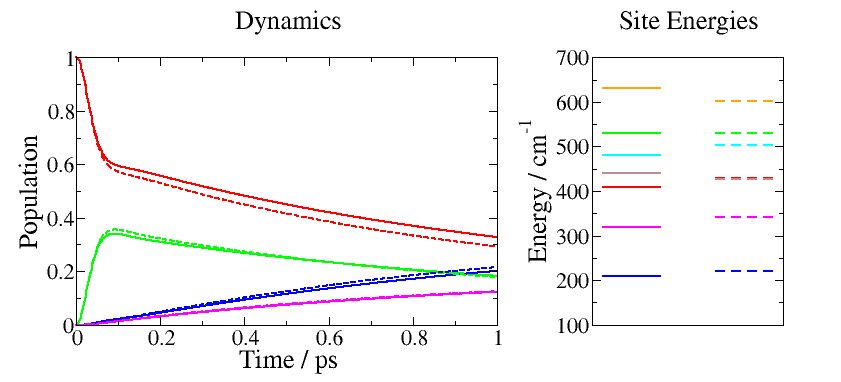}
\caption[Dynamics for three realizations of static disorder.]{\small{Dynamics for 3 realizations of static disorder. The left-hand column gives the dynamics, and the right-hand column shows the site energies for these realizations. The solid line in each case is the situation in which static disorder is ignored, and the dashed line indicates its presence.}} 
\label{realizations}
\end{center}
\end{figure}

We show firstly the results for 3 realizations of static disorder in Fig. \ref{realizations}: alongside each set of results is a diagram showing the energy levels of the different sites for this realization, compared to the energy levels in the absence of disorder.

Each individual set of site energies gives dynamics that differ a little from the case where disorder is neglected: this is a general observation that can be made for any one realization that is taken, although in some cases (as in Fig. \ref{realizations} c), the differences are slightly more pronounced.

\begin{figure}[!ht]
\begin{center}
\includegraphics[width=13.5cm, keepaspectratio=true]{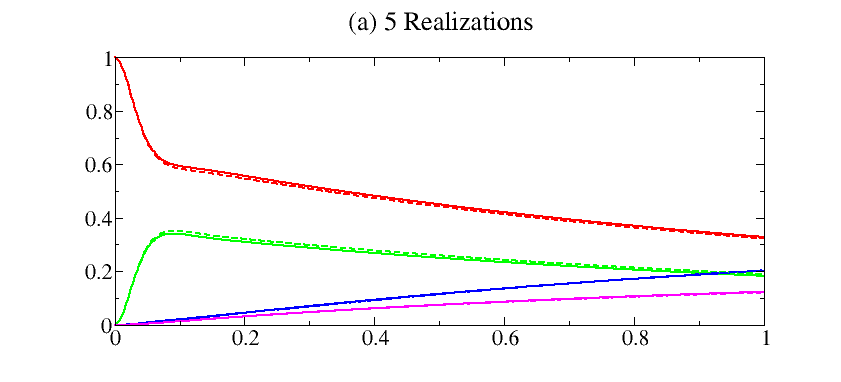}
\includegraphics[width=13.5cm, keepaspectratio=true]{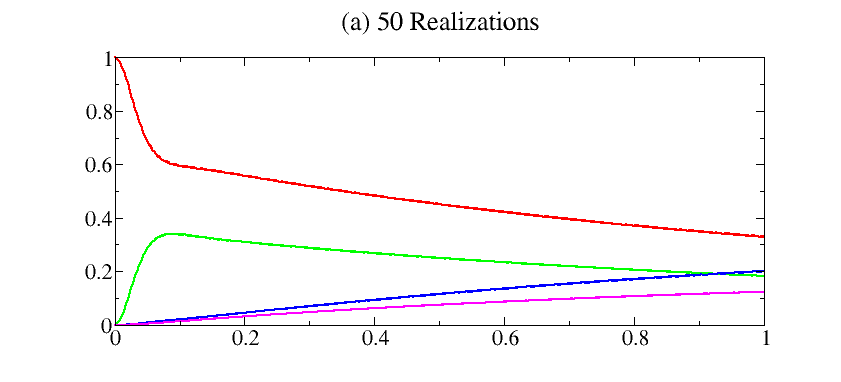}
\caption[Average dynamics for numbers of realizations of static disorder.]{\small{Average dynamics (dashed lines) for (a) 5 and (b) 50 realizations of static disorder. The solid lines in each case show the dynamics without disorder.}}
\label{static_disorder}
\end{center}
\end{figure}

In a real physical application, there will be a great number of FMO complexes, and the dynamics will be an average over many realizations. Fig. \ref{static_disorder} compares the average dynamics for different numbers of realizations to the dynamics in the absence of disorder. A relatively small number of realizations is required for the average dynamics to converge.

\subsection{Discussion}

The results presented in this Section illustrate that inhomogeneous disordering effects on the dynamics of the FMO complex are minimal. This can be explained by examining the energy level diagrams in the right-hand column of Fig. \ref{realizations}.

In the results we have observed, the populations of bacteriochlorophylls 5, 6 and 7 remain close to zero, and only the other four have appreciable populations throughout the simulation.

The electronic energy separations of the four sites involved in the transfer are larger than the differences induced by static disorder, so that the important transition energies are not likely to be very different for two different realizations.

Thus, while there will generally be small differences for the dynamics of these two realizations, the most important features will be the same when comparing the two, and these small differences are not important in the FMO complex because of the effective averaging of the dynamics.\\[4cm]


\section{Conclusions}

In the two subsections of this Chapter, conclusions have been drawn separately about the effects of a structured spectral density and of static disorder due to an inhomogeneous environment. Here, we will make some comments about the implications for the FMO complex.

Firstly, it was found that the structure of the spectral density is largely unimportant at 300 K, and that the temperature and reorganization energy were far more important in determining the rate of energy transfer.

Secondly, the effects of static disorder characterized by a physically reasonable FWHM were found to be very little, in addition to which the average behaviour converged with only a small number of realizations.

This leads to a conclusion which may seem surprising at first: that is, the electronic energy transfer dynamics in the FMO complex are largely insensitive to the fine-structure of the environment. Similarly, the disorder inherent in any experiment has little effect, giving a very robust transfer dynamics.

The energy transfer dynamics of the FMO complex are very stable with respect to changes in its vibrational environment. Perhaps this is not a coincidence, but rather a result of the complex having evolved to perform its job so robustly.


{\renewcommand{\baselinestretch}{1}\normalsize
\chapter{Conclusions And Further Work}\label{conclusions}
}

In this project, we have investigated the physics of the Fenna-Matthews-Olson complex. The Hierarchical Equations of Motion were utilized to find numerically exact excitation energy transfer dynamics for the full 24-site trimer, with a fully converged bath correlation function. These are the first exact results that have been computed for this system.

We have also applied the approximate Redfield and F\"orster methods (the latter of which neglects electronic coherence) to the problem, and found that at room temperature, F\"orster theory described the features of the dynamics well, leading to the conclusion that quantum-mechanical coherence is, for the most part, irrelevant for the efficiency of transfer.

Having shown that F\"orster theory is applicable at this temperature, we then applied it to the problem of altering the effect of the environment on the system. It was found that the effects of adding structure to the spectral density of the bath phonons coupled to the electronic system or of including static disorder were minimal, and that the bath reorganization energies and the temperature were the most important parameters for determining the energy transfer dynamics.

There has been much excitement recently about the importance of coherence in photosynthetic systems \cite{coherence,ishizaki_review,nalbach_thorwart,quant_coherence_assists,shim_rebentrost_spectral}, and the incoherent nature of F\"orster theory has led to its dismissal as an appropriate method for describing transfer \cite{i&f_redfield,fleming_lightharv}. However, the results of this work show that the search for an adequate description of electronic energy transfer need not necessarily require that the effects of coherence be included.

Before moving on to discuss potential further work, it is quite satisfying to note that the energy transfer dynamics in the FMO complex, structurally characterized nearly 40 years ago \cite{fenna_matthews_olson}, can be modelled in a very satisfactory way by a theory that is even more venerable \cite{forster1}.

\section{Further Work}\label{further_work}

The different facets of this work suggest a number of directions for further research:
\begin{itemize}
\item The availability of numerically exact results for larger and more complex systems than previously available allows effective benchmarking of approximate methods: the current glut of such methods, giving different results, means that it is a great advantage to be able to test them against a method (the HEOM) that can be used for many different systems and parameter regimes.

\item Having seen that we need not requite that coherence effects be included in an accurate prediction of energy transfer dynamics, methods can be sought that are more accurate than the F\"orster theory, and with greater ranges of applicability, while still being incoherent.

\item An approximate method that gives accurate results at lower temperatures might allow for a more comprehensive test of the effects of a structured spectral density at these temperatures.
\end{itemize}

With such investigations, perhaps we will be able to gain more insight into how important, if at all, quantum-mechanical interference effects are in biology.


\appendix

\fancyhead{}
\renewcommand{\chaptermark}[1]{\markright{\MakeUppercase{Appendix \thechapter.\ #1}}{}}
\fancyhead[LO,RE]{\rightmark}



{\renewcommand{\baselinestretch}{1}\normalsize
\chapter[Appendix]{}
\section{The Bath Correlation Function}\label{QFDT}
}

The following correlation functions may be defined, with the tilde representing the interaction picture as in Chapter \ref{chapter_heom} and $\hat{q}_{j\alpha}$ being bath coordinates as in Section \ref{open_qsys}:
\begin{subequations}
\begin{equation}\label{C+}
C_{j\alpha}^{+}(t) = \left<\tilde{q}_{j\alpha}(t)\tilde{q}_{j\alpha}(0)\right>_{\beta},
\end{equation}
\begin{equation}\label{C-}
C_{j\alpha}^{-}(t) = \left<\tilde{q}_{j\alpha}(0)\tilde{q}_{j\alpha}(t)\right>_{\beta}.
\end{equation}
\end{subequations}

From this, using the fact that $\tilde{\xi}_{j}(t) = \sum_{\alpha}g_{j\alpha}\tilde{q}_{j\alpha}(t)$, and the independence of the bath degrees of freedom, it can be seen that:
%
\begin{equation}\label{alpha_corr}
\alpha_{j}(t) = \left<\tilde{\xi}_{j}(t)\tilde{\xi}_{j}(0)\right>_{\beta} = \sum_{\alpha}g_{j\alpha}C_{j\alpha}^{+}(t).
\end{equation}

The symmetrized and antisymmetrized correlation functions are given by (pp. 122-124 of \cite{weiss}):
\begin{subequations}
\begin{equation}\label{symmetrized}
S_{j\alpha}(t) = \frac{1}{2}\left(C_{j\alpha}^{+}(t) + C_{j\alpha}^{-}(t)\right),
\end{equation}
\begin{equation}\label{antisymmetrized}
A_{j\alpha}(t) = \frac{1}{2i}\left(C_{j\alpha}^{+}(t) - C_{j\alpha}^{-}(t)\right).
\end{equation}
\end{subequations}

From the definitions \eqref{C+} and \eqref{C-}:
\begin{align}
C_{j\alpha}^{-}(t) & = \frac{1}{Z_{B}}tr_{B}[\tilde{q}_{j\alpha}(0)\tilde{q}_{j\alpha}(t)e^{-\beta\hat{H}_{B}}]\nonumber\\
& = \frac{1}{Z_{B}}tr_{B}[\tilde{q}_{j\alpha}(0)e^{-\beta\hat{H}_{B}}e^{+\beta\hat{H}_{B}}\tilde{q}_{j\alpha}(t)e^{-\beta\hat{H}_{B}}]\nonumber\\
& = \frac{1}{Z_{B}}tr_{B}[\tilde{q}_{j\alpha}(0)e^{-\beta\hat{H}_{B}}e^{i\hat{H}_{B}/\hbar\cdot(-i\beta\hbar)}\tilde{q}_{j\alpha}(t)e^{-i\hat{H}_{B}/\hbar\cdot(-i\beta\hbar)}]\nonumber\\
& = \frac{1}{Z_{B}}tr_{B}[\tilde{q}_{j\alpha}(0)e^{-\beta\hat{H}_{B}}\tilde{q}_{j\alpha}(t-i\beta\hbar)]\nonumber\\
& = \left<\tilde{q}_{j\alpha}(t-i\beta\hbar)\tilde{q}_{j\alpha}(0)\right>_{\beta}\nonumber\\
& \equiv C_{j\alpha}^{+}(t-i\beta\hbar).\label{imagtime}
\end{align}

Now, taking the Fourier transforms of the correlation functions,
\begin{equation}\label{fourier}
F[\omega] = \int_{-\infty}^{\infty}F(t)e^{i\omega t}dt.
\end{equation}

\noindent and using Eqn. \eqref{imagtime} gives:
\begin{align}
C_{j\alpha}^{-}[\omega] & = \int_{-\infty}^{\infty}C_{j\alpha}^{-}(t)e^{i\omega t}dt\nonumber\\
& = \int_{-\infty}^{\infty}C_{j\alpha}^{+}(t-i\beta\hbar)e^{i\omega t}dt\nonumber\\
& = \int_{-\infty}^{\infty}C_{j\alpha}^{+}(u)e^{i\omega(u+i\beta\hbar)}du\nonumber\\
& = \int_{-\infty}^{\infty}C_{j\alpha}^{+}(t)e^{i\omega t}e^{-\beta\hbar\omega}dt\nonumber\\
& = e^{-\beta\hbar\omega}C_{j\alpha}^{+}[\omega].\label{detailedbalance}
\end{align}

The Fourier transform of the antisymmetrized correlation function is then (using \eqref{antisymmetrized} and \eqref{detailedbalance}) given by $A_{j\alpha}[\omega] = \frac{1}{2i}\left(1-e^{-\beta\hbar\omega}\right)C_{j\alpha}^{+}[\omega]$, or:
\begin{equation}\label{corrfunc_anti}
C_{j\alpha}^{+}[\omega] = \frac{2iA_{j\alpha}[\omega]}{1-e^{-\beta\hbar\omega}}.
\end{equation}

$A_{j}[\omega]$ can then be found using Eqn. \eqref{bath_position} for $\tilde{q}_{j\alpha}(t)$ and Eqn. \eqref{antisymmetrized}:
\begin{align}
A_{j\alpha}[\omega] & = \frac{1}{2i}\int_{-\infty}^{\infty}dt\cdot e^{i\omega t} \left< [\tilde{q}_{j\alpha}(t),\tilde{q}_{j\alpha}(0)]\right>_{\beta}\nonumber\\
& = \frac{1}{2i}\int_{-\infty}^{\infty}dt\cdot e^{i\omega t} \left< -\frac{i\hbar}{m_{j\alpha}\omega_{j\alpha}}\sin(\omega_{j\alpha}t)\right>_{\beta}\nonumber \\
& = -\frac{\hbar}{4im_{j\alpha}\omega_{j\alpha}}\int_{-\infty}^{\infty}dt\cdot\left(e^{i(\omega+\omega_{j\alpha})t} - e^{i(\omega-\omega_{j\alpha})t}\right)\nonumber \\
& = \frac{\pi\hbar}{2im_{n\alpha}\omega_{j\alpha}}\left(\delta(\omega-\omega_{j\alpha})-\delta(\omega+\omega_{j\alpha}) \right).\label{antisymm_delta}
\end{align}

Inserting this expression into \eqref{alpha_corr} and using Eqns. \eqref{corrfunc_anti} and \eqref{spectral_density}:
\begin{align}
\alpha_{j}[\omega] & = \frac{2\pi}{1-e^{-\beta\hbar\omega}}\sum_{\alpha}\frac{\hbar}{2m_{j\alpha}\omega_{j\alpha}}\left(\delta(\omega-\omega_{j\alpha})-\delta(\omega+\omega_{j\alpha}) \right)\nonumber \\
& = \frac{2\pi J_{j}(\omega)}{1-e^{-\beta\hbar\omega}}.
\end{align}

Here, the inclusion of two delta functions means that the spectral density is defined for negative frequencies as well as positive, and that $J_{j}(\omega) = -J_{j}(-\omega)$. By Fourier inversion:
\begin{equation}\label{bath_corrfunc_infty}
\alpha_{j}(t) = \int_{-\infty}^{\infty}d\omega \frac{J_{j}(\omega)e^{-i\omega t}}{1-e^{-\beta\hbar\omega}}.
\end{equation}

This expression can be rearranged to give Eqn. \eqref{bath_corrfunc}, making use of the fact that $J_{j}(\omega)$ is an odd function of $\omega$.


\newpage{\renewcommand{\baselinestretch}{1}\normalsize
\section{The Pad\'e Approximant}\label{pade_approximant}
}

In this appendix we find a rational function approximation to the Bose-Einstein function $f_{Bose}(x) = (1-e^{-x})^{-1}$. A similar procedure must be followed to find an approximation to the Fermi-Dirac function $f_{Fermi}(x) = (1+e^{x})^{-1}$, and after the derivation is outlined, the results for both are given. Firstly, we take:
\begin{equation}
\frac{1}{1-e^{-x}} = \frac{1}{2}\frac{e^{\sfrac{x}{2}} - e^{-\sfrac{x}{2}}}{e^{\sfrac{x}{2}} - e^{-\sfrac{x}{2}}} + \frac{1}{2}\frac{e^{\sfrac{x}{2}} + e^{-\sfrac{x}{2}}}{e^{\sfrac{x}{2}} - e^{-\sfrac{x}{2}}} = \frac{1}{2} + \frac{1}{2}\coth(x/2).
\end{equation}

Then, the continued-fraction representation of $\tanh(y)$ is given by the expression below \cite{pade2}, where the notation is explained using the third expression:
\begin{equation}
\tanh(y) = \frac{y}{1+}~\frac{y^2}{3+}~\frac{y^2}{5+\dots} := \frac{y}{1+\frac{y^2}{3+\dots}}.
\end{equation}

Inverting this expression gives $\coth(y)$:
\begin{equation}
\coth(y) = \frac{1}{y} + \frac{y}{3+}~\frac{y^2}{5+}~\frac{y^2}{7+\dots},
\end{equation}
\noindent or,
\begin{align}
\frac{1}{1-e^{-x}} & = \frac{1}{2} + \frac{1}{x} + x\cdot \frac{\sfrac{1}{4}}{3+}~\frac{\sfrac{x^2}{4}}{5+}~\frac{\sfrac{x^2}{4}}{7+\dots} \nonumber\\
& = \frac{1}{2} + \frac{1}{x} + x\cdot \frac{\sfrac{1}{4}}{b_{1}+}~\frac{\sfrac{x^2}{4}}{b_{2}+}~\frac{\sfrac{x^2}{4}}{b_{3}+\dots},
\end{align}

where $b_{n} = 2n + 1$. The $M^{th}$ convergent of the continued fraction is \cite{pade2}:
\begin{equation}
\mathcal{C}_{M}(x^{2}) = \frac{\sfrac{1}{4}}{b_{1}+}~\frac{\sfrac{x^2}{4}}{b_{2}+}~\dots~\frac{\sfrac{x^2}{4}}{b_{M}} := \frac{\mathcal{A}_{M}(x^{2})}{\mathcal{B}_{M.}(x^{2})}
\end{equation}

$\mathcal{A}_{M}(x^{2})$ and $\mathcal{B}_{M}(x^{2})$ are polynomials in $x^{2}$. Each function, $f_{M}(x^{2})$, is linked to $f_{M-1}(x^{2})$ and to $f_{M-2}(x^{2})$ by a recursion relation, and it is found \cite{pade1} that $\mathcal{A}_{2N}(x^{2})$ is of order $N-1$, while $\mathcal{B}_{2N}(x^{2})$ is of order $N$.

Thus, $\mathcal{C}_{2N}(x^{2})$ is a rational function whose numerator has degree $N-1$ and whose denominator has degree $N$. It can be rewritten:
\begin{equation}\label{convergent}
\mathcal{C}_{2N}(x^{2}) = \frac{\prod_{j=1}^{N-1}\left(x^{2} + \zeta_{j}^{2}\right)}{\prod_{j=1}^{N}\left(x^{2} + \xi_{j}^{2}\right)} = \sum_{j=1}^{N}\frac{2\eta_{j} x}{x^{2} + \xi_{j}^{2}}.
\end{equation}

The roots of the numerator polynomial are then $\lbrace -\zeta_{j}^{2}\rbrace$ and those of the denominator are $\lbrace -\xi_{j}^{2}\rbrace$. The $\eta_{j}$ coefficients ($j=1,2,\dots,N$) are found by decomposing the first expression into partial fractions \cite{pade2}:
\begin{equation}\label{etacoeffs}
\eta_{j} = \frac{1}{2}\left( (x^{2} + \xi_{j}^{2})\mathcal{C}_{2N}(x^{2}\right)_{x^{2} = -\xi_{j}^{2}} = \frac{1}{2}N b_{N+1}\frac{\prod_{k=1}^{N-1}\left( \zeta_{k}^{2} - \xi_{j}^{2}\right)}{\prod_{k\ne j}^{N}\left( \xi_{k}^{2} - \xi_{j}^{2}\right)}.
\end{equation}

Now, a straightforward method is required to find the roots of the two polynomials. It is shown in \cite{continued_fraction,pade2} that the convergent of a continued fraction can be expressed as,
\begin{equation}
\mathcal{C}_{M}(x^{2}) = \left(\underline{\underline{D}} + \frac{1}{2}ix\underline{\underline{E}}\right)^{-1}_{11},
\end{equation} 

\noindent with the elements of $\underline{\underline{D}}$, $D_{mn} = b_{m}\delta_{mn}$ and of $\underline{\underline{E}}$, $E_{mn} = \delta_{m,n\pm 1}$, with each of $m$ and $n$ running from $1$ to $M$.

The roots of $\mathcal{B}_{M}(x^{2})$ are the poles of $\mathcal{C}_{M}(x^{2})$, and so are the roots of $\det\left( \underline{\underline{D}} + \frac{1}{2}ix\underline{\underline{E}}\right) = 0$.

Thus, for $\mathcal{B}_{2N}(x^{2})$, the matrix $\underline{\underline{\Lambda}} = \underline{\underline{D}}^{-1}\underline{\underline{C}}\;\underline{\underline{D}}^{-1}$ is defined, with elements as in Eqn. \eqref{lambda_matrices}, whose eigenvalues are $\lbrace\pm 2/\xi_{1},\pm 2/\xi_{2},\dots,\pm 2/\xi_{N}\rbrace$ \cite{pade2}.

The roots of $\mathcal{A}_{2N}(x^{2})$ can be similarly found by using $\mathcal{C}_{2N}(x^{2})^{-1}$, whose poles are these roots. A matrix $\underline{\underline{\tilde{\Lambda}}}$ is defined, whose elements are also given in Eqn. \eqref{lambda_matrices}, and whose eigenvalues are $\lbrace 0,\pm 2/\zeta_{1},\pm 2/\zeta_{2},\dots,\pm 2/\zeta_{N-1}\rbrace$.

Thus, the convergent $\mathcal{C}_{2N}(x^{2})$ can be found using \eqref{convergent} and \eqref{etacoeffs}, so that the Pad\'e approximant for the Bose-Einstein function is:
\begin{equation}\label{bose_function}
f_{Bose}(x) = \frac{1}{2} + \frac{1}{x} + x\mathcal{C}_{2N}(x^{2}).
\end{equation}
If a similar analysis is carried out for the Fermi-Dirac function, then the only difference in defining the coefficient is that here $b_{n} = 2n - 1$, and that:
\begin{equation}\label{fermi_function}
f_{Fermi}(x) = \frac{1}{2} - x\mathcal{C}_{2N}(x^{2}).
\end{equation}
\\[5cm]


\newpage{\renewcommand{\baselinestretch}{1}\normalsize
\section{Supplementary Information}\label{supplementary}
}

\subsection{System Hamiltonians}

The system Hamiltonian for the 7-site FMO is that of Adolphs and Renger \cite{adolphs_renger} (energies in cm$^{-1}$):
\begin{equation}\label{7site_Hsys}
\begin{pmatrix}
410 & -87.7 & 5.5 & -5.9 & 6.7 & -13.7 & -9.9 \\
-87.7 & 530 & 30.8 & 8.2 & 0.7 & 11.8 & 4.3 \\
5.5 & 30.8 & 210 & -53.5 & -2.2 & -9.6 & 6.0 \\
-5.9 & 8.2 & -53.5 & 320 & -70.7 & -17.0 & -63.3 \\
6.7 & 0.7 & -2.2 & -70.7 & 480 & 81.1 & -1.3 \\
-13.7 & 11.8 & -9.6 & -17.0 & 81.1 & 630 & 39.7 \\
-9.9 & 4.3 & 6.0 & -63.3 & -1.3 & 39.7 & 440
\end{pmatrix}
\end{equation}

For the 24-site FMO, the Hamiltonian is given in \cite{zofe_fmo2}, and has the form:
\begin{equation}\label{24site_full}
\begin{pmatrix}
\underline{\underline{h}}_{A} & \underline{\underline{h}}_{B} & \underline{\underline{h}}_{B}^{\dagger} \\
\underline{\underline{h}}_{B}^{\dagger} & \underline{\underline{h}}_{A} & \underline{\underline{h}}_{B} \\
\underline{\underline{h}}_{B} & \underline{\underline{h}}_{B}^{\dagger} & \underline{\underline{h}}_{A}
\end{pmatrix}
\end{equation}

The matrix $\underline{\underline{h}}_{A}$ gives the intra-monomer couplings and site energies. The off-diagonal elements (couplings) are (energies in cm$^{-1}$):
\begin{equation}\label{24site_Hsys}
\begin{pmatrix}
 & -80.3 & 3.5 & -4.0 & 4.5 & -10.2 & -4.9 & 21.0 \\
 -80.3 &  & 23.5 & 6.7 & 0.5 & 7.5 & 1.5 & 3.3 \\
 3.5 & 23.5 &  & -49.8 & -1.5 & -6.5 & 1.2 & 0.7 \\
 -4.0 & 6.7 & -49.8 &  & -63.4 & -13.3 & -42.2 & -1.2 \\
 4.5 & 0.5 & -1.5 & -63.4 &  & 55.8 & 4.7 & 2.8 \\
 -10.2 & 7.5 & -6.5 & -13.3 & 55.8 &  & 33.0 & -7.3 \\
 -4.9 & 1.5 & 1.2 & -42.2 & 4.7 & 33.0 &  & -8.7 \\
 21.0 & 3.3 & 0.7 & -1.2 & 2.8 & -7.3 & -8.7 & 
\end{pmatrix}
\end{equation}

The diagonal elements (site energies) found using the Olbrich \emph{et al.} (Olb, \cite{olbrich}) and Schmidt am Busch \emph{et al.} (SaB, \cite{schmidt_am_busch}) studies are (energies in cm$^{-1}$):
\begin{equation}\label{24site_sites}
\begin{matrix}
Site & \vline & 1 & 2 & 3 & 4 & 5 & 6 & 7 & 8 \\ \hline
\text{Olb} & \vline & 186 & 81 & 0 & 113 & 65 & 89 & 492 & 218 \\
\text{SaB} & \vline & 310 & 230 & 0 & 180 & 405 & 320 & 270 & 505
\end{matrix}
\end{equation}\\[3cm]

The matrix $\underline{\underline{h}}_{B}$ gives the inter-monomer couplings (energies in cm$^{-1}$):
\begin{equation}\label{24site_inter}
\begin{pmatrix}
1.0 & 0.3 & -0.6 & 0.7 & 2.3 & 1.5 & 0.9 & 0.1 \\
1.5 & -0.4 & -2.5 & -1.5 & 7.4 & 5.2 & 1.5 & 0.7 \\
1.4 & 0.1 & -2.7 & 5.7 & 4.6 & 2.3 & 4.0 & 0.8 \\
0.3 & 0.5 & 0.7 & 1.9 & -0.6 & -0.4 & 1.9 & -0.8 \\
0.7 & 0.9 & 1.1 & -0.1 & 1.8 & 0.1 & -0.7 & 1.3 \\
0.1 & 0.7 & 0.8 & 1.4 & -1.4 & -1.5 & 1.6 & -1.0 \\
0.3 & 0.2 & -0.7 & 4.8 & -1.6 & 0.1 & 5.7 & -2.3 \\
0.1 & 0.6 & 1.5 & -1.1 & 4.0 & -3.1 & -5.2 & 3.6
\end{pmatrix}
\end{equation}\\[12cm]

\subsection{Structured Spectral Density}

The structured spectral density in Fig. \ref{klein_specdens} is given by Eqn. \eqref{kl_spectral}, the parameters for which are given by the table below \cite{kleinekathoefer}:
\begin{equation}\label{klein_spectrum_data}
\begin{matrix}
k & \vline & c_{jk} / \text{cm$^{-2}$} & \vline & \omega_{jk} / \text{cm$^{-1}$} & \vline & \gamma_{jk} / \text{cm$^{-1}$} \\ \hline
1  & \vline & 7091.9081 & \vline & 0.0000    & \vline & 86.4772\\
2  & \vline & 2478.5629 & \vline & 0.0000    & \vline & 11.5643\\
3  & \vline & 617.8502  & \vline & 1813.8341 & \vline & 16.3525\\
4  & \vline & 2961.7657 & \vline & 1752.8329 & \vline & 22.4542\\
5  & \vline & 642.2013  & \vline & 1692.3597 & \vline & 15.1094\\
6  & \vline & 833.1906  & \vline & 1600.5955 & \vline & 21.9446\\
7  & \vline & 1108.6926 & \vline & 1453.4383 & \vline & 67.8879\\
8  & \vline & 4038.4676 & \vline & 817.5591  & \vline & 898.2804\\
9  & \vline & 957.3336  & \vline & 1109.2920 & \vline & 47.6728\\
10 & \vline & 683.7415  & \vline & 721.8440  & \vline & 51.0171\\
11 & \vline & 593.0216  & \vline & 605.5993  & \vline & 19.7612\\
12 & \vline & 1160.2597 & \vline & 366.4331  & \vline & 50.8997\\
13 & \vline & 97.4045   & \vline & 972.4901  & \vline & 17.8089\\
14 & \vline & 454.0769  & \vline & 501.7510  & \vline & 47.7413\\
15 & \vline & 725.2817  & \vline & 243.7265  & \vline & 59.1580
\end{matrix}
\end{equation}

The spectral density of Fig. \ref{highfreq_1ps} (b) (dashed line) is given by dropping the last five entries in Eqn. \eqref{klein_spectrum_data} and scaling all $\eta_{jk}$ values by a factor of 1.1194 (to give the correct reorganization energy).



\begin{thebibliography}{99}\addcontentsline{toc}{chapter}{Bibliography}
\bibliographystyle{unsrt}

\fancyhead{}
\renewcommand{\sectionmark}[1]{\markright{\thesection.\ #1}}
\fancyhead[LO,RE]{\rightmark}


\bibitem{blankenship}
R. E. Blankenship, \emph{Molecular Mechanisms Of Photosynthesis} (Blackwell Science, Oxford, 2002)
\bibitem{renger_review}
T. Renger, \emph{Photosynth. Res.}, \textbf{102}, 471 (2009)
\bibitem{fleming_lightharv}
Y. Cheng and G. R. Fleming, \emph{Annu. Rev. Phys. Chem.}, \textbf{60}, 241 (2009)
\bibitem{coherence}
G. S. Engel, T. R. Calhoun, E. L. Read, T. Ahn, T. Man\v cal, Y. Cheng, R. E. Blankenship and G. R. Fleming, \emph{Nature}, \textbf{446}, 782 (2007)
\bibitem{coherence2}
E. Collini, C. Y. Wong, K. E. Wilk, P. M. G. Curmi, P. Brumer and G. D. Scholes, \emph{Nature}, \textbf{463}, 644 (2010)
\bibitem{ishizakifleming}
A. Ishizaki and G. R. Fleming, \emph{Proc. Natl. Acad. Sci.}, \textbf{106}, 17255 (2009)
\bibitem{schulten_LH2}
J. Str\"umpfer and K. Schulten, \emph{J. Chem. Phys.}, \textbf{131}, 225101 (2009)
\bibitem{2d_fmo}
B. Hein, C. Kreisbeck, T. Kramer and M. Rodr\'iguez, \emph{New. J. Phys.}, \textbf{14}, 023018 (2012)
\bibitem{fenna_matthews_olson}
R. E. Fenna, B. W. Matthews, J. M. Olson and E. K. Shaw, \emph{J. Mol. Biol.}, \textbf{84}, 231 (1974)
\bibitem{fenna_matthews}
R. E. Fenna and B. W. Matthews, \emph{Nature}, \textbf{258}, 573 (1975)
\bibitem{eight_site}
D. E. Tronrud, J. Wen, L. Gay and R. E. Blankenship, \emph{Photosynth. Res.}, \textbf{100}, 79 (2009)
\bibitem{schmidt_am_busch}
M. Schmidt am Busch, F. M\"uh, M. E. Madjet and T. Renger, \emph{J. Phys. Chem. Lett.}, \textbf{2}, 93 (2011)
\bibitem{adolphs_renger}
J. Adolphs and T. Renger, \emph{Biophys. J.}, \textbf{91}, 2778 (2006)
\bibitem{fluor_lifetime}
R. J. W. Louwe and T. J. Aartsma, \emph{J. Phys. Chem. B}, \textbf{101}, 7221 (1997)
\bibitem{tanimura1}
Y. Tanimura and R. Kubo, \emph{J. Phys. Soc. Jpn.}, \textbf{58}, 101 (1989)
\bibitem{tanimura2}
A. Ishizaki and Y. Tanimura, \emph{J. Phys. Soc. Jpn.}, \textbf{74}, 3131 (2005)
\bibitem{influence}
R. P. Feynman and F. L. Vernon Jr., \emph{Ann. Phys. (New York)}, \textbf{24}, 118 (1963)
\bibitem{feynman}
R. P. Feynman and A. R. Hibbs, \emph{Quantum Mechanics And Path Integrals}, Emended Edition (Dover, New York, 2010)
\bibitem{density_matrix}
U. Fano, \emph{Rev. Mod. Phys.}, \textbf{29}, 74 (1957)
\bibitem{heom1}
A. Ishizaki and G. R. Fleming, \emph{J. Chem. Phys.}, \textbf{130}, 234111 (2009)
\bibitem{irreversible}
A. Royer, in \emph{Lecture Notes In Physics: Irreversible Quantum Dynamics}, pp. 47-63 (Springer, London, 2003)
\bibitem{dirac}
P. A. M. Dirac, \emph{The Principles Of Quantum Mechanics}, 4th Edition (Oxford University Press, 1958)
\bibitem{quantum_langevin}
G. W. Ford and M. Kac, \emph{J. Stat. Phys.}, \textbf{46}, 803 (1987)
\bibitem{kubo}
R. Kubo, M. Toda and N. Hashitsume, \emph{Statistical Physics II: Nonequilibrium Statistical Mechanics}, pp. 22-27 (Springer, Berlin, 1985)
\bibitem{filtering1}
Q. Shi, L. Chen, G. Nan, R. Xu and Y. Yan, \emph{J. Chem. Phys.}, \textbf{130}, 084105 (2009)
\bibitem{proton_transfer}
L. Chen and Q. Shi, \emph{J. Chem. Phys.}, \textbf{130}, 134505 (2009)
\bibitem{spin_boson}
M. Thoss, H. Wang and W. H. Miller, \emph{J. Chem. Phys.}, \textbf{115}, 2991 (2001)
\bibitem{weiss}
U. Weiss, \emph{Quantum Dissipative Systems}, 2nd Edition (World Scientific, Singapore, 1999)
\bibitem{variational}
D. P. S. McCutchen, N. S. Dattani, E. M. Gauger, B. W. Lovett and A. Nazir, \emph{Phys. Rev. B}, \textbf{84}, 081305 (2011)
\bibitem{redfield1}
A. G. Redfield, \emph{IBM J. Res. Dev.}, \textbf{1}, 19 (1957)
\bibitem{forster_original}
Th. F\"orster, in \emph{Modern Quantum Chemistry, Part III: Action Of Light And Organic Crystals}, pp. 120-137 (Academic Press, New York, 1965)
\bibitem{forster1}
Th. F\"orster, \emph{Ann. der Phys.}, \textbf{437}, 55 (1948)
\bibitem{forster2}
Th. F\"orster, \emph{Discuss. Faraday Soc.}, \textbf{27}, 7 (1959)
\bibitem{zofe1}
J. Roden, W. T. Strunz and A. Eisfeld, \emph{J. Chem. Phys.}, \textbf{134}, 034902 (2011)
\bibitem{redfield_forster}
M. Yang and G. R. Fleming, \emph{Chem. Phys.}, \textbf{282}, 163 (2002)
\bibitem{ishizaki_review}
A. Ishizaki, T. R. Calhoun, G. S. Schlau-Cohen and G. R. Fleming, \emph{Phys. Chem. Chem. Phys.}, \textbf{12}, 7319 (2010)
\bibitem{leegwater_klug}
J. A. Leegwater, J. R. Durrant and D. R. Klug, \emph{J. Phys. Chem. B}, \textbf{101}, 7205 (1997)
\bibitem{i&f_redfield}
A. Ishizaki and G. R. Fleming, \emph{J. Chem. Phys.}, \textbf{130}, 234110 (2009)
\bibitem{tanimura_redfield}
A. Ishizaki and Y. Tanimura, \emph{Chem. Phys.}, \textbf{347}, 185 (2007)
\bibitem{modified_redfield}
M. Schr\"oder, M. Schreiber, U. Kleinekath\"ofer, \emph{J. Lumin.}, \textbf{125}, 126 (2007)
\bibitem{dexter}
D. L. Dexter, \emph{J. Chem. Phys.}, \textbf{21}, 836 (1953)
\bibitem{RET}
G. D. Scholes, \emph{Annu. Rev. Phys. Chem.}, \textbf{54}, 57 (2003)
\bibitem{silbey}
S. Jang, Y. Jung, R. J. Silbey, \emph{Chem. Phys.}, \textbf{275}, 319 (2002)
\bibitem{mukamel}
S. Mukamel, \emph{Principles Of Nonlinear Optical Spectroscopy}, pp. 212-220, 250-251, 2nd Edition (Oxford University Press, 1995)
\bibitem{cumulant}
R. Kubo, \emph{J. Phys. Soc. Jpn.}, \textbf{17}, 1100 (1962)
\bibitem{zofe_fmo1}
G. Ritschel, J. Roden, W. T. Strunz and A. Eisfeld, \emph{New. J. Phys.}, \textbf{13}, 113034 (2011)
\bibitem{zofe_fmo2}
G. Ritschel, J. Roden, W. T. Strunz, A. Aspuru-Guzik and A. Eisfeld, \emph{J. Phys. Chem. Lett.}, \textbf{2}, 2912 (2011)
\bibitem{numerical_recipes}
W. H. Press, B. P. Flannery, S. A. Teukolsky and W. T. Vetterling, \emph{Numerical Recipes (FORTRAN Version)} (Cambridge University Press, 1989)
\bibitem{filtering2}
Q. Shi, L. Chen, G. Nan, R. Xu and Y. Yan, \emph{J. Chem. Phys.}, \textbf{130}, 164518 (2009)
\bibitem{complex_analysis}
H. A. Priestley, \emph{Introduction To Complex Analysis} (Clarendon Press, Oxford, 1985)
\bibitem{pade1}
J. Hu, R. Xu and Y. Yan, \emph{J. Chem. Phys.}, \textbf{133}, 101106 (2010)
\bibitem{pade2}
J. Hu, M. Luo, F. Jiang, R. Xu and Y. Yan, \emph{J. Chem. Phys.}, \textbf{134}, 244106 (2011)
\bibitem{meier_tannor}
C. Meier and D. J. Tannor, \emph{J. Chem. Phys.}, \textbf{111}, 3365 (1999)
\bibitem{pade3}
J. Ding, J. Xu, J. Hu, R. Xu and Y. Yan, \emph{J. Chem. Phys.}, \textbf{135}, 164107 (2011)
\bibitem{chromophore_solvent}
G. R. Fleming and M. Cho, \emph{Annu. Rev. Phys. Chem.}, \textbf{47}, 109 (1996)
\bibitem{olbrich}
C. Olbrich, T. L. C. Jansen, J. Liebers, M. Aghtar, J. Str\"umpfer, K. Schulten, J. Knoester and U. Kleinekath\"ofer, \emph{J. Phys. Chem. B}, \textbf{115}, 8609 (2011)
\bibitem{static_disorder}
D. R. Buck, S. Savikhin and W. S. Struve, \emph{J. Phys. Chem. B}, \textbf{101}, 8395 (1997)
\bibitem{kleinekathoefer}
C. Olbrich, J. Str\"umpfer, K. Schulten and U. Kleinekath\"ofer, \emph{J. Phys. Chem. Lett.}, \textbf{2}, 1771 (2011)
\bibitem{read_fleming}
E. L. Read, G. S. Schlau-Cohen, G. S. Engel, J. Wen, R. E. Blankenship and G. R. Fleming, \emph{Biophys. J.}, \textbf{95}, 847 (2008)
\bibitem{nalbach_thorwart}
P. Nalbach, D. Braun and M. Thorwart, \emph{Phys. Rev. E}, \textbf{84}, 041926 (2011)
\bibitem{quant_coherence_assists}
J. Str\"umpfer, M. \c Sener and K. Schulten, \emph{J. Phys. Chem. Lett.}, \textbf{3}, 536 (2012)
\bibitem{shim_rebentrost_spectral}
S. Shim, P. Rebentrost, S. Valleau and A. Aspuru-Guzik, \emph{Biophys. J.}, \textbf{102}, 649 (2012)
\bibitem{continued_fraction}
T. Ozaki, \emph{Phys. Rev. B}, \textbf{75}, 035123 (2007)
\end{thebibliography}
\end{document}